\documentclass[12pt]{article}

\usepackage{pst-pdf}
\usepackage{graphicx}
\usepackage{amssymb}
\usepackage{amsmath}
\usepackage{srcltx}
\usepackage{color}
\usepackage{cite}
\usepackage{url}
\usepackage{ntheorem}
\usepackage{psfrag}
\usepackage[bookmarks,colorlinks=false]{hyperref}

\newcommand{\htt}{h_{tt}}
\newcommand{\hrr}{h_{rr}}
\newcommand{\hpsipsi}{h_{\psi \psi}}
\newcommand{\hthetatheta}{h_{\theta\theta}}
\newcommand{\htpsi}{h_{t\psi}}
\newcommand{\htphi}{h_{t\phi}}
\newcommand{\hphiphi}{h_{\varphi \varphi}}
\newcommand{\hphipsi}{h_{\varphi \psi}}

\newcommand{\Sing}{\mathrm{Sing}}

\newtheorem{Theorem} {\sc  Theorem\rm} [section]

\newtheorem{Lemma} [Theorem] {\sc  Lemma\rm}
\newtheorem{Proposition} [Theorem] {\sc  Proposition\rm}
\newtheorem{prop} [Theorem] {\sc  Proposition\rm}

\theorembodyfont{\upshape}

\newtheorem{remark}[Theorem]{\sc  Remark\rm}

\theoremsymbol{\ensuremath{\diamondsuit}}
\newcommand{\fcoco}{\small}
\theorembodyfont{\fcoco}\theoremseparator{}\theoremindent0.5cm

\theoremstyle{nonumberplain}\theorembodyfont{\fcoco}
\theoremseparator{}\theoremindent0.5cm

\DeclareFontFamily{OT1}{rsfs}{}
\DeclareFontShape{OT1}{rsfs}{m}{n}{ <-7> rsfs5 <7-10> rsfs7 <10-> rsfs10}{}
\DeclareMathAlphabet{\mycal}{OT1}{rsfs}{m}{n}

{\catcode `\@=11 \global\let\AddToReset=\@addtoreset}
\AddToReset{equation}{section}

\newcounter{mnotecount}[section]

\AddToReset{figure}{section}
\renewcommand{\themnotecount}{\thesection.\arabic{mnotecount}}

\newcommand{\mnotex}[1]
{\protect{\stepcounter{mnotecount}}$^{\mbox{\footnotesize
$
\bullet$\themnotecount}}$ \marginpar{
\raggedright\tiny\em
$\!\!\!\!\!\!\,\bullet$\themnotecount: #1} }

\newcommand{\ub}[1]{\underbrace{#1}}

\newcommand{\jlcax}[1]{}
\newcommand{\eean}{\nonumber\end{eqnarray}}

\newcommand{\ii}{\mathrm{i}}

\newcommand{\kk}[1]{}

\newcommand{\Span}{\mathrm{Span}}

\newcommand{\beq}{\begin{equation}}

\newcommand{\T}{\Bbb T}

\newcommand{\FS}       
                  {F}

\newcommand{\HS} 
       {H_{\mbox{\scriptsize volume}}}

{\ptc{this should be removed in the oberwolfach version}}%

\newcommand{\mcA}{{\mycal A}}%
\newcommand{\eeal}[1]{\label{#1}\end{eqnarray}}
\newcommand{\C}{{\mathbb C}}
\newcommand{\bed}{\begin{deqarr}}
\newcommand{\eed}{\end{deqarr}}
\newcommand{\bedl}[1]{\begin{deqarr}\label{#1}}
\newcommand{\eedl}[2]{\arrlabel{#1}\label{#2}\end{deqarr}}

\newcommand{\mcU}{{\mycal U}}

\newcommand{\bel}[1]{\begin{equation}\label{#1}}
\newcommand{\bea}{\begin{eqnarray}}
\newcommand{\bean}{\begin{eqnarray}\nonumber}
\newcommand{\beal}[1]{\begin{eqnarray}\label{#1}}
\newcommand{\eea}{\end{eqnarray}}

\newcommand{\qed}{\hfill $\Box$ \medskip}
\newcommand{\proof}{\noindent {\sc Proof:\ }}
\newcommand{\be}{\begin{equation}}
\newcommand{\eeq}{\end{equation}}
\newcommand{\ee}{\end{equation}}
\newcommand{\beqa}{\begin{eqnarray}}
\newcommand{\eeqa}{\end{eqnarray}}
\newcommand{\beqan}{\begin{eqnarray*}}
\newcommand{\eeqan}{\end{eqnarray*}}
\newcommand{\ba}{\begin{array}}
\newcommand{\ea}{\end{array}}

\newcommand{\scri}{{\mycal I}}%
\newcommand{\scrip}{\scri^{+}}%
\newcommand{\scrim}{\scri^{-}}%

\newcommand{\mnote}[1]
{\protect{\stepcounter{mnotecount}}$^{\mbox{\footnotesize
$
\bullet$\themnotecount}}$ \marginpar{
\raggedright\tiny\em
$\!\!\!\!\!\!\,\bullet$\themnotecount: #1} }

\newcommand{\warn}[1]
{\protect{\stepcounter{mnotecount}}$^{\mbox{\footnotesize
$
\bullet$\themnotecount}}$ \marginpar{
\raggedright\tiny\em
$\!\!\!\!\!\!\,\bullet$\themnotecount: {\bf Warning:} #1} }

\newcommand{\R}{\mathbb R}

\newcommand{\eq}[1]{(\ref{#1})}

\newcommand{\Mext}{M_\ext}
\newcommand{\ext}{\mathrm{ext}}

\newcommand{\ptc}[1]{\mnote{{\bf ptc:}#1}}

\newcommand{\beqar}{\begin{deqarr}}
\newcommand{\eeqar}{\end{deqarr}}

\newcommand{\beaa}{\begin{eqnarray*}}
\newcommand{\eeaa}{\end{eqnarray*}}

\newcommand{\Proof}{\proof}

\newcommand{\sign}{\mathrm{sign}}

\begin{document}
\title{On the global structure of the Pomeransky-Senkov black holes}

\author{Piotr T.~Chru\'sciel\thanks{The authors are
grateful to the Mittag-Leffler Institute, Djursholm, Sweden,
for hospitality and financial support during part of work on
this paper. PTC was supported in part by the Polish Ministry of
Science and Higher Education grant Nr N N201 372736, and by the
EPSRC Science and Innovation award to the Oxford Centre for
Nonlinear PDE (EP/E035027/1). AGP is supported by the Research Centre of
Mathematics of the University of Minho (Portugal) through the ``Funda\c{c}\~ao para a Ci\^encia e a Tecnologia'' (FCT) Pluriannual
Funding Program. Finally we thank the ``Albert Einstein Institut f\"ur Gravitationsphysik'' in Golm (Germany), for hospitality during
the completion of some parts of this work and financial support.}
\\
LMPT, F\'ed\'eration Denis Poisson, Tours \\ Mathematical
Institute and Hertford College, Oxford \\
\\
Julien Cortier \\ Institut de Math\'ematiques et de
Mod\'elisation de Montpellier\\ Universit\'e Montpellier 2
\\
\\
Alfonso Garc\'{\i}a-Parrado G\'omez-Lobo\\
Centro de Matem\'atica, Universidade do Minho\\
4710-057 Braga, Portugal}
\maketitle{}
\date{}
\begin{abstract}
We construct  analytic extensions of the Pomeransky-Senkov
metrics with multiple Killing horizons and asymptotic regions.
We show that, in our extensions, the singularities associated
to an obstruction to differentiability of the metric lie beyond
event horizons. We analyze the topology of the non-empty
singular set, which turns out to be parameter-dependent. We
present numerical evidence for stable causality of the domain
of outer communications. The resulting global structure is
somewhat reminiscent of that of Kerr space-time.
\end{abstract}
%

\tableofcontents

\section{Introduction}
\label{s18IX0.1}

In~\cite{PS} a family of five-dimensional vacuum
black-hole-candidate metrics has been presented:%
\footnote{We use $(\psi,\varphi)$ where Pomeransky \& Senkov
use $(\varphi,\psi)$.}
\begin{eqnarray}
ds^2 &= & \frac{2 H(x,y)k^2}
{(1-\nu)^2(x-y)^2}\left(\frac{dx^2}{G(x)}-
\frac{dy^2}{G(y)}\right)  -2\frac{ J(x,y)}{H(y,x)}d\varphi d\psi\nonumber\\
&&-
\frac{H(y,x)}{H(x,y)}(dt+\Omega)^2-\frac{F(x,y)}{H(y,x)}d\psi^2+
\frac{F(y,x)}{H(y,x)}d\varphi^2\; ,
\label{eq:line-element1}
\end{eqnarray}
where
\begin{eqnarray*}
H(x,y)=\lambda ^2+2 \nu  \left(1-x^2\right) y \lambda +2
   x \left(1-\nu ^2 y^2\right) \lambda -\nu ^2+\nu
   \left(-\lambda ^2-\nu ^2+1\right) x^2 y^2+1,\\
F(x,y)=\frac{2 k^2}{(x-y)^2 (1-\nu)^2}
\left(\left(1-y^2\right) \left(\left((1-\nu
   )^2-\lambda ^2\right) (\nu +1)+\right.\right.\\
\left.\left.+y \lambda  \left(-\lambda ^2-3
   \nu ^2+2 \nu +1\right)\right) G(x)+\right.\\
\left.+\left(-(1-\nu ) \nu
   \left(\lambda ^2+\nu ^2-1\right) x^4+\lambda  \left(2 \nu
   ^3-3 \nu ^2-\lambda ^2+1\right) x^3+\left((1-\nu )^2-\lambda
   ^2\right) (\nu +1) x^2+\right.\right.\\
\left.\left.+\lambda  \left(\lambda ^2+(1-\nu
   )^2\right) x+2 \lambda ^2\right) G(y)\right),\\
J(x,y)=\frac{2 k^2 \left(1-x^2\right) \left(1-y^2\right)
   \lambda  \sqrt{\nu } \left(\lambda ^2+2 (x+y) \nu  \lambda
   -\nu ^2-x y \nu  \left(-\lambda ^2-\nu
   ^2+1\right)+1\right)}{(x-y) (1-\nu )^2},\\
G(x)=\left(1-x^2\right) \left(\nu  x^2+\lambda  x+1\right)
 \;,
\end{eqnarray*}
and where $\Omega$ is a 1-form given by
$$
\Omega=M(x,y)d\psi+P(x,y)d\varphi
 \;,
$$
with
\begin{eqnarray*}
M(x,y)
&=& \frac{2 k\lambda\sqrt{(\nu +1)^2-\lambda
   ^2} (y+1) (-\lambda +\nu -2 \nu  x+\nu  x ((\lambda
   +\nu -1) x+2) y-1)}{(1-\lambda +\nu ) H(y,x)}
   \\
&=:& \frac{\sqrt{(\nu +1)^2-\lambda
   ^2}\hat M(x,y)}{ (1-\lambda +\nu )  H(y,x)}
\\
P(x,y)
 &= &\frac{2k\lambda  \sqrt{\nu } \sqrt{(\nu
   +1)^2-\lambda ^2} \left(x^2-1\right) y}{H(y,x)}
\\
 &=: &\frac{2\sqrt \nu \hat P(x,y)}{H(y,x)}
   \;,
\end{eqnarray*}
where $\hat P$ and $\hat M$ are polynomials in all variables.

\begin{figure}
\begin{center}
\psfrag{fvar}{$\nu$}
\includegraphics[width=.4\textwidth]{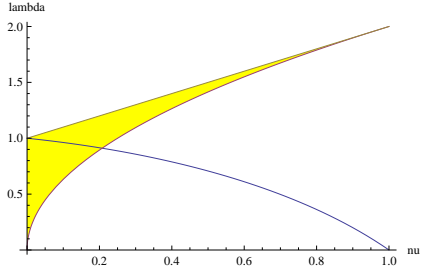}
 \caption{The parameters
$\nu$ and $\lambda$ belong to the yellow-shaded region
bounded by the vertical axis and by the two increasing graphs.
The decreasing graph is the function $\lambda_k(\nu)$ of the proof of
 Proposition~\ref{P4V.2}. \label{Udef4V09}}
\end{center}
\end{figure}

The parameter $k$ is assumed to be in $\R^*$, while the
parameters $\lambda$ and $\nu$ have been restricted
in~\cite{PS} to belong to the set%
\footnote{Strictly speaking, $\nu=0$ is allowed in~\cite{PS}.
It is shown there that this corresponds to Emparan-Reall
metrics (compare Appendix~\ref{AERPS}), which have already been
analysed elsewhere~\cite{ChCo}, and so we only consider
$\nu>0$.}
\bel{3IX.1}  \mcU:=\{(\nu,\lambda):\ \nu\in (0,1)\;,\
2\sqrt{\nu}\le \lambda< 1 +\nu\}
  \;,
\ee
which is the region between the two increasing curves of
Figure~\ref{Udef4V09}. The coordinates $x$, $y$, $\phi$,
$\psi$, $t$ vary within the ranges $-1\leq x\leq 1$,
$-\infty<y<-1$, $0\leq\phi\leq 2\pi$, $0\leq\psi\leq 2\pi$ and
$-\infty<t<\infty$.

Studies of the properties of the metric
(\ref{eq:line-element1}) are available in the literature: some
physical properties of this solution can be found
in~\cite{ElvangRodriguez},  and a study of some of its
geodesics has been performed in~\cite{Durkee}. However, there
remain unanswered questions dealing with global properties of
the solution: causality, existence and topology of event
horizons, absence of naked singularities, and structure of the
singular set.

One thus wishes to understand the global structure of the
corresponding space-time, and its possible extensions. In
particular, one wishes to analyze the zeros of the denominators
of the metric functions, and the zeros of the metric functions
themselves, and their consequences for the space-time. The
zeros of $G(y)$ are obvious, given by $y= \pm 1$ and
$$
y_h:= -\frac{\lambda-\sqrt{\lambda^2-4\nu}}{2\nu} \;, \quad y_c:=
-\frac{\lambda+\sqrt{\lambda^2-4\nu}}{2\nu} \;.
$$
These quantities are real for values of $\lambda$, $\nu$ belonging to $\mcU$ and we have
$y_c\leq y_h$. All these considerations lead naturally to the definition
\bel{Om} \Omega_0\equiv\{(x,y,\nu,\lambda) \in \mathbb{R}^4\ ;\
-1\leq x\leq 1\;,\ y_c\leq y< -1\;,\ 0<\nu<1\;,\
2\sqrt{\nu}\leq \lambda < 1+ \nu \}
 \;,
\ee
see Figure~\ref{F8IX.2}. One then wishes to know the following:
\begin{figure}[h]
\begin{center}
\includegraphics[width=.5\textwidth]{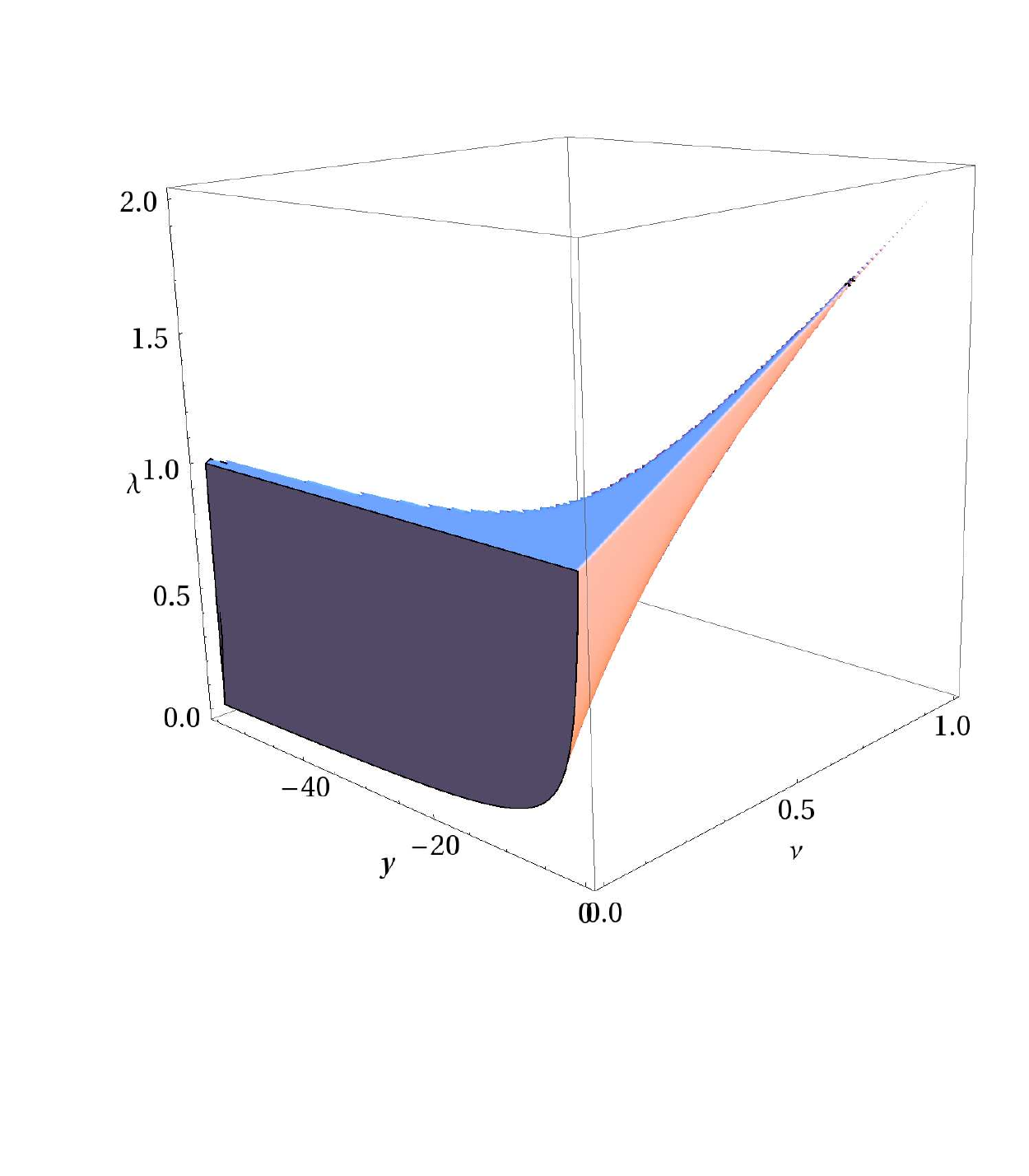}
 \caption{\label{F8IX.2} Points in $\Omega_0$ with $y_c<y<y_h$; the $x$ variable has been suppressed.
}
\end{center}
\end{figure}

\begin{enumerate}
 \item Do the denominators of the metric functions have
     zeros in $ \Omega_0$?

\item One expects that the hypersurface $\{y=y_h\}$ is an
    event horizon, and that the hypersurface $\{y=y_c\}$ is
    a Cauchy horizon. Is this the case?

\item Are the space-times so defined extendible?

\item Do they represent suitably regular black-hole
    spacetimes, as implicit in~\cite{PS}, so that e.g. the
    usual classification theory~\cite{HY,HY3,ChCo} applies?

\item  Are the conditions defining the set $\Omega_0$ the
    only possibility for the existence of regular black
    hole spacetimes or is it possible to select different
    values for the parameters and the coordinates such that
    regular black holes are present too?

\end{enumerate}

\begin{figure}
\begin{center}
\includegraphics[width=.45\textwidth]{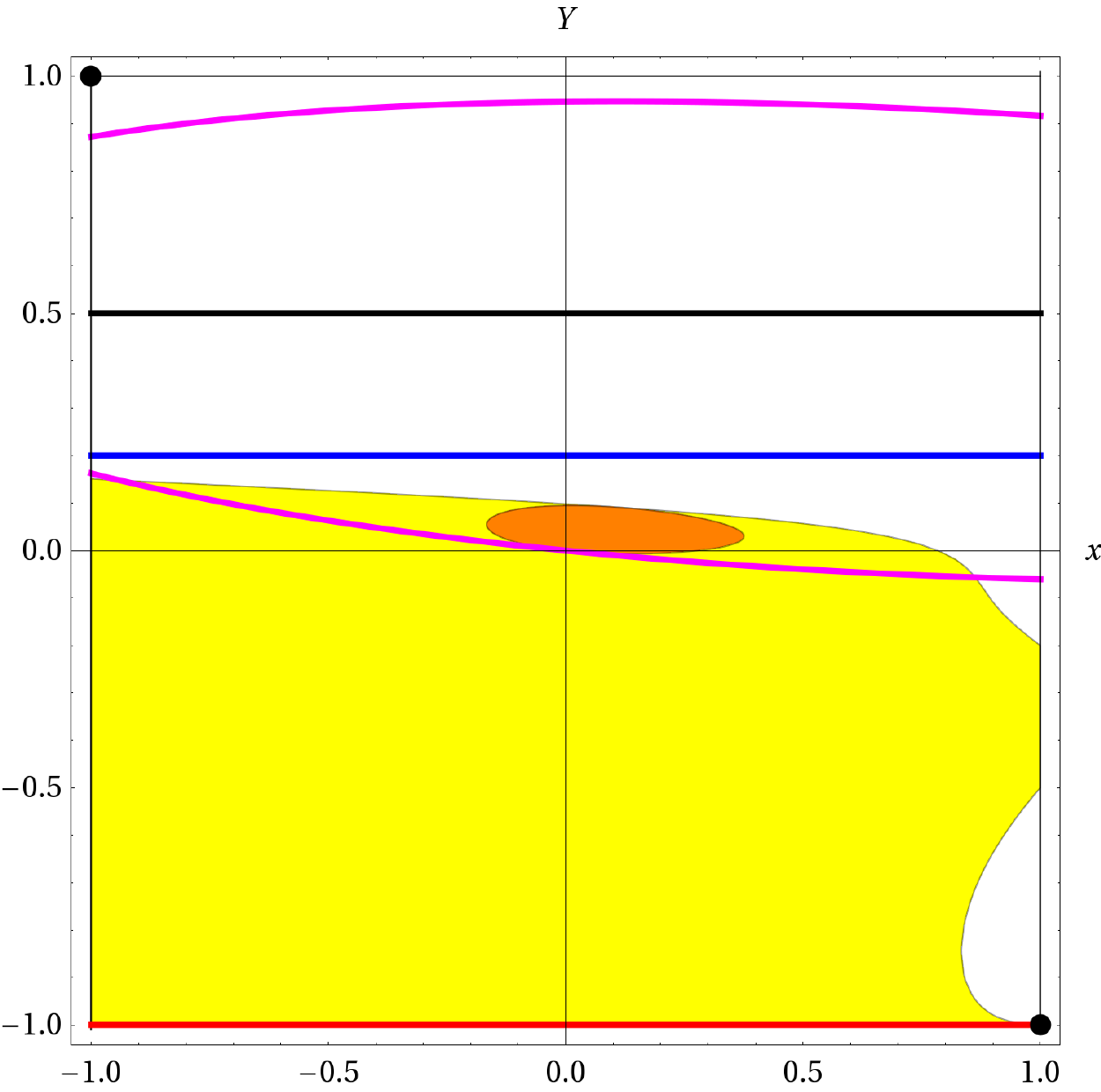}
\includegraphics[width=.45\textwidth]{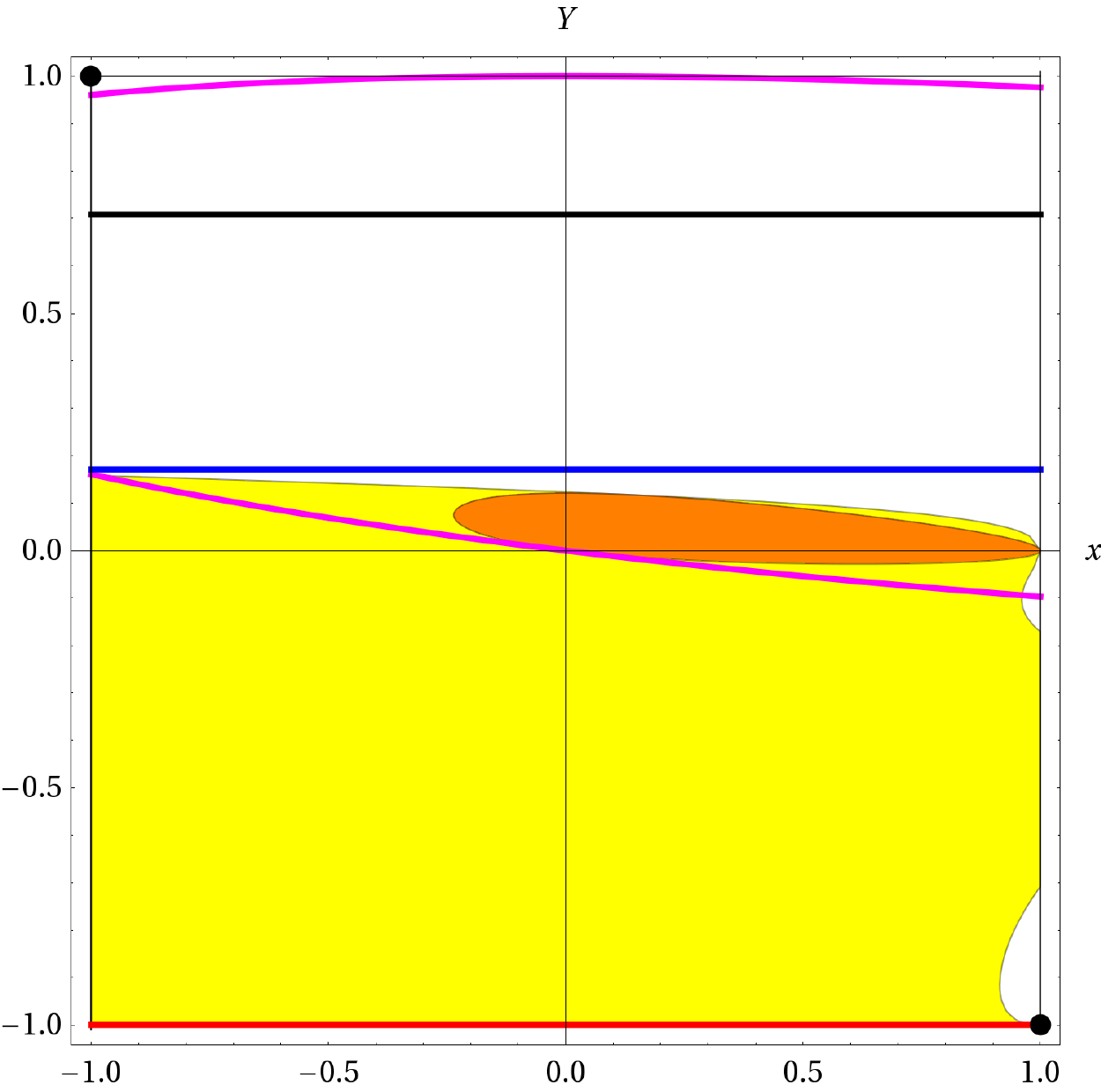}
 \caption{\label{Fsummary}
Some global features of the solutions in $(x,Y=-1/y)$ coordinates: a)
the $\T^3$ singularity with $\nu=.1$ and $\lambda=.7$;
 b) the borderline case $\lambda=1-\nu$
 with $\nu=.121$.
 The boundaries $x=\pm 1$ and $Y=\pm 1$ are rotation axes for suitable Killing vectors, with
 a conical singularity at $Y=-1$ for generic values of $k$, except for the black dot
 at $(-1,1)$ which is the infinity of an asymptotically Minkowskian region,
 and the black dot at $(1,-1)$, which is the infinity of
 a second, non asymptotically Minkowskian, end. The singular set $\{H(x,y)=0\}$ is
the boundary of the orange (darker shaded) region, where $H(x,y)<0$.
 The strong-causality violations occur in the yellow (lightly shaded) region,
where $\det g_{AB}<0$ (see Section~\ref{ss8IX.1}). The black (upper) thick horizontal line
 corresponds to the location of the event horizon, the blue (lower) thick
horizontal line corresponds to an interior Killing horizon and the magenta curves to the ergosurface, $\{H(y,x)=0\}$. These pictures indicate that when $\lambda+\nu<1$ the ergosurface consists of two disconnected {\em rings}
$\R\times S^1\times S^2$ (the {\em inner} ergosurface, intersecting the causality
violating region, and the {\em outer} ergosurface).
 When $\lambda+\nu=1$ the {\em outer} ergosurface becomes
 a ``pinched" $\R\times S^1\times S^2$.}
\end{center}
\end{figure}

\begin{figure}[h]
\begin{center}
\includegraphics[width=.45\textwidth]{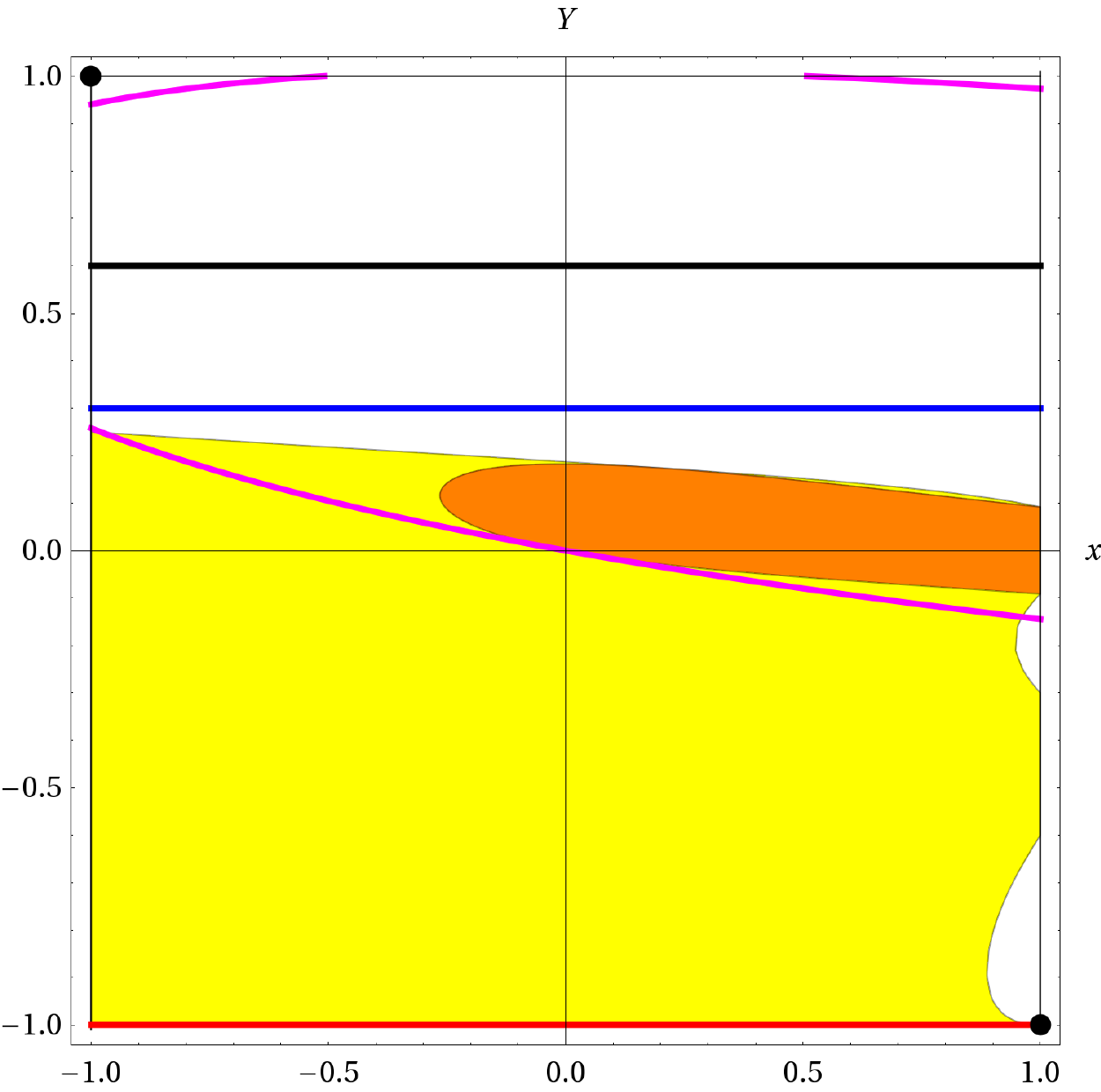}
 \caption{\label{Fsummary2} The global structure in $(x,Y=-1/y)$ coordinates
when $\lambda+\nu>1$: the $S^1\times S^2$ singularity with $\nu=.18$ and $\lambda=.9$
and the ergosurface which now has three connected components. The {\em outer} ergosurface
is the union of two $\R\times S^3$'s  and the inner ergosurface is a ring
$\R\times S^2\times S^1$. Color-coding as in Figure~\ref{Fsummary}.
}
\end{center}
\end{figure}

The aim of this paper is to answer some of those questions: We
show that the metric is smooth and Lorentzian in the range of
coordinates and parameters defined by $\Omega_0$.
We show that
the non-empty set $\{(x,y,\nu,\lambda): H(x,y)=0\}$ does not
intersect $\Omega_0$, and that the metric cannot be $C^2$
continued across this set. We show that there are always
singularities of the Kretschmann scalar somewhere on this set,
and report numerical studies that suggest blow up of the
Kretschmann scalar everywhere there. We construct
Kruskal-Szekeres type extensions of the metric across the
Killing horizons $y=y_c$ and $y=y_h$, and we show that the set
$\{y=y_h\}$ forms the boundary of the domain of outer
communications  (d.o.c.) in our extensions. We present
numerical evidence for stable causality of  the d.o.c.. We
construct an extension of the metric across ``the set
$\{y=-\infty\}$", to a region which contains singularities,
causality violations, and another asymptotic end. In the
extended space-time the singular set has topology $\R\times
\T^3$, or $\R\times S^1 \times S^2$, or a ``pinched $\R\times
S^1 \times S^2$", depending upon the values of the parameters.
Three representative $(x,Y)$ coordinate plots, where $Y=-1/y$
(see Section~\ref{S11X09.1}), illustrating the behavior of the
metric are presented in Figures~\ref{Fsummary}
and~\ref{Fsummary2}. {It should be kept in mind that those
figures do \emph{not} provide a representation of the global
structure of the associated space-time, as the metric is
singular at $Y=-1/y_c$ and $Y=-1/y_h$ in the $(x,Y)$
coordinates: the space-time is constructed from the
$(x,Y)$--coordinates representation by continuation in
appropriate new coordinates, across every non-degenerate
Killing horizon, to three distinct new regions. The global
structure of the resulting analytic extension resembles that of
the Kerr space-time, see Section~\ref{S29X.1}.}
\begin{figure}[h]
\begin{center}
\includegraphics[width=.45\textwidth]{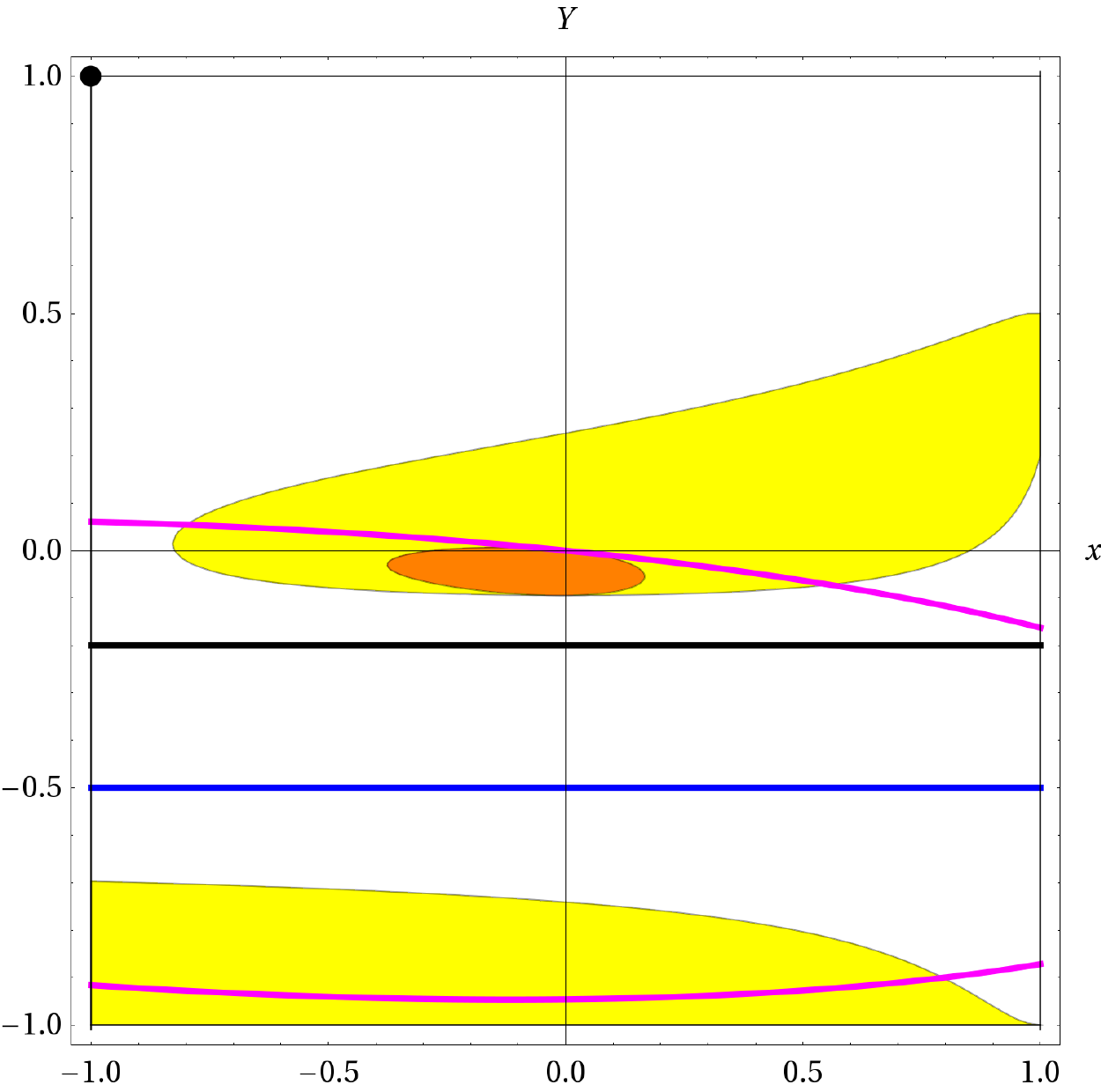}
\includegraphics[width=.45\textwidth]{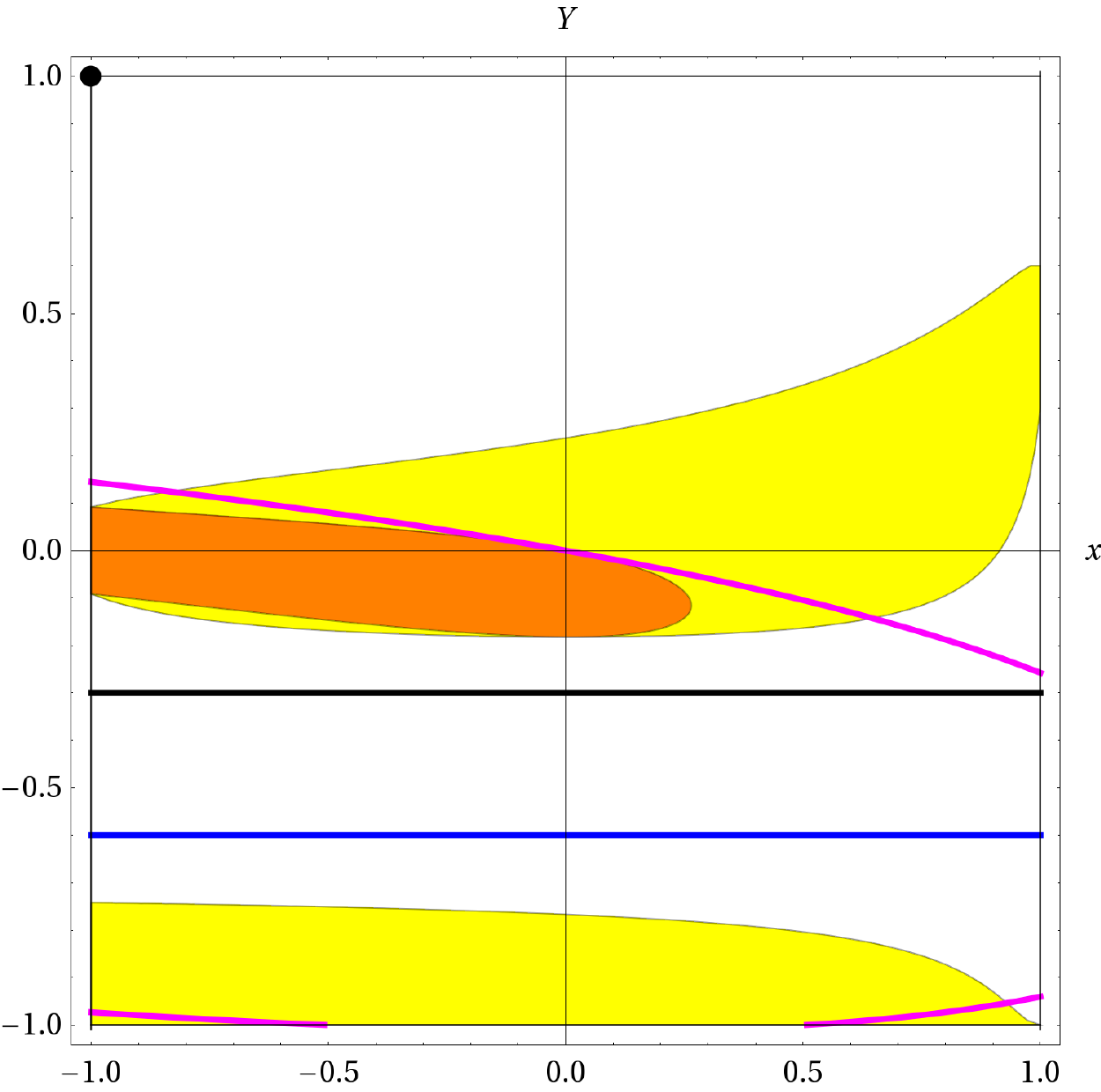}
\caption{\label{Fsummary3}
Killing horizons, ergosurfaces, causality violating regions, and singularities in
spacetimes  with $\nu$ having the same
value as in the spacetimes depicted in the left Figure~\ref{Fsummary} and in
Figure~\ref{Fsummary2} (color coding as in previous figures), but with $\lambda$ here equal
to $-\lambda$ there.
In the current spacetimes the singularity is surrounded by the causality violating region.
}
\end{center}
\end{figure}

Similarly to the analysis of the Emparan-Reall black rings
in~\cite{CC}, one would like to prove that the  d.o.c. is
globally hyperbolic, that it is $I^+$-regular in the sense
of~\cite{ChCo}, and that the extensions described in
Section~\ref{S29X.1} are maximal. One would also like to
understand better the nature of the asymptotic end associated
with $(x=1,Y=-1)$. All those questions require further studies.

An essential tool in the research reported on here was the
tensor manipulation package {\sc xAct}~\cite{xAct}.

\section{Generalities}
 \label{S26X}

In this section we establish some generic properties of the PS
familiy of solutions, some of which will be needed in the rest
of the paper. First of all, the requirement that the metric
(\ref{eq:line-element1}) is real-valued and well-defined
enforces
\bel{11IV10.5}
 0\le \nu\ne 1 \ \mbox{ and } \ -( \nu +1)\le \lambda < \nu +1
  \;.
\ee

Next, we look for the points in the $(x,y)$ plane in which the
signature is Lorentzian. It follows from
(\ref{eq:orthogonal-frame}) that, away from the sets $\{x=y\}$,
$\{H(x,y)=0\}$, $\{H(y,x)=0\}$, $\{F(y,x)=0\}$,  $\{G(x) =0\}$
and $\{ G(y)=0\}$, the metric signature is
\bean \lefteqn{
\big(-\sign(H(y,x)H(x,y))\;,-\sign(G(y)H(x,y))\;,}
 &&
\\
 &&
 -\sign(G(x)H(x,y)F(y,x)G(y))\;,
  \sign(F(y,x)H(y,x)),\sign(G(x)H(x,y))\big)
 \;.
 \nonumber
\\
 &&
 \label{eq:signature-formula}
\eea
However, it follows from \eq{4IX.1} and \eq{7IX.4} that the
signature of the metric can  change at most at $\{G(x)=0\}$,
$\{G(y)=0\}$, $\{x=y\}$, $\{H(y,x)=0\}$ and  $\{H(x,y)=0\}$. It
has been pointed out to us by H.~Elvang (private communication)
that the singularity of the metric at $\{H(y,x)=0\}$ is an
artifact of the parameterisation, see Section~\ref{SHE} for
details, and therefore no signature change can occur across
this set. Similarly, zeros of $F(y,x)$  in
\eqref{eq:signature-formula} (if any) are an artifact of the
choice of frame in (\ref{eq:orthogonal-frame}).
%
%

A numerical study of (\ref{eq:signature-formula}) indicates
that, under the conditions of equation (\ref{11IV10.5}), the
signature is never Lorentzian if either $|x|>1$, $|y|>1$ or
$|x|<1$ and $|y|<1$. Therefore, it is likely that the
coordinates $x$, $y$ must vary within the set $\mathcal D$
defined by
\begin{equation}
\mathcal D\equiv
(\{|x|<1\}\cap\{|y|>1\})\cup(\{|x|>1\}\cap\{|y|<1\})
\label{eq:define-d}
\end{equation}
including possibly parts of its boundaries, which would e.g.
correspond to lower dimensional orbits of the isometry group.

At this stage it is useful to recall the Lichnerowicz theorem,
which asserts in space-time dimension four that the only
stationary, with one asymptotically flat end, well-behaved
solution of the Einstein equations is Minkowski space-time. In
retrospect, this theorem can be viewed as a simple consequence
of the positive energy theorem, regardless of dimension, on
those manifolds, without boundary and with one asymptotically
flat end, on which the rigid positivity of mass holds: Indeed,
as is well known, the ADM mass equals the Komar mass; but the
latter is zero as the divergence of the Komar boundary
integrand is zero. Since the positive energy theorem is true
for all manifolds in dimension five~\cite{SchoenCatini}, there
are no non-trivial, five-dimensional, spatially compact
asymptotically flat metrics containing only one asymptotic end.
As the existence of another asymptotically flat end  leads to
an event horizon, any non-trivial such solution must either
contain event horizons, or regions with non-asymptotically-flat
failure of spatial compactness
(compare~\cite{manderson:stationary}). If we decide that the
latter is undesirable (``naked singularities"), we conclude
that the only solutions of interest are those with horizons. It
is known that within the current class of metrics the existence
of a horizon implies existence of a Killing prehorizon (cf.,
e.g.,~\cite{CarterJMP,CC}); the existence of a Killing horizon
in the interesting solutions follows then from~\cite{ChGstatic}
under the usual global conditions.

Now, Killing horizons require real zeros of the polynomials
$G(x)$ and/or $G(y)$, keeping in mind that the sets $y=\pm 1$
and $x=\pm 1$ are expected to be axes of rotation.
One is then led to require that the polynomial
\begin{equation}
p(\xi)\equiv \nu\xi^2+\lambda\xi+1\;,
\label{eq:define-xi}
\end{equation}
where $\xi$ represents either the variable $x$ or $y$, has real
zeros, denoted by $\xi_{-}\leq\xi_{+}$; this will be the case
if
$$
 2\sqrt{\nu}\le |\lambda|
$$
holds, which we assume
henceforth. We may distinguish now two alternative
possibilities: the case with $0\leq \nu<1$ and the case with
$1<\nu$. In the first possibility it can be shown  that
$|\xi_{\pm}|>1$ whereas the second possibility leads to
$|\xi_{\pm}|<1$ (see Proposition \ref{prop:p-roots} of Appendix
\ref{sec:properties-metric}). Combining this with
(\ref{eq:define-d}) one is led to consider the following
coordinate ranges:

\begin{itemize}
 \item $0<\nu<1$, $|x|< 1$, $|y|>1$. This is the case which we are going to study in this paper, see below
for further explanations.
\item $0<\nu<1$, $|x|>1$, $|y|<1$. We have not been able to
    exclude the possibility of existence of well behaved
    asymptotically flat solutions in this range of
    parameters and coordinates.  
\item $1<\nu$, $|x|< 1$, $|y|> 1$. The transformation
    of Proposition~\ref{prop:ps-invariance} below implies
    that this case is  equivalent to the previous one.
\item $1<\nu$, $|x|>1$, $|y|<1$. Again, using the
    transformation (\ref{eq:ps-symmetry}) we deduce that
    this case is equivalent to the case of the first bullet
    point and hence the considerations there also apply
    here.
\end{itemize}

Now, as already pointed out by Pomeransky and Senkov in
\cite{PS}, and analyzed in detail in
Section~\ref{subsec:asymptotic} below, the point $x=-1$,
$y=-1$, corresponds to an asymptotically flat region for  any
$\lambda$, $\nu$ such that the PS metric is defined.
Anticipating the analysis in Section~\ref{S11X09.1}, the
introduction of the new variable $Y\equiv -1/y$ leads to a
polynomial function $Y^2 H(x,-1/Y,\lambda,\nu)$ that vanishes
at the point $x=0$, $Y=0$, for any $\lambda$, $\nu$; and it
turns out that this point corresponds to a naked singularity of
the metric for all parameters for which the metric is defined.
This, together with the requirement of absence of naked
singularities within the domain of outer communications, leads
to the condition of existence of a negative root of $G(y)$;
equivalently
$$ 0<\lambda< 1 +\nu
 \;.
$$
We have therefore recovered the set \eqref{3IX.1}.

We will say that $(\nu,\lambda)$ are \emph{admissible} if
$(\nu,\lambda)\in \mcU$. Unless otherwise stated, only
admissible values of $\lambda$, $\nu$ will be considered in
this paper.

We finish this section by noting that the PS metrics are
$C^2$--inextendible across $\{H(x,y)=0\}$: this can be seen by
inspection of the norm $g(\partial_t,\partial_t)$ of the
Killing vector $\partial_t$:
$$
 g(\partial_t, \partial_t) = g_{tt} = - \frac{H(y,x)}{H(x,y)}
 \;.
$$
We will see shortly, in Section~\ref{ss7IX.5}, that $H(y,x)$
does not vanish on the set $\{H(x,y)=0\}$. This shows that
$g_{tt}$ is unbounded near this set, and standard arguments
(see, e.g., \cite[Section~4.2]{CC}) show $C^2$-inextendibility
of the metric across the zero level set of $H(x,y)$. Evidence
of a curvature singularity there will be presented in
Section~\ref{SKretschmann}. As already mentioned, points near $x=y=-1$,
with $y\le-1\le x$ and $(x,y)\ne (-1,-1)$, belong to an
asymptotically flat region, where $H(x,y)$ is positive, and
where the signature is Lorentzian. For reasons just explained,
in the associated domain of outer communications we must thus
have $\{H(x,y)>0\}$

To summarise, we want to understand the geometry of the PS
metric in those connected components of the region
\bel{1XI.3}
 \{ x\in [-1,1]\;,\ y\not \in (-1,1)\;, \ H(x,y)>0\}
\ee
which contain $(1,1)$ and $(-1,-1)$ in their closures
and assume admissible values of $(\nu,\lambda)$. This is the region which shall be considered in this paper.



%
%
%

\section{The function $H$}
 \label{sfunH}

The function $H$ appears in the denominators of
 \eq{eq:line-element1} as $H(x,y)$ and $H(y,x)$. We start by
 eliminating the possibility that both functions vanish
 simultaneously.

\subsection{$H(x,y)=0$, $H(y,x)=0$}
 \label{ss7IX.5}

First of all note the algebraic property 
\begin{equation}
H(x, y)-H(y, x)=2 \lambda  (-1 + \nu  x y) (x -y) (-1 + \nu)
 \;.
\label{h-prop}
\end{equation}
If $H(x,y)=0$, $H(y,x)=0$ then (\ref{h-prop}) entails the
alternatives
\begin{equation}
\lambda=0\quad\mbox{or}\quad x=y\quad\mbox{or}\quad y=\frac{1}{\nu x}\quad \mbox{or}\quad \nu=1.
\end{equation}
In the region $x\ne y$ only the third alternative is compatible
with admissible $(\lambda, \nu)$.  If we impose this condition
on $H(x,y)=0$, $H(y,x)=0$ we get
$$
\frac{(\nu -1) \left((\nu +1)^2-\lambda^2\right)}{\nu }=0,
$$
which is again not compatible with the ranges of $\lambda$,
$\nu$ imposed by Pomeransky and Senkov:
\bel{12XII8.2x}
 0 < \nu < 1\;, \qquad
 2\sqrt \nu \le \lambda < 1 +\nu
 \;.
\ee

\subsection{$H(y,x)=0$}
 \label{SHE}

It follows from \eq{4IX.1} that the zeros of $H(y,x)$ only
occur in the denominators of the components of the metric
induced on the planes
$\Span\{\partial_\varphi,\partial_\psi\}$. It has been pointed
out to us by H.~Elvang (private communication) that this is an
artifact of the parameterization of the metric, as those
components can be rewritten in a way which makes it clear that
their regularity is not affected by zeros of $H(y,x)$: 
\begin{eqnarray}
 g{}_{\varphi}{}_{\varphi}
  &=&
  \frac{2k^2(1-x^2)\Theta _{\varphi \varphi }(x, y)}
{H(x,y)(-1 + \nu)^{2}(x -y)^{2}}
 \;,
 \nonumber
\\
 \quad g{}_{\varphi}{}_{\psi}
 &=&
 \frac{2k^2\lambda\sqrt{\nu}(x^2-1)(1+y)
\Theta_{\varphi \psi }(x, y)}{H(x,y)(-1 + \nu)^{2}(x -y)}
 \;,
 \nonumber
\\
 g{}_{\psi}{}_{\psi}
 &=&
 \frac{2k^2(1+y)\Theta_{\psi \psi }(x, y)}
{(1 - \lambda +\nu)(-1 + \nu)^{2}H(x, y)(x-y)^{2}}
 \;,
\label{eq:metric-angular}
\end{eqnarray}
where $\Theta _{\varphi \varphi }(x, y)$,   $\Theta _{\psi \psi
}(x, y)$  and $\Theta _{\varphi \psi }(x, y)$ are polynomials
in $x$, $y$, $\nu$ and $\lambda$:
\begin{eqnarray}
 \nonumber
&&\Theta _{\varphi \varphi }(x, y)=
(-1 + \nu)^{2}(1 + \nu  x^{2}y^{2})(1 + \nu y^{2})
(1 + \nu )+\lambda(-1 + \nu)\times\nonumber\\
&&\big((-1 +x^{2})\nu y(-1 + y^{2}) (-1 + \nu) + x
(1 + \nu y^{2})(-3 -\nu + \nu y^{2} (1 + 3 \nu ))\big) -\nonumber\\
&&-\lambda ^3 \left(\nu x^{2} y (-1 + y^{2})-x
(-1 + \nu  y^{2})^{2} + \nu  y(1 -y^{2})\right)+\lambda^2 \left(1 + 2 x^{2} +
\right.\nonumber\\
&& \left.+\nu\left(-1 + y^{2}(1 -3 \nu+x^{2} (-3 + \nu )) +
\nu  y^{4} (2 \nu  + x^{2} (-1 + \nu ))\right)\right)\;,
\end{eqnarray}
\begin{eqnarray}
&&\hspace{-.8cm}\Theta_{\varphi \psi }(x, y)=
(1 - y)\left((-1 + \nu  x y)(-1 + \nu ^2) + \lambda ^2 (1 + \nu  x y)\right) + \nonumber\\
&&\phantom{xxxxxxxx} +2 \lambda (-x (1 + \nu  y (-2 + \nu )) + y (1 + \nu(-2 + \nu  y))).
\end{eqnarray}
The explicit form of $\Theta _{\psi \psi }(x, y)$, calculated
with {\sc Mathematica} and omitted because of its length, is
available upon request. Here we only report that near $x=y=1$
we have
\bel{badnewsforpsi}
 \Theta _{\psi \psi }(x, y) = -(y-1) (\nu -1)^2 (\lambda -\nu -1) (\lambda +\nu
   +1)^3 + O\big((x-1)^2 + (y-1)^2)\big)
   \;,
\ee
while on $\{y=1\}$ it holds that
\bean
 \Theta _{\psi \psi }(x, 1) &=&
 4 (x-1)^2 \lambda ^2 (\nu -1)^2 \left(x^2 \nu
   (\lambda +\nu -1)-\lambda +\nu -1\right)
\\
 &=&
   -4 (x-1)^2 \lambda ^2 (1-\nu )^3 (\lambda +\nu
   +1)+O\left((x-1)^3\right)
   \;.
\eeal{badnewsforpsi2}
Since
$$
 H(1,1)= (1-\nu) (\lambda +\nu +1)^2 >0
 \;,
$$
we see that $g_{\psi\psi}$ is negative for $y=1>x$ sufficiently
near $x=y=1$, and hence the periodic Killing vector
$\partial_\psi$ is timelike there.  This shows existence of
causality violations in that region. In fact, the zeros of
$\Theta _{\psi \psi }(x, 1) $ are $x=\pm1 $ and
$$
 x =\pm  \frac{\sqrt{\lambda -\nu
   +1}}{\sqrt{\nu  (\lambda +\nu -1)}}
   \;.
$$
This is real only for $\lambda+ \nu >1$,  but then always
larger than one in absolute value. We will prove that $H(x,1)$
is positive for admissible $\lambda$, hence causality is
violated throughout a neighbourhood of $\{x\in [-1,1]\;,\
y=1\}$.

\subsection{$H(x,y)=0$: formulation of the problem}
 \label{ssFormulation}

The question of zeros of $H(x,y)$ appears to be considerably
more difficult. In this section we wish to prove that the
polynomial $H(x,y)$ defined by:
$$ H(x,y):=1+ \lambda^2 - \nu^2 + 2\lambda \nu(1-x^2)y +
2x\lambda(1-\nu^2y^2) + x^2y^2\nu(1-\lambda^2-\nu^2)$$
does not vanish on the set $\Omega_0$ defined in \eq{Om}. Every
slice of $\Omega_0$ at fixed $\nu$ and $\lambda$ corresponds to
the region outside the ``interior", presumably Cauchy, horizon
of the metric with parameters $\nu$ and $\lambda$.
Equivalently, we want to show that the set
\bel{7IX.1} \mathcal{H}\equiv\{(x,y,\lambda,\nu)\in
\mathbb{R}^4 : (\nu,\lambda)\in \mcU\;,\ y_c(\nu,\lambda)\le y
< -1\;,\ H(x,y)=0\} \ee
does not intersect $\Omega_0$.

\medskip
We start by writing $H$ as  a polynomial in $x$,
$$
 H(x,y)=(1+ \lambda^2 - \nu^2 + 2\lambda \nu y)\ +\
2\lambda(1-\nu^2y^2)x\ +\ \nu
 y\Big(y(1-\lambda^2-\nu^2)-2\lambda\Big)x^2\;,
$$
and note that the coefficient of $x^2$ does not vanish in the
ranges of interest:
\begin{Lemma}
 \label{L5IX.1}
 The highest order coefficient $\nu
y\Big(y(1-\lambda^2-\nu^2)-2\lambda\Big)$ of the polynomial in
$x$, $H(x,y)$, does not vanish in the domain of interest
$\{(y,\nu,\lambda) : (\nu,\lambda) \in \mcU\;,\ y \in
[y_c(\nu,\lambda),-1)\}$.
\end{Lemma}

\proof For $\nu y \ne 0$ the condition $\nu
y\Big(y(1-\lambda^2-\nu^2)-2\lambda\Big)=0$ is equivalent to
the condition
%
$$y=y_0(\nu,\lambda):=
\frac{-2\lambda}{\nu^2+\lambda^2-1}\;.
$$
So, for all $(\nu,\lambda) \in \mcU$ such that
$\nu^2+\lambda^2<1$, $y_0(\nu,\lambda)$ is positive, hence
outside the region of interest for the parameter $y$. The
simultaneous equalities $\nu^2+\lambda^2=1$ and $y=y_0$ are
also impossible on $\Omega_0$. We claim that in the case
$\nu^2+\lambda^2>1$ we have the inequality $y_0<y_c$: Indeed,
we first note that
%
$$
y_0(\nu,1+\nu)=-\frac{1}{\nu}=y_c(\nu,1+\nu)\;.
$$
Next, the derivatives of $y_0$ and $y_c$ with respect to
$\lambda$ read
%
$$
\frac{\partial y_0}{\partial \lambda}(\nu,\lambda)=
\frac{2(1+\lambda^2-\nu^2)}{(\nu^2+\lambda^2-1)^2}\;,\quad
\frac{\partial y_c}{\partial \lambda}(\nu,\lambda)=
-\frac{\sqrt{\lambda ^2-4 \nu }+\lambda }{2 \nu\sqrt{\lambda ^2-4 \nu }}.
$$
The first expression is positive for all $\{0<\nu<1\;,
\nu^2+\lambda^2\ne 1\}\subset \mcU$, thus $\lambda \mapsto
y_0(\nu,\lambda)$ is increasing on $(\sqrt{1-\nu^2},1+\nu)$,
while the second expression is negative when
$\lambda\in(2\sqrt{\nu},1+\nu)$  so $\lambda \mapsto
y_c(\nu,\lambda)$ is decreasing on $(2\sqrt{\nu},1+\nu)$.
This enables us to conclude that $y_0(\nu,\lambda)$ is always
outside the region $(y_c(\nu,\lambda),-1)$, and therefore that the
dominant coefficient of $H(x,y)$ is nonzero for all $(\nu,\lambda)
\in \mcU$.
\qed

\medskip

Then, the discriminant of the second-order polynomial in $x$,
$H(x,y)$, is the function $\Delta_x$, given by:
%
\be \Delta_x(y,\nu,\lambda) =
4\left[\lambda^2(1-\nu^2y^2)^2-\nu y(1+ \lambda^2 - \nu^2 +
2\lambda \nu y)(y(1-\lambda^2-\nu^2)-2\lambda)\right]\;. \ee
If $\Delta_x(y,\nu,\lambda)\geq 0$ for some values of
$y,\nu,\lambda$, then the equation $H(x,y)=0$ has two roots
(counting multiplicity) $x=x_\pm (y,\nu,\lambda)$ which have
the expression:
%
\bel{xpm} x_\pm (y,\nu,\lambda)=\frac{-\lambda(1-\nu^2y^2) \pm
\sqrt{W(y,\nu,\lambda)}}{\nu y
(y(1-\lambda^2-\nu^2)-2\lambda)}\;,\ee
with
$$
 W:=\frac14\Delta_x
 \;.
$$
Note that the denominator has no zeros in the range of interest
by Lemma~\ref{L5IX.1}.

In what follows, we begin with considering the restriction to
the particular cases $y=-1$, and $x=-1$. We then pass to a
study of the set $\mathcal{A}$, defined as the collection of
all $(y,\nu,\lambda)$, $ y_c\le y\le -1$, $(\nu,\lambda)\in
\mcU$, such that $W(y,\nu,\lambda)$ is non-negative, in
particular its connectedness. We eventually conclude, using the
continuity of the functions $x_\pm$ on $\mathcal{A}$, that for
all such $(y,\nu,\lambda)$, the corresponding roots
$x_\pm(y,\nu,\lambda)$ lie outside the required interval $-1\le
x \le 1$.

\subsection{$H(x,y)=0$, $y=-1$}
\begin{Lemma}
 \label{L5IX.3}
There exist values of $(\nu,\lambda) \in \mcU$ such that
$x_+(-1,\nu,\lambda)< -1$, and thus $x_-(-1,\nu,\lambda)< -1$
as well.
\end{Lemma}

\proof Let us try values of $\nu$ and $\lambda$ in the allowed
ranges such that $\nu^2 + \lambda^2 =1$. This is possible for
$\nu$ small enough, namely $\nu \in \left(0,\sqrt{5}-2\right]$.
We first have that
%
%
$$
 W(-1,\nu,\sqrt{1-\nu^2})=(1-\nu^2)\big((1-\nu^2)^2+4\nu(\nu-\sqrt{1-\nu^2})\big)
$$
is positive. Then we compute $x_\pm$, and we get:
%
\be x_\pm(-1,\nu,\sqrt{1-\nu^2}) + 1 = \frac{-1+\nu^2 + 2\nu \pm
\sqrt{(1-\nu^2)^2+4\nu(\nu-\sqrt{1-\nu^2})}}{2\nu}\ ; \ee
The numerator of the right-hand side term of the last equality
for $x_+$ is negative, as can be seen from the following
equivalent inequalities:
%
\beaa & 1-\nu^2-2\nu & >  \sqrt{(1-\nu^2)^2+4\nu(\nu-\sqrt{1-\nu^2})}\\
\Leftrightarrow & (1-\nu^2)^2-4\nu(1-\nu^2)+4\nu^2 & >  (1-\nu^2)^2+4\nu^2-4\nu\sqrt{1-\nu^2} \\
\Leftrightarrow & -4\nu(1-\nu^2) & >  -4\nu\sqrt{1-\nu^2} \eeaa
the last one being of course true for $\nu \in (0,1)$.
 \qed
\subsection{$H(x,y)=0$, $x=-1$}
\begin{Lemma}
 \label{L5IX.2}
There is no solution for
$H(x,y)=0$ when $x=-1$.
\end{Lemma}

\proof We can first write
$$
H(-1,y)=(1+\nu-\lambda)\big(\nu(1+\lambda-\nu)y^2+1-\nu-\lambda\big)\;.
$$
The first factor is positive from the definition of $\mcU$. We show
that the second factor cannot vanish for any value of the parameters
$y,\nu,\lambda$ in the allowed ranges. Indeed, this second factor is
quadratic in $y$, the coefficient $\nu(1+\lambda-\nu)$ is positive,
so that the roots are
$$
y_\pm=\pm \sqrt{\frac{\nu+\lambda-1}{\nu(1+\lambda-\nu)}}\;,
$$
provided that $\nu+\lambda\ge 1$, otherwise there is no root
and $H(-1,y)$ is indeed positive. But the required condition
$\lambda<1+\nu$ is equivalent, for $\nu \in (0,1)$,
to $\nu+\lambda-1 < \nu(1+\lambda-\nu)$,
so that both solutions $y_\pm$ above are larger than $-1$, thus
out of the authorized range for the coordinate $y$.
\qed

In other words, this lemma expresses that no connected component of
the set $\mathcal{H}$ can intersect the hypersurface $\{x=-1\}$ for
values of $y$ smaller than $-1$.

Recall that $\mathcal{A}$  is the set of points
$(y,\nu,\lambda)$, $(\nu,\lambda)\in \mcU$, $y_c\le y\le -1$,
such that solutions $x \in \mathbb{R}$ of the equation
$H(x,y)=0$ do exist. We will show shortly that $\mathcal{A}$ is
connected. Then, since $x_+$ and $x_-$ are continuous functions
we deduce that $x_+(\mathcal{A})$ and $x_-(\mathcal{A})$ are
connected subsets of $\mathbb{R}$ and hence they must be
intervals. On the other hand, by Lemma~\ref{L5IX.2} we have
$x_+(\mathcal{A})\cap\{-1\}=\emptyset$,
$x_-(\mathcal{A})\cap\{-1\}=\emptyset $ and hence either
$x_+(\mathcal{A})\subset (-\infty ,-1)$ or
$x_+(\mathcal{A})\subset (-1,\infty)$; and similarly
$x_-(\mathcal{A})\subset (-\infty ,-1)$ or
$x_-(\mathcal{A})\subset (-1,\infty)$. The alternatives
$x_+(\mathcal{A})\subset (-1,\infty)$ and
$x_-(\mathcal{A})\subset (-1,\infty)$ can be ruled out because
in Lemma~\ref{L5IX.3} we have proved that, for the particular
case of $y=-1$, $\lambda=\sqrt{1-\nu^2}$, and for $\nu$ small
enough, both solutions $x_\pm(-1,\nu,\sqrt{1-\nu^2})$ satisfy
$x<-1$. Therefore necessarily $x_-(\mathcal{A})\subset (-\infty
,-1)$, $x_+(\mathcal{A})\subset (-\infty ,-1)$ which entails
$x_-(\mathcal{A})\cap (-1,1)=\emptyset $, $x_+(\mathcal{A})\cap
(-1,1)=\emptyset $. From this result we conclude that
$\mathcal{H}$ does not intersect $\Omega_0$.

The aim of the next section is to establish the connectedness
of $\mathcal{A}$, needed to complete the proof.

\subsection{Connectedness of $\mcA$}

Throughout this section, we assume that $(\nu,\lambda)\in
\mcU$. Recall that if $H(x,y)=0$ then $x=x_\pm$, where
%
\begin{eqnarray*}
x_\pm={\frac {\lambda-\lambda\,{\nu}^{2}{y}^{2}\pm\sqrt {W(
y,\nu,\lambda) }}{\nu\,y \left(({\lambda}^{2}+{\nu}^{
2}-1)y+2\lambda \right) }} \;,
\end{eqnarray*}
and where
%
\beaa
  W( y,\nu,\lambda)&:=& {\lambda}^{2}{\nu}^{4}{y}^{4}+
  2\,\lambda\,{\nu}^{2} \left(
 {\lambda}^{2}+{\nu}^{2}-1 \right) {y}^{3}
\\
 &&
  -\nu\, \left(1 -2\,\nu\,{\lambda}^{2}-{\lambda}^{4}+{\nu}^{4}-2\,{\nu}^
 {2}  \right)
  {y}^{2}
   +2\,\nu\,\lambda\, \left(  1+{\lambda}^{2}-{\nu}^{2} \right)
        y
        +{\lambda}^{2}
 \;.
\eeaa
For large $y$, whether positive or negative, $W$ is positive so
zeros of $H(x,y)$ exist. Our task here is to prove that the
region $\{(\nu,\lambda)\in \mcU,\; x\in [-1,1],\; y_c\le y\le
-1\}$ does not contain any of them. As just explained, this
will follow from:

\begin{Theorem} \label{T4V.1}
The set
\bel{8IX.-1}
 \mcA:=\{(y,\nu,\lambda):\ W( y,\lambda,\nu)\ge 0\;,\ (\lambda,\nu)\in \mcU\;,\ y_c(\nu,\lambda)\le y \le -1\}
\ee
is connected.
\end{Theorem}

\proof We show in Lemma~\ref{L2V.0} below that $W(-1,\nu,\lambda)>0$
on $\mcU$. Next, Proposition~\ref{P4V.2} establishes that for all
$(\lambda,\nu)\in \mcU$, the set $\{y \in [y_c,-1] :\ W(
y,\lambda,\nu)\ge 0\}$ is connected, which readily implies the
result.
\qed

We supply now the details:

\begin{Lemma}
\label{L2V.0} $W(-1,\nu,\lambda)>0$ on $\mcU$.
\end{Lemma}
\proof
We have
%
$$
W(-1,\nu,\lambda)=  \ub{\left( 1+
\nu\,{\lambda}^{2}-2\,\nu\,\lambda-{\nu}^{2}
\right)}_{=:P(\nu,\lambda)}
 \ub{\left( {\nu}^{3}-\nu-2\,\nu\,\lambda+{\lambda}^{2} \right)}_{=:Q(\nu,\lambda)}
 \;.
$$
The equation $P=0$ is solved by
%
$$
\nu_\pm= \frac 12 \left(\,{\lambda}^{2}-2\lambda\pm \,\sqrt
{({\lambda}^{2}-2\,{\lambda})^{2}+ 4  }\right) \;.
$$
Clearly $\nu_-<0$ and $\nu_+>[(\lambda-1)^2+1]/2$.   For
$\lambda\in[0,2)$ we have
$$
 \nu_-(\lambda) < \lambda -1 < \frac 14 \lambda^2 <
 \nu_+(\lambda)\;.
$$
Indeed, the first inequality is equivalent to positivity of
$3\lambda^2-4\lambda+4 = 2\lambda^2 + (\lambda-2)^2$.  The
second one is always true for $\lambda \neq 2$, while the last
inequality follows easily from the already indicated inequality
$\nu_+>[(\lambda-1)^2+1]/2$.
Then, we notice that:
$$(\nu,\lambda) \in \mcU \quad \Longleftrightarrow \quad \lambda \in (0,2)\;,\
\lambda-1<\nu \leq\lambda^2/4\;.
$$
Hence, for all $(\nu,\lambda) \in \mcU$, we have
$\nu_-(\lambda)<\nu<\nu_+(\lambda)$, and we conclude that $P>0$
on $\mcU$.

Next, we have
%
$$
Q= \left( \lambda-\nu+\sqrt {{\nu}^{2}-{\nu}^{3}+\nu} \right)
\left( \lambda-\nu-\sqrt {{\nu}^{2}-{\nu}^{3}+\nu} \right)
 \;,
$$
and note that the polynomial ${\nu}^{2}-{\nu}^{3}+\nu$ vanishes
at $0$ and at $(1\pm \sqrt 5)/2$. We want to show that $Q$ is
positive on $\mcU$, this proceeds as follows: Straightforward
algebra shows that, for $\nu>0$, the inequality
\bel{5V.5}
 2\sqrt{\nu}>\nu+\sqrt {{\nu}^{2}-{\nu}^{3}+\nu}
\ee
is equivalent to
$$
\left( ({\nu}+1)^2 +8 \right)  \left( \nu-1 \right) ^{2}>0\;.
$$
So \eq{5V.5} holds for $\nu\in [0,1)$. But the right-hand-side
of \eq{5V.5} is the larger root of $Q$, and we conclude that
the roots of $Q$ do not intersect the graph of $\nu\mapsto
2\sqrt{\nu}$ in the range of interest.
Next, \eq{5V.5} also shows that $Q$ is positive on this graph
for small positive $\nu$. Since $Q$ does not change sign on
$\mcU$, it is positive on $\mcU$.
\qed

We continue with:
\begin{Proposition}
 \label{P4V.2}
For every admissible values of $\lambda$ and $\nu$, the set
$\{y \in [y_c,-1]:\ W(y,\nu,\lambda)\ge 0,\ (\nu,\lambda) \in
\mcU \}$ is connected.
\end{Proposition}

\proof We start by a study of the variations of $y \mapsto
W(y,\nu,\lambda)$ on $[y_c(\nu,\lambda),-1]$, for all
$(\nu,\lambda) \in \mcU$. To do so, we first compute the
derivatives of $W$, with respect to $y$, up to third order.
We have
%
\beaa
 \frac{\partial W}{\partial y}(y,\nu,\lambda)&=&
 2\nu\Big(2\nu^3\lambda^2y^3+3\nu\lambda(\nu^2+\lambda^2-1)y^2
\\
&& +\big(\lambda^4-\nu^4-1+2\nu(\nu+\lambda^2)\big)y
+ \lambda(1+\lambda^2-\nu^2)\Big)\;, \\
\frac{\partial^2 W}{\partial y^2}(y,\nu,\lambda)&=&
2\nu\Big(6\nu^3\lambda^2y^2+6\nu\lambda(\nu^2+\lambda^2-1)y+\lambda^4-\nu^4-1+2\nu(\nu+\lambda^2)\Big)\;,
\\
\frac{\partial^3 W}{\partial y^3}(y,\nu,\lambda)&=&
12\nu^2\lambda\big(2\nu^2\lambda y+  \nu^2+\lambda^2-1
\big)\;.\eeaa
Since $\nu\lambda \ne 0$ for all allowed $\nu$ and $\lambda$,
we see that $\frac{\partial^3 W}{\partial y^3}(y,\nu,\lambda)$
vanishes at $y=y_3(\nu,\lambda)$, where
%
\bel{y_3}
y_3(\nu,\lambda):=\frac{1-\nu^2-\lambda^2}{2\nu^2\lambda}\ \;
,\ee
and therefore the function $y\mapsto \frac{\partial^2
W}{\partial y^2}(y,\nu,\lambda)$ reaches its minimum there,
equal to:
%
%
\bel{minW_yy} \min_{y \in \mathbb{R}}\frac{\partial^2
W}{\partial y^2}(y,\nu,\lambda)=(1+\nu+\lambda)(1+\nu-\lambda)
\underbrace{\big(\lambda^2(3-2\nu)-(1-\nu)^2(3+2\nu)\big)}_{k(\nu,\lambda)}\;.\ee
The   sign of the minimum is determined by  the sign of the
third factor $k(\nu,\lambda)$ in (\ref{minW_yy}), since the
first  two factors are positive for  $(\nu,\lambda)\in \mcU$.
{We start by supposing that $k(\nu,\lambda)\geq 0$.} This
corresponds to values of $\nu$ and $\lambda$ such that $\lambda
\geq
    \lambda_k(\nu)$, where
%
\bel{lk}
 \lambda_k(\nu):=(1-\nu)\sqrt{\frac{3+2\nu}{3-2\nu}} \;,
\ee
see Figure~\ref{Udef4V09}. In this range of parameters the
function $y\mapsto W(y,\nu,\lambda)$ is therefore convex.
Connectedness of $\{y \in [y_c,-1]:\ W(y,\nu,\lambda)\ge 0,\
(\nu,\lambda) \in \mcU \}$ in the case $k(\nu,\lambda)\ge 0$,
will be a consequence of the following:
\begin{Lemma}
 \label{L26V.1}
For all $(\nu, \lambda) \in \mcU$,
$W(y_c(\nu,\lambda),\nu,\lambda)$ is negative.
\end{Lemma}
This result, together with the convexity of $y\mapsto
W(y,\nu,\lambda)$ and with the Lemma \ref{L2V.0}, shows that
the function $y\mapsto W(y,\nu,\lambda)$ is negative on
$[y_c,y_*)$, and then positive on $(y_*,-1]$ for some $y_* \in
(y_c,-1)$, hence the proposition \ref{P4V.2} is proved for all
$(\nu,\lambda)\in \mcU$ such that $k(\nu,\lambda)\ge 0$.

We now turn to the proof of the Lemma:

\proof We have
\beaa
W(y_c(\nu,\lambda),\nu,\lambda)&=&-\frac{1-\nu}{2\nu}\Big(\lambda^6+\lambda^5\sqrt{\lambda^2-4\nu}-2\lambda^4
(1+3\nu) -2 \lambda^3\sqrt{\lambda^2-4\nu}(1+2\nu)
\\
 && + \lambda^2
(1+7\nu+9\nu^2-\nu^3) +\lambda\sqrt{\lambda^2-4\nu}
 (1+5\nu+3\nu^2-\nu^3)
\\
 &&
 -2\nu(1-\nu)(1+\nu)^2\Big)\;. \eeaa
The occurrence of $\sqrt{\lambda^2-4\nu}$ above leads us to
introduce a change of variables $(\nu,\lambda) \rightarrow
(\nu,\eta)$, with $\eta \geq 0$, defined as
$$\lambda=2\sqrt{\nu}\cosh{\eta}\;.$$
Then the expression simplifies remarkably as a polynomial in
$e^\eta$:
$$
W\big(y_c(\nu,2\sqrt{\nu}\cosh{\eta}),\nu,2\sqrt{\nu}\cosh{\eta}\big)=-(1-\nu)e^{-2\eta}(e^{4\eta}-\nu)(1-e^{2\eta}\nu)^2
\;.
$$
The factors are all positive for $\nu \in (0,1)$ and $\eta \geq
0$, except the last factor $(1-e^{2\eta}\nu)^2$ which can
vanish for $\eta=-\ln{\nu}/2$, which corresponds precisely to
$\lambda=1+\nu$, hence not for $(\nu,\lambda) \in \mcU$. This
finishes the proof of the lemma.
\qed

{We now turn attention to the case $k(\nu,\lambda) < 0$.}
Equivalently, $\lambda<\lambda_k(\nu)$. Note that
    $$
     \lambda_k'(\nu)=-{\frac {3-4\,{\nu}^{2}+6\,\nu }{ \left( 3- 2\,\nu \right) \sqrt {9- 4\,{\nu}^{2}}}}
 \;,
 $$%
 %
 which is negative for $\nu\in (0,1)$. Since
    $\lambda_k(0)=1$ and  $\lambda_k(1)=0$, the  inequality
    $\lambda<\lambda_k(\nu)$ is compatible with the allowed
    ranges of the parameters $\nu$ and $\lambda$ only when
    $\nu \in \left(0,\nu_0\right)$, where $\nu_0\approx
    0.207$ is the unique solution in $(0,1)$ of the
    equation $2\sqrt{\nu}=\lambda_k(\nu)$;  see
Figure~\ref{Udef4V09}.   In other words, it suffices to
consider $0<\nu \leq \nu_0$. Recall that this case corresponds
to a negative minimum for $y\mapsto\frac{\partial^2 W}{\partial
y^2}(y,\nu,\lambda)$, obtained for $y=y_3(\nu,\lambda)$ (see
equation (\ref{y_3})). To analyse the position of $y_3$
compared to $-1$  we write:
\be
y_3(\nu,\lambda)+1=\frac{1-\nu^2-\lambda^2+2\nu^2\lambda}{2\nu^2\lambda}\;.\ee
The numerator is a polynomial in $\lambda$ of degree-two which
has, for all $\nu$, two real roots $\lambda_\pm(\nu)=\nu^2\pm
\sqrt{1-\nu^2+\nu^4}$.

Now, for all $\nu$ in $\left(0,\nu_0\right)$, one has the chain
of inequalities
%
\be \lambda_-(\nu) < 0<2\sqrt{\nu} <\lambda_k(\nu) < 1
<\lambda_+(\nu) \ ;\ee
Indeed, to obtain the first inequality we note that for $\nu\in
(0,1)$ we have $1-\nu^2+\nu^4>\nu^4$, and negativity of
$\lambda_-(\nu)$, for all $\nu \in (0,\nu_0)$, follows. The
third inequality has already been established, $\nu_0$ being
precisely the value at which the inequality is saturated. The
fourth follows from the fact that $\lambda_k$ is decreasing.
The last inequality can be proved by noting that $1-\nu^2+\nu^4
= (1-\nu^2)^2+\nu^2>(1-\nu^2)^2$.

Since $y_3(\nu,\lambda)+1$ is positive for $\lambda$ between
$\lambda_-(\nu) $ and $\lambda_+(\nu)$, we obtain that
$y_3(\nu,\lambda)+1$ is positive for all $\nu \in
\left(0,\nu_0\right)$ and $\lambda \in
\left[2\sqrt{\nu},\lambda_k(\nu)\right)$. Thus
$$
 y_3>-1
$$
in the range of  parameters of interest. This implies that
$y\mapsto\frac{\partial^3 W}{\partial y^3}(y,\nu,\lambda)$ is
negative on $\left(-\infty,y_3(\nu,\lambda)\right)$, and in
particular on $\left[y_c,-1\right]$. Hence
$y\mapsto\frac{\partial W}{\partial y }(y,\nu,\lambda)$ is
concave on $\left[y_c,-1\right]$, and therefore lies above its
arc. So it reaches its minimum on this interval either at
$y=y_c(\nu,\lambda)$, or at $y=-1$. But from Lemma \ref{L2V.2},
which will be proved shortly, we get that $\partial_y
W(-1,\nu,\lambda)$ is non-negative for $(\nu,\lambda)\in \mcU$,
$\lambda<\lambda_k(\nu)$. Then, we have again two cases:
\begin{itemize}
\item If $\partial_y W(y_c,\nu,\lambda)$ is non-negative,
    then $\partial_y W(y,\nu,\lambda)$ is non-negative for
    all $y \in [y_c,-1]$, therefore $y \mapsto
    W(y,\nu,\lambda)$ is increasing on this interval, and
    the set $\{y \in [y_c,-1]:\ W(y,\nu,\lambda)\ge 0\}$ is
    connected, and contains $-1$.
\item If $\partial_y W(y_c,\nu,\lambda)$ is negative, then,
    since it is concave, the function $y \mapsto \partial_y
    W(y,\nu,\lambda)$ is negative on $[y_c,y_*)$ and
    non-negative on $[y_*,-1]$ for some $y_*(\nu,\lambda)
    \in (-1/\nu,-1]$. Thus, $y \mapsto W(y,\nu,\lambda)$ is
    decreasing on $[y_c,y_*)$, and increasing on
    $[y_*,-1]$. From Lemma \ref{L26V.1}, this implies that
    $W(y,\nu,\lambda)$ is negative at least on $[y_c,y_*)$,
    then increasing on $[y_*,-1]$. We can therefore
    conclude again that the set $\{y \in [y_c,-1]:\
    W(y,\nu,\lambda)\ge 0\}$ is connected. Thus, in order
    to finish the proof of the proposition, and hence of
    the theorem, the following lemma remains to be proved:
\end{itemize}

\begin{Lemma}
 \label{L2V.2}
For $0<\nu<\nu_0$ and
$2\sqrt{\nu}\leq\lambda\leq\lambda_k(\nu)$ the function
 $\frac{\partial W}{\partial y}(-1,\nu,\lambda)$ is
non-negative.
\end{Lemma}

\proof
The result is clear by inspection of the graph in
Figure~\ref{L28}, a possible formal proof proceeds as follows:
\begin{figure}[h]
\begin{center}
\includegraphics[width=.7\textwidth]{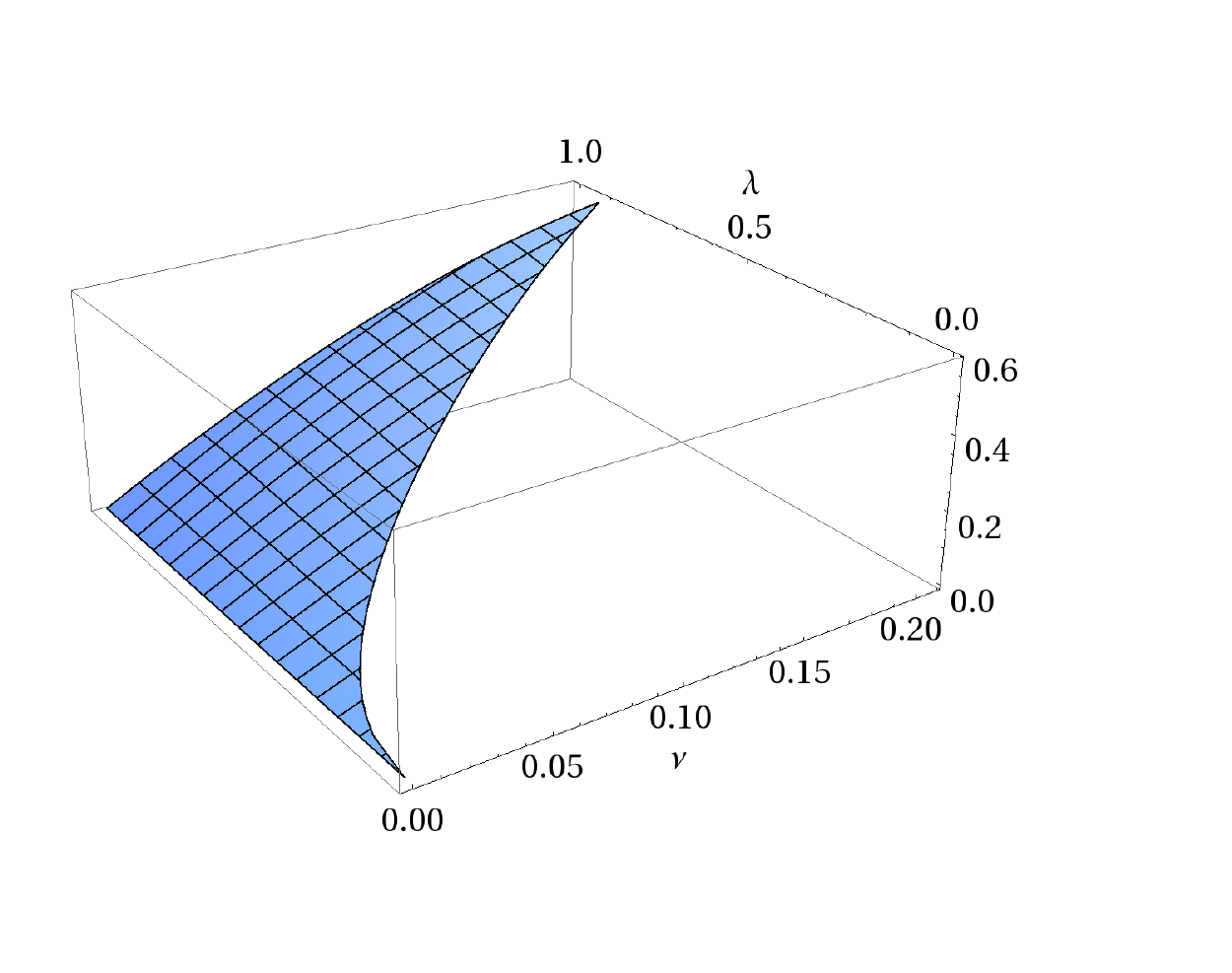}
 \caption{\label{L28}
 The graph of $\partial_y W$ over the set $\{0<\nu<\nu_0\;,\
 2\sqrt{\nu}\leq\lambda\leq\lambda_k(\nu)\}$.}
\end{center}
\end{figure}
At $\lambda=2\sqrt{\nu}$ we have
%
\bel{W_y1} \frac{\partial W}{\partial y}(-1,\nu,2\sqrt{\nu})=
2\nu(1-\sqrt{\nu})^3(1+5\sqrt{\nu}+12\nu+24\nu^{3/2}+15\nu^2+7\nu^{5/2})
 \;
 ,
\ee
which is clearly positive for all $0< \nu <1 $.  We continue by
noting that
%
\bel{W_yl} \frac{\partial^2 W}{\partial \lambda\partial
y}(-1,\nu,\lambda)= 2\nu\big(1-3\nu-\nu^2+3\nu^3
-4\nu(1+\nu^2)\lambda +3(1+3\nu)\lambda^2-4\lambda^3\big)\;
,\ee
%
%
\bel{W_yllxyz} \frac{\partial^3 W}{\partial \lambda^2\partial
y}(-1,\nu,\lambda)= -4\nu\big(6\lambda^2 -3(1+3\nu)\lambda
+2\nu(1+\nu^2)\big)\;.\ee
For all $0<\nu<1$ the right-hand-side of (\ref{W_yllxyz}) has
two real roots,
%
\be
\lambda=\frac{1+3\nu}{4} \pm
\frac{\sqrt{3(27\,{\nu}^{2}+2\,\nu+3-16\,{\nu}^{3})}}{12}=:\lambda_{0\pm}(\nu)
 \;.
\ee 
Then, we have the inequalities $\lambda_{0-}(\nu)<2\sqrt{\nu}$,
and $\lambda_{0+}(\nu)<\lambda_k(\nu)$ for $0<\nu<\nu_0$,
whereas the difference $2\sqrt{\nu}-\lambda_{0+}(\nu)$ is
positive on $\left(0,\nu_1\right)$ and negative on
$\left(\nu_1,\nu_0\right)$ for some $\nu_1 \in
\left(0,\nu_0\right)$. Indeed we have, for $\nu \in (0,1)$,
%
$$\lambda_{0-}(\nu)=\frac{1+3\nu}{4}-\frac{\sqrt{3}}{12}\sqrt{3+2\nu+27\nu^2-16\nu^3}<
 \frac{1+3\nu}{4}-\frac{\sqrt{3}}{12}\sqrt{3}= \frac{3\nu}{4} < 2\nu < 2\sqrt{\nu}\;.$$
Next, we prove that $\nu \mapsto \lambda_{0+}(\nu)-2\sqrt{\nu}$
is decreasing on $(0,\nu_0)$. Indeed,
%
$$ \frac{d}{d\nu}(\lambda_{0+}(\nu)-2\sqrt{\nu})=\frac14
\left(3-\frac{4}{\sqrt{\nu}}+\frac{1/3 +9\nu -8\nu^2}{\sqrt{1+2\nu/3
+ 9\nu^2 -16 \nu^3/3}}\right)\;;$$
For $\nu \in (0,1)$, we have
$$\frac{1/3 +9\nu
 -8\nu^2}{\sqrt{1+2\nu/3 + 9\nu^2 -16 \nu^3/3}}< 1/3 +9\nu -8\nu^2
 \;,
$$
so that we obtain:
\bel{L0p} \frac{d}{d\nu}(\lambda_{0+}(\nu)-2\sqrt{\nu})<\frac14
\left(\frac{10}{3}-\frac{4}{\sqrt{\nu}}+9\nu-8\nu^2\right)\;. \ee
Moreover, $1/3 +9\nu -8\nu^2$ is less than $5$ for $\nu \in
(0,1/4)$, while $3-\frac{4}{\sqrt{\nu}}$ is less than $-5$ for all
$\nu \in (0,1/4) \supset (0,\nu_0)$.
Therefore, $\frac{d}{d\nu}(\lambda_{0+}(\nu)-2\sqrt{\nu})$ is
negative for $\nu \in (0,1/4)$, in particular for $\nu \in
(0,\nu_0)$.

\begin{figure}[h]
\begin{center}
\includegraphics[width=.5\textwidth]{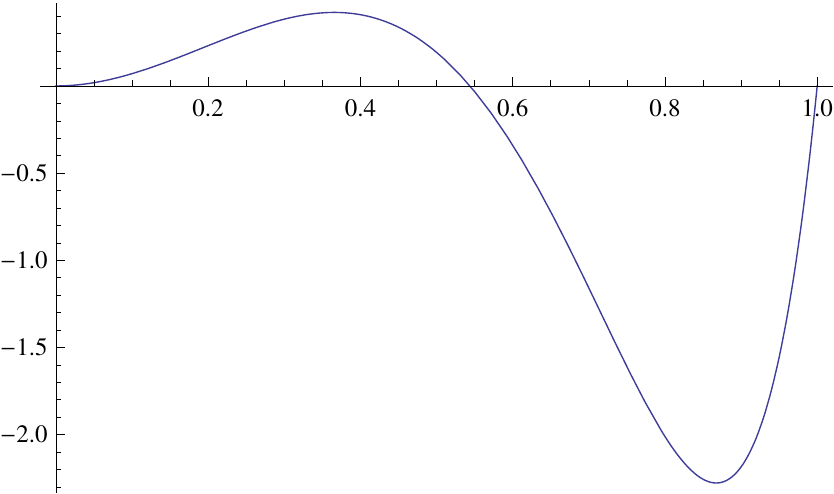}
 \caption{\label{FWln}The function $\nu\mapsto \frac{\partial^2 W}{\partial \lambda\partial
y}(-1,\nu,\lambda_k(\nu))$; the non-triviial zero is  at, approximately, $.544$.}
\end{center}
\end{figure}

We further note that the function $\nu \mapsto
\lambda_{0+}(\nu)-2\sqrt{\nu}$ takes the value $1/2$ at
$\nu=0$, and that it is negative at $\nu=1/8 < \nu_0$:
$$\lambda_{0+}(1/8)-2\sqrt{1/8}=\frac{11}{32}-\frac{1}{\sqrt{2}}+
\frac14 \sqrt{\frac{233}{192}} < \frac18 \left(\frac{11}{4} -
4\sqrt{2} + \sqrt{5} \right) < 0\;.$$
This proves the existence of $\nu_1 \in (0,\nu_0)$ (and more
precisely $\nu_1 \in (0,1/8)$) such that $\lambda_{0+}(\nu)>
2\sqrt{\nu}$ for $\nu \in (0,\nu_1)$, and $\lambda_{0+}(\nu)<
2\sqrt{\nu}$ for $\nu \in (\nu_1,\nu_0)$. Moreover, integrating
the inequality (\ref{L0p}), we obtain
%
$$\lambda_{0+}(\nu)<\frac12 + \frac56 \nu + \frac98 \nu^2 - \frac23
\nu^3\;,$$
therefore $\lambda_{0+}(\nu) < 5/8$ for all $\nu \in (0,1/8)$,
whereas $\lambda_k(\nu) > 3/4 > 5/8$ for all $\nu \in (0,1/4)$.
Then, since $2\sqrt{\nu}< \lambda_k(\nu)$ for all $\nu \in
(0,\nu_0)$,
we obtain, combining the previous remarks, that
$\lambda_{0+}(\nu)<\lambda_k(\nu)$ for all $\nu \in (0,\nu_0)$.
So we have again two cases:
\begin{itemize}
\item if $\nu \in \left(0,\nu_1\right)$: then $\lambda
    \mapsto \frac{\partial^2 W}{\partial \lambda\partial
    y}(-1,\nu,\lambda)$ is increasing on
    $\left[2\sqrt{\nu},\lambda_{0+}(\nu)\right]$, then
    decreasing on
    $\left[\lambda_{0+}(\nu),\lambda_k(\nu)\right]$;
\item if $\nu \in \left[\nu_1,\nu_0\right)$: then $\lambda
    \mapsto \frac{\partial^2 W}{\partial \lambda\partial
    y}(-1,\nu,\lambda)$ is decreasing on
    $\left[2\sqrt{\nu},\lambda_k(\nu)\right]$.
\end{itemize}
%
So, to make sure that the function $\lambda \mapsto
\frac{\partial^2 W}{\partial \lambda\partial
y}(-1,\nu,\lambda)$ is positive on
$\left[2\sqrt{\nu},\lambda_k(\nu)\right]$, we only need to show
that it is positive for $\lambda=2\sqrt{\nu}$ and for
$\lambda=\lambda_k(\nu)$, this for all $\nu \in
\left(0,\nu_0\right)$. We have in fact:
%
%
$$\frac{\partial^2
W}{\partial \lambda\partial
y}(-1,\nu,2\sqrt{\nu})=2\nu(1-\sqrt{\nu})^2
 (1+2\sqrt{\nu}+12\nu-18\nu^{3/2}-13\nu^2-8\nu^{5/2})
 \;,
$$
and we can write this, with $\nu=s^2$:
%
%
\beaa \lefteqn{ \frac{\partial^2 W}{\partial \lambda\partial
y}(-1,s^2,2s)
  =
   2s^2(1-s)^2
    \times
    }
    &&
\\
 &&
    \Big((1-8s^3)+(2s-8s^3)+(s^2-2s^3)+(4s^2-16s^4)+7s^2+3s^4-8s^5\Big)
    \;,
\eeaa
with each term positive  for $s \in (0,1/2)$, i.e. for $\nu \in
(0,1/4)$.
Then we have
$$
 \frac{\partial^2 W}{\partial \lambda\partial y}(-1,\nu,\lambda_k(\nu))
 =\frac{8\nu}{3-2\nu}
  \Big((1-\nu)(3+4\nu-5\nu^2-3\nu^3)+\lambda_k(\nu)(-3+\nu+3\nu^2-5\nu^3+2\nu^4)
  \Big)
\;.
$$
%
Since $0 < \lambda_k(\nu) < 1$ on $(0,1/4)$, and since the factor
$3-\nu-3\nu^2+5\nu^3-2\nu^4$ is positive for this range of $\nu$, we
have
$$\frac{\partial^2 W}{\partial \lambda\partial
y}(-1,\nu,\lambda_k(\nu))>\frac{8\nu}{3-2\nu}(2\nu-6\nu^2-3\nu^3+5\nu^4)\;,$$
still positive for $\nu \in (0,1/4)$. The plot of $\frac{\partial^2
W}{\partial \lambda\partial y}(-1,\nu,\lambda_k(\nu))$ can be found
in Figure~\ref{FWln}.

It follows that the function $\lambda \mapsto \frac{\partial
W}{\partial y}(-1,\nu,\lambda)$ is increasing on
$\left[2\sqrt{\nu},\lambda_k(\nu)\right]$ for all $\nu \in
\left(0,\nu_0\right)$. This, together with (\ref{W_y1}) above,
finishes the proof of Lemma~\ref{L2V.2}.
\qed

\section{Extensions across Killing horizons}
 \label{PSExt}

In this section, we write $H$, $F$, $J$, etc. to mean,
respectively, $H(x,y)$, $F(x,y)$, $J(x,y)$,  whereas $\hat{H}$,
$\hat{F}$ refer respectively to $H(y,x)$,
$F(y,x)$.\\

We write
$$
 G(y) = \nu (1-y^2) (y-y_h)(y-y_c)
 \;,
$$
and we assume $y_h\ne y_c$. In the calculations that follow we
suspend the convention $y_h>y_c$, used elsewhere in this paper,
so that the analysis below applies both to the smaller and to
the larger roots of $G$.

We want to construct explicitly a Kruskal-Szekeres-type
extension of the metric at $y=y_h$ and $y=y_c$. An identical
calculation applies at both values of $y$, so  in the
calculations that follow the reader can think of the symbol
$y_h$ as representing either $y_h$ or $y_c$.

 We define new coordinates (inspired from the
extension of the Kerr metrics, see \cite{CarterKerr}, and of
the Emparan-Reall metrics, see \cite{CC}) via the equations
\bean du &=& dt + \frac{\sigma}{y-y_h}dy\; , \\
      dv &=& dt - \frac{\sigma}{y-y_h}dy\; ,
 \eeal{1X.1}
and new angular coordinates:
\bean d\hat{\psi} &=& d\psi - a dt\; ,\\
      d\hat{\varphi } &=& d\varphi  - b dt\; ,
 \eeal{1X.2}
where $a$, $b$, $\sigma$ are ($\nu$-- and $\lambda$--dependent)
constants, to be chosen shortly. In terms of the new
coordinates $(u,v,\hat{\psi},\hat{\varphi },x)$ the original
coordinate differentials $(t,y,\psi,\varphi ,x)$ read:
\beaa dt &=& \frac{du+dv}{2}\; ,\\
dy &=& \frac{y-y_h}{2\sigma}(du-dv)\; ,\\
d\psi &=& d\hat{\psi} + a \frac{du+dv}{2}\; ,\\
d\varphi  &=& d\hat{\varphi } + b \frac{du+dv}{2}\; . \eeaa
It is convenient to write
$$
 \Omega_\psi = \frac {\hat \Omega_\psi}{\hat H}\;, \quad
 \Omega_\varphi  = \frac {\hat \Omega_\varphi }{\hat H}\;,
$$
where $\hat \Omega_\psi$ and $\hat \Omega_\varphi $ are
polynomials in $x$ and $y$, with coefficients which are
rational functions of $\nu$ and $\lambda$. One can now write
the coefficients of the metric expressed in the coordinates
$(u,v,\hat{\psi},\hat{\varphi },x)$:
\beaa
 g_{uu}=g_{vv}
  &=&
 \frac14\Bigg(-\frac{\hat{H}}{H}\big(1+a\Omega_\psi + b
 \Omega_\varphi \big)^2-a^2\frac{F}{\hat{H}} - 2ab \frac{J}{\hat{H}}
 + b^2 \frac{\hat{F}}{\hat{H}}
\\
 & &
 \hspace{2cm} - \frac{2k^2 H(y-y_h)}{\sigma^2 \nu (1-\nu)^2
 (1-y^2) (x-y)^2 (y-y_c)}\Bigg) \;,
\\
&=&
 \frac14\Bigg(-\frac{\hat{H} +2a\hat \Omega_\psi +2 b
\hat \Omega_\varphi  }{H}  +a^2 g_{\psi\psi} + 2ab g_{\varphi  \psi} + b^2 g_{\varphi  \varphi } \\
& &  \hspace{2cm} - \frac{2k^2 H(y-y_h)}{\sigma^2 \nu (1-\nu)^2
(1-y^2) (x-y)^2 (y-y_c)}\Bigg) \;,\\
g_{uv} &=&
 \frac14\Bigg(-\frac{\hat{H} +2a\hat \Omega_\psi +2 b
 \hat \Omega_\varphi  }{H}  +a^2 g_{\psi\psi} + 2ab g_{\varphi  \psi} + b^2 g_{\varphi  \varphi }
\\
& & \hspace{2cm} + \frac{2k^2 H(y-y_h)}{\sigma^2 \nu (1-\nu)^2
(1-y^2) (x-y)^2 (y-y_c)}\Bigg) \;,\\
g_{u\hat{\psi}}=g_{v\hat{\psi}} &=& -\frac12 \Bigg(
\frac{\hat{H}}{H}\big(1+a\Omega_\psi
 + b\Omega_\varphi \big)\Omega_\psi + a\frac{F}{\hat{H}} + b\frac{J}{\hat{H}} \Bigg)
\\
  &= &
\frac12\Big(  a g_{\psi\psi} + b g_{\psi\varphi } -
\frac{\hat{\Omega}_{\psi}}{H} \Big)
 \;,  \\
g_{u\hat{\varphi }}=g_{v\hat{\varphi }} &=& -\frac12 \Bigg(
\frac{\hat{H}}{H}\big(1+a\Omega_\psi + b\Omega_\varphi
\big)\Omega_\varphi  + a\frac{J}{\hat{H}} -
b\frac{\hat{F}}{\hat{H}} \Bigg)
\\
  &= &
\frac12 \Big(  a g_{\varphi \psi} + b g_{\varphi \varphi } -
 \frac{\hat{\Omega}_{\varphi }}{H} \Big)
 \;,
\\
g_{\hat{\psi}\hat{\psi}} &=& -\frac{\hat{H}}{H}\Omega_\psi^2 -
\frac{F}{\hat{H}}=
g_{ {\psi} {\psi}} \;,\\
g_{\hat{\varphi }\hat{\varphi }} &=&
-\frac{\hat{H}}{H}\Omega_\varphi ^2 + \frac{\hat{F}}{\hat{H}}=
g_{ {\varphi } {\varphi }}\;,\\
g_{\hat{\psi}\hat{\varphi }} &=& -\frac{\hat{H}}{H}\Omega_\psi
\Omega_\varphi  - \frac{J}{\hat{H}}=
g_{ {\psi} {\varphi }}\;,\\
g_{xx} &=& \frac{2k^2 H}{(1-\nu)^2(x-y)^2G(x)}\;,
 \eeaa
whereas the other components vanish. Recall that the potential
singularity of the coefficients above at zeros of $\hat H$ is
an artifact of the parameterization of the metric, as explained
in Section~\ref{SHE}, regardless of the values of $a$ and $b$,
and that we have proved that there are no zeros of $H$ in the
region of interest. It should now be clear that in the new
coordinate system the metric coefficients are analytic
functions of all their arguments near $y=y_h$.

The Jacobian of the transformation relating  the two coordinate
systems  reads
$$
\frac{\partial(u,v,\hat{\psi},\hat{\varphi },x)}{\partial(t,y,\psi,\varphi ,x)}=-\frac{2\sigma}{y-y_h}\;,
$$
so that the determinant of the metric in the coordinates
$(u,v,\hat{\psi},\hat{\varphi },x)$ is
$$
\det\left(g_{(u,v,\hat{\psi},\hat{\varphi },x)}\right)=-\frac{4k^8 H^2
(y-y_h)^2}{\sigma^2 (1-\nu)^6 (x-y)^8}\;.
$$
To get rid of the zero of the  determinant  at $y=y_h$, the
usual calculation is to introduce exponential coordinates
$$ \hat{u}:=e^{\gamma u}\;,\ \hat{v}:=e^{-\gamma v}\;,$$
hence
$$ d\hat{u}=\gamma\,\hat{u}\,du\;,\ d\hat{v}=-\gamma\,\hat{v}\,dv\;.$$
We then express the metric coefficients in the  coordinates
$(\hat{u},\hat{v},\hat{\psi},\hat{\varphi },x)$. We first
notice that, for $y>y_h$,
$$\hat{u}\hat{v}=e^{\gamma (u-v)}=\exp\left(2\gamma \int^y
 \frac{\sigma}{y-y_h}dy \right)=e^{2\gamma \sigma \ln (y-y_h)} = (y-y_h)^{2\gamma\sigma}\;,
$$
with an appropriate choice of the integration constant for the
second equality above. The choice
$$
2\gamma \sigma=1\;.
$$
leads to
\bel{uvy} \hat{u}\hat{v}=y-y_h\;, \ee
which can be used to define $y$ as a function of the
exponential coordinates $\hat{u}$, $\hat{v}$. The Jacobian of
the last coordinate transformation is
$$
\frac{\partial
(\hat{u},\hat{v},\hat{\psi},\hat{\varphi },x)}{\partial(u,v,\hat{\psi},\hat{\varphi },x)}=-\gamma^2
\hat{u}\hat{v}=-\frac{y-y_h}{4\sigma^2}\;,
$$
and the determinant of the metric in the exponential
coordinates has no zeros near $y=y_h$:
$$
\det\left(g_{(\hat{u},\hat{v},\hat{\psi},\hat{\varphi },x)}\right)=
-\frac{64k^8 \sigma^2 H^2}{(1-\nu)^6 (x-y)^8}\;.
$$
The metric coefficient in the exponential coordinates read:
$$
 g_{\hat{u}\hat{u}}=\frac{g_{uu}\hat{v}^2}{\gamma^2
 \hat{u}^2\hat{v}^2}\;,\quad g_{\hat{v}\hat{v}}=\frac{g_{vv}\hat{v}^2}{\gamma^2
 \hat{u}^2\hat{v}^2}\;,\quad g_{\hat{u}\hat{v}}=\frac{g_{uv}}{\gamma^2
 \hat{u}\hat{v}}\;,\quad g_{\hat{u}\hat{\psi}}=\frac{g_{u\hat{\psi}}\,\hat{v}}{\gamma
 \hat{u}\hat{v}}\;, \quad \mbox{etc.}
$$
If we replace $\hat{u}$, $\hat{v}$ by their values in terms of
the original coordinates in the first three expressions we get
$$
g{}_{\hat{u}}{}_{\hat{u}}=\frac{e^{-2 \gamma  t}
g{}_{u}{}_{u}}{\gamma ^{2}(-y_h + y)}=g{}_{\hat{v}}{}_{\hat{v}}\;,\quad
g{}_{\hat{u}}{}_{\hat{v}}=\frac{g{}_{u}{}_{v}}{(-y_h + y)\gamma ^{2} }
$$

So, from   (\ref{uvy}), to establish regularity of the new
metric coefficients we need to check that
\begin{itemize}
  \item $g_{uu}=g_{vv}$ have a zero of order 2 at $y=y_h$,
      and that
  \item $g_{uv}$, $g_{u\hat{\psi}}=g_{v\hat{\psi}}$, and
      $g_{u\hat{\varphi }}=g_{v\hat{\varphi }}$ all vanish
      at $y=y_h$.
\end{itemize}
(If we just seek an extension through the future event horizon,
then the conditions are
\begin{itemize}
  \item $g_{uu}=g_{vv}$ and $g_{uv}$ all vanish at $y=y_h$,
      and that
  \item $g_{u\hat{\psi}}=g_{v\hat{\psi}}$, and
      $g_{u\hat{\varphi }}=g_{v\hat{\varphi }}$ all vanish
      at $y=y_h$.)
\end{itemize}
More precisely, we wish to determine the  parameters $a$, $b$
and $\sigma$ so that the conditions required above are
fulfilled. We start by solving the linear system in $a$ and
$b$:
\beaa g_{u\hat{\psi}}|_{y=y_h} &=& 0 \\
g_{u\hat{\varphi }}|_{y=y_h} &=& 0\;,
 \eeaa
which we write as (all functions evaluated at $y=y_h$):
\beaa a \left(-\frac{\hat{H}}{H}\Omega_\psi^2
-\frac{F}{\hat{H}}\right) +
b\left(-\frac{\hat{H}}{H}\Omega_\psi \Omega_\varphi
-\frac{J}{\hat{H}}\right) &=&
\frac{\hat \Omega_\psi}{H} \\
a \left(-\frac{\hat{H}}{H}\Omega_\psi \Omega_\varphi
-\frac{J}{\hat{H}}\right) +
b\left(-\frac{\hat{H}}{H}\Omega_\varphi ^2
 +\frac{\hat{F}}{\hat{H}}\right)
  &=&
  \frac{\hat{\Omega}_\varphi }{H} \;.
\eeaa
The determinant of this system reads \beaa \Delta_{a,b} &=&
\left(-\frac{\hat{H}}{H}\Omega_\psi^2
-\frac{F}{\hat{H}}\right)\left(-\frac{\hat{H}}{H}\Omega_\varphi
^2 +\frac{\hat{F}}{\hat{H}}\right)-
\left(-\frac{\hat{H}}{H}\Omega_\psi
\Omega_\varphi  -\frac{J}{\hat{H}}\right)^2 \\
&=& \frac{1}{H}\left(F\Omega_\varphi  ^2 - \hat{F}\Omega_\psi
^2 - 2J\Omega_\varphi  \Omega_\psi\right) -
\frac{F\hat{F}+J^2}{\hat{H}^2}\;, \eeaa
Then, from the identity (\ref{jidentity}), at $y=y_h$ we have
\bel{13VI.1}
 F\hat F = - J^2
 \;,
\ee
and so the last term $- {F\hat{F}+J^2}/{\hat{H}^2}$ vanishes.
Next, if we view the remaining part of $\Delta_{a,b}$ as a
second-order polynomial in $\Omega_\varphi $, it has
discriminant $4\Omega_\psi ^2(J^2+\hat{F}F) $, which vanishes
again for $y=y_h$. This leads to the simpler expression for the
determinant of the system in $a,b$:
$$
\Delta_{a,b}=\frac{F}{H}\left(\Omega_\varphi  -
\frac{J\Omega_\psi}{F}\right)^2\;.
$$
Next, assuming that $\Delta_{a,b}$ does not vanish at $y=y_h$,%
\footnote{In fact, the vanishing or not of this determinant is
irrelevant, insofar as we check that the values of $a$ and $b$
that are calculated below give the answer we need.}
we can write the expressions of $a$ and $b$ solving the system:
\beaa
 a &=&  \frac{\hat{F}\Omega_\psi +
 J\Omega_\varphi }{H\Delta_{a,b}}\;,\\
 b &=&  \frac{J\Omega_\psi -
 F\Omega_\varphi }{H \Delta_{a,b}}
 \;.
 \eeaa
Using \eq{13VI.1}, this can be rewritten as
\beaa
 a &=& \frac{J}{F\Omega_\varphi  - J\Omega_\psi}\;,\\
 b &=& -\frac{F}{ F\Omega_\varphi -J\Omega_\psi }\;.
\eeaa
We insert in this expression the explicit values of
$\Omega_\psi$ and $\Omega_\varphi $:
\begin{eqnarray*}
&&a=\frac{H(y_h,x) J(x, y_h)(-1 + \lambda -\nu )}{2k\lambda\sqrt{(1+\nu)^2-\lambda^2}}\times\\
&&\bigg(y_h\sqrt{\nu}F(x, y_h)(-1 + x^{2}) (-1 + \lambda -\nu )+\\ && +J(x, y_h) (1 + y_h)\big(-1-\lambda+ \nu  + 2 \nu  x (-1 + y_h) + y_h \nu x^{2}
(-1 + \lambda  + \nu )\big)\bigg)^{-1},\\
&& b=\frac{-F(x,y_h)H(y_h, x)(-1 + \lambda -\nu )}{2k\lambda\sqrt{(1+\nu)^2-\lambda^2}}\times\\
&&\bigg(y_h \sqrt{\nu}F(x, y_h) (-1 + x^{2})(-1 + \lambda-\nu )+\\
&&+J(x, y_h)(1 + y_h)\big(-1 -\lambda  + \nu  + 2 \nu  x (-1 + y_h) + y_h \nu x^{2}
(-1 + \lambda  + \nu )\big)\bigg)^{-1}.
\end{eqnarray*}

We need to check that $a$ and $b$ are $x$--independent.  For
this, we found it convenient
to replace $\lambda$ by a parameter $t\in \R$ defined as (we
hope that a conflict of notation with the time coordinate $t$
will not confuse the reader)
\bel{deft13VI.1}
 \lambda = 2 \sqrt{\nu} \cosh t
 \;.
\ee
With this redefinition we have
$$
 y_h = - \frac{e^{-t}}{\sqrt \nu}\;,
 \quad
 y_c = - \frac{e^t}{\sqrt \nu}\;.
$$
Thus, the transition from $y_h$ to $y_c$ is obtained by
changing $t$ to its negative.

Using {\sc Mathematica}, the expressions above, evaluated at
$y=y_h$, are indeed $x$--independent as desired, and take the
form
\beaa
 a &=&  \frac{(1- \sqrt{\nu} e^{-t})(1- \sqrt{\nu} e^{t})  }{2k\sqrt{1+\nu^2-2\nu \cosh(2t)}}\;, \\
 b &=&  \frac{( \sqrt{\nu}-e^{-t}) ( \sqrt{\nu}-e^{ t}) (1+e^{2t} {\nu})}{2k\sqrt{\nu}(1+e^{2t})\sqrt{1+\nu^2-2\nu
 \cosh(2t)}}\;.
\eeaa
Note that the value of $a$ is the same both for both horizons,
but that of $b$ is not.

We continue by checking that the remaining metric coefficients
vanish with these values of $a$ and $b$. First, one finds
directly that $1 + a\Omega_\psi + b\Omega_\varphi $ vanishes at
$y=y_h$. Next, the last term in $g_{uu}|_{y=y_h}$ vanishes,
while the remaining term in the expression of $g_{uu}|_{y=y_h}$
reads
$$
-\frac{a^2 F + 2abJ-b^2\hat{F}}{4\hat{H}}  =
\frac{F(J^2+F\hat{F})}{4\hat{H}(F\Omega_\varphi -J\Omega_\psi)^2}\;.
$$
This vanishes at $y=y_h$ by (\ref{13VI.1}). The calculation
also shows that both $g_{uu}$  and $g_{uv}$   vanish at
$y=y_h$.

Next, a {\sc Mathematica} calculation 
shows that $g_{uu}$ will have a second order zero at $y_h$ if
and only if the constant $\sigma$
 equals
$$
\sigma=\pm \frac{2k\sqrt{\nu}(e^t+\sqrt{\nu})\coth(t)}{(1-\nu)(1-e^t
\sqrt{\nu})}\;;
$$
%

Reexpressing everything in terms of the original parameters,
our calculations in this section can be summarized as follows:
The metric functions $g_{uu}$ and $g_{uv}$  vanish at $y=y_h$
when the parameters $a$ and $b$ take the values
\begin{eqnarray}
 a&=&\frac{1-\lambda  + \nu}{2k\sqrt{(1 + \nu)^{2}-\lambda ^2   }}
  \;,\label{eq:ablambdanu1}
\\
 b&=&\frac{\left((\nu-1)\sqrt{\lambda^2-4\nu}+\lambda(1+\nu)\right)(1+\nu-\lambda)}
 {4k\lambda\sqrt{\nu((1+\nu)^2-\lambda^2)}}\label{eq:ablambdanu2}
 \;.
\end{eqnarray}
With this choice, $g_{uu}$ has a second order zero at $y=y_h$
if and only if $\sigma$ takes the value
\begin{equation}
\sigma ^2=\frac{2 k^2 \lambda ^2 \left(\lambda ^2
   \left(\nu ^2+1\right)-\lambda  \left(\nu ^2-1\right)
   \sqrt{\lambda ^2-4 \nu }-2 \nu  (\nu +1)^2\right)}{(\nu
   -1)^2 \left(\lambda ^2-4 \nu \right) (-\lambda +\nu
   +1)^2}.
\label{eq:sigmasquare}
\end{equation}
At $y=y_c$ the analogous analysis leads to the same value of
$a$, while $b$ and $\sigma$ are now
\begin{eqnarray}
 b&=&\frac{\left((-\nu+1)\sqrt{\lambda^2-4\nu}+\lambda(1+\nu)\right)(1+\nu-\lambda)}
 {4k\lambda\sqrt{\nu((1+\nu)^2-\lambda^2)}}\;,\label{eq:byc}\\
 \sigma ^2&=&\frac{2 k^2 \lambda ^2 \left(\lambda ^2
   \left(\nu ^2+1\right)+\lambda  \left(\nu ^2-1\right)
   \sqrt{\lambda ^2-4 \nu }-2 \nu  (\nu +1)^2\right)}{(\nu
   -1)^2 \left(\lambda ^2-4 \nu \right) (-\lambda +\nu
   +1)^2}.\label{eq:sigmayc}
\end{eqnarray}
%
We note that the values obtained for the constants
$a$, $b$ and $\sigma$ are all well defined under the
assumptions of eq. (\ref{11IV10.5}) together with
$\lambda^2-4\nu>0$ ($y_h$ and $y_c$ real and distinct) and
hence the extension through the Killing horizons will remain
valid for any member of the PS family of solutions whose
parameters meet these requirements.

\subsection{Degenerate case}

 \label{xPSExt}

%
It was shown in~\cite{KunduriLuciettiReall} that the
near-horizon limit of the degenerate PS solutions admits smooth
extensions. Here we check that the method presented there for
extending across a degenerate horizon applies to the PS
metrics, and not only their near-horizons limits.

The first step is to define new coordinates
$\hat{t},\hat{\psi},\hat{\varphi}$, not to be confused with the
hatted coordinates defined previously in this section in the
non-degenerate case,

We now follow for our case the method developed
in~\cite{KunduriLuciettiReall}. In the original coordinates
$(t,y,\psi,x,\varphi)$, the PS metrics do not satisfy the
requirement that the coefficients $g_{t\psi}$, $g_{t\varphi}$
and $g_{tt}$ vanish at the degenerate horizon
$\{y=y_0:=-\frac{1}{\sqrt{\nu}}\}$, with cancelation at order
two for $g_{tt}$. So the first step is to define new
coordinates $\hat{t},\hat{\psi},\hat{\varphi}$, not to be
confused with the hatted coordinates defined previously in this
section in the non-degenerate case:
$$\hat{t}=t\ ,\ \hat\psi=\psi - a_\psi t\ ,\ \hat\varphi=\varphi-a_\varphi t\;,$$
where $a_\psi$ and $a_\varphi$ are constants. Written in these
coordinates, the conditions for the metric coefficients
$g_{\hat t \hat\psi}$ and $g_{\hat t \hat\varphi}$ to vanish at
$y_0$ read again:
\beaa a_\psi \left(-\frac{\hat{H}}{H}\Omega_\psi^2
-\frac{F}{\hat{H}}\right) +
a_\varphi\left(-\frac{\hat{H}}{H}\Omega_\psi \Omega_\varphi
-\frac{J}{\hat{H}}\right) &=&
\frac{\hat \Omega_\psi}{H} \\
a_\psi \left(-\frac{\hat{H}}{H}\Omega_\psi \Omega_\varphi
-\frac{J}{\hat{H}}\right) +
a_\varphi\left(-\frac{\hat{H}}{H}\Omega_\varphi ^2
 +\frac{\hat{F}}{\hat{H}}\right)
  &=&
  \frac{\hat{\Omega}_\varphi }{H} \;,
\eeaa
where all the functions are evaluated at
$y=y_0=-\frac{1}{\sqrt{\nu}}$. Here, since
$\lambda=2\sqrt{\nu}$, we obtain that
\beaa a_\psi &=& \frac{1-\nu}{2k(1+\sqrt{\nu})^2}\\
a_\varphi &=& \frac{1-\nu^2}{4k\sqrt{\nu}(1+\sqrt{\nu})^2}
\eeaa
are solutions, i.e. make the coefficients $g_{\hat t \hat\psi}$ and $g_{\hat t \hat\varphi}$ vanish at $y_0$. 
The metric is then of the form of equation (41)
in~\cite{KunduriLuciettiReall}, where $R$ is replaced by
$y-y_0$, the indices $i,j=1,2$ refer to the angular coordinates
$\hat\psi$ and $\hat\varphi$, and $t$ is replaced by $\hat t$. Explicitely we have:
\bel{metricKLR}
g=g_{\hat t \hat t}d\hat t ^2 + 2g_{\hat t i}d\hat t d\hat\varphi^i + g_{ij}d\hat\varphi^i d\hat\varphi^j
+ g_{yy}dy^2 + g_{xx}dx^2\;,\ee
with
\bel{metricKLRconditions} g_{\hat t\hat t}=f_t(y,x)(y-y_0)^2,\ \ g_{\hat t i}=f_i(y,x)(y-y_0),\ \ g_{yy}=\frac{h(y,x)}{(y-y_0)^2}\;,\ee
for some functions $f_t$, $f_\psi$, $f_\varphi$ and $h$,
bounded at $y=y_0$ in their first variable.

\bigskip

Then, in order to remove the singularity at the horizon $y=y_0$, we define new coordinates $(v,z,\Psi,\Phi)$, such that:

%
\beaa
\hat t &=& v-\frac{a_0}{z-y_0}+f(z,x)
 \;,
\\
y &=& z
 \;,
\\
 \hat\psi &=& \Psi + b^\psi \ln (z-y_0) 
 \;,
\\
 \hat\varphi &=& \Phi + b^\varphi \ln (z-y_0) 
 \;,
\eeaa
where $a_0$, $b^\psi$ and $b^\varphi$ are constants. Our aim is
to find values of those constants for which the PS metrics, written in the coordinates $(v,z,\Psi,\Phi)$ , are regular at $z=y_0$.
In fact, we choose   $f(z,x)=a_1 \ln(z-y_0)$, where $a_1$ is a
constant, as in~\cite{KunduriLuciettiReall}.
Thus, the coordinate transformations
read
\beaa
d\hat{t} &=& dv + \left(\frac{a_0}{(z-y_0)^2}+\frac{a_1}{z-y_0}\right)dz\\
dy &=& dz\\
d\hat\psi &=& d\Psi + \frac{b^\psi}{z-y_0}dz\\
d\hat\varphi &=& d\Phi + \frac{b^\varphi}{z-y_0}dz\;. \eeaa
Using these coordinate transformations along with the
equalities~(\ref{metricKLR}) and ~(\ref{metricKLRconditions}),
we compute the metric coefficients in the new coordinate system
$(v,z,\Psi,x,\Phi)$:
$$g_{vv}=g_{\hat t\hat t}\ ,\ \ g_{vz}=(a_0 + (z-y_0)a_1) f_t + b^i f_i \ ,\quad  g_{vi}=g_{\hat t i}\ ,$$
$$ g_{zi}=\Big((a_0 + (z-y_0)a_1) f_i
+ g_{ij}b^j\Big)\frac{1}{z-y_0}\ ,$$
for $i=1,2$ referring to the variables $\Psi$ and $\Phi$, and
$$g_{zz}=\Big( (a_0 + (z-y_0)a_1)^2 f_t + 2 (a_0 + (z-y_0)a_1) b^i f_i + g_{ij}b^i b^j + h\Big)\frac{1}{(z-y_0)^2}\;.$$
This shows that the metric coefficients, to be smooth at $z=y_0$, must satisfy:
$$(z-y_0)^2 g_{zz}|_{z=y_0}=0\;,\ \frac{\partial}{\partial z}((z-y_0)^2 g_{zz})|_{z=y_0}=0\;,\ (z-y_0)g_{zi}|_{z=y_0}=0$$
for $i=1,2$.
Therefore we can derive the conditions that the constants
$a_0$, $a_1$, $b^\psi$ and $b^\varphi$ should satisfy to yield
a smooth metric at $z=y_0$. Those read:
\beaa \label{Extdegen}
a_0 f_\psi  + g_{\psi i}b^i &=& 0\\
a_0 f_\varphi  + g_{\varphi i}b^i &=& 0\\
h + a_0 ^2 f_t  + 2 a_0 b^i f_i + g_{ij}b^i b^j &=& 0\\
\partial_z h + a_0 ^2 \partial_z f_t + 2 a_0 b^i \partial_z f_i + \partial_z g_{ij} b^i b^j + 2 a_1 a_0 f_t + 2 a_1 b^i f_i &=& 0\;.
\eeaa
Before making any attempt to solve this system, note that we
can slightly simplify it:
\beaa \label{Extdegen2}
a_0 f_\psi  + g_{\psi i}b^i &=& 0\\
a_0 f_\varphi  + g_{\varphi i}b^i &=& 0\\
h + a_0 ^2 f_t  + a_0 b^i f_i &=& 0\\
\partial_z h + a_0 ^2 \partial_z f_t + 2 a_0 b^i \partial_z f_i + \partial_z g_{ij} b^i b^j -2\frac{a_1}{a_0}h &=& 0\;.
\eeaa

Then, we start by looking for solutions in
$(a_0,b^\psi,b^\varphi)$ of the first three equations of the
system~(\ref{Extdegen2}) above, and we begin with the special
case $x=0$. The calculations are then tractable using
\textsc{Mathematica}, and we obtain the following two triplets
of solutions:
$$(a_0,b^\psi,b^\varphi) \in \left\{\left(\frac{4k}{(1-\sqrt{\nu})^2},0,-1\right),\left(-\frac{4k}{(1-\sqrt{\nu})^2},0,1\right)\right\}\;.$$
Next, we insert these values in the left-hand side of the three first equalities in~(\ref{Extdegen2}) (this time for any
value of $x$), and we still obtain zero at the corresponding right-hand sides.
Moreover, the value of $a_1$ is imposed by the fourth equation
and the values of $a_0$, $b^\psi$ and $b^\varphi$. Finally, we
obtain two solutions:
\begin{eqnarray*}
 \lefteqn{
(a_0,b^\psi,b^\varphi,a_1) \in
\left\{\left(\frac{4k}{(1-\sqrt{\nu})^2},0,-1,-\frac{4k
\sqrt{\nu}(1+\nu)}{(1-\sqrt{\nu})^2(1-\nu)}\right),\right.
}
 &&
\\
&&
\phantom{xxxxxxxxxxxxxxxx}
\left.
\left(-\frac{4k}{(1-\sqrt{\nu})^2},0,1,\frac{4k
\sqrt{\nu}(1+\nu)}{(1-\sqrt{\nu})^2(1-\nu)}\right)\right\}\;.
\end{eqnarray*}
It is important to note that these are \emph{not} functions of
$x$ but constants, as desired. We may ask whether other
solutions exist for this system. But the determinant of the
linear system of the first two equations in the variables
$b^\psi$ and $b^\varphi$, at fixed $a_0$, is
$$\det\big((g_{ij})_{1\leq i,j \leq 2}\big)|_{y=y_0}=\frac{32k^4 \nu (1+\sqrt{\nu})(1-x^2)}{(1-\sqrt{\nu})^3 (1+\nu + 4x\sqrt{\nu} + x^2 (1+\nu))}\;,$$
and this last expression is always positive for any allowed values
of $\nu$ and $x$, except at the axis $x=\pm 1$ where it
vanishes. Hence, one can obtain each of $b^\psi$ and
$b^\varphi$ in terms of $a_0$ from the two first equations.
Next, the third equation, when replacing the $b^i$'s, becomes a
quadratic equation in $a_0$, hence admitting no more than two
solutions. Finally, $a_1$ is uniquely determined by the fourth
equation from $a_0$ and the $b^i$'s.

\medskip

We conclude that, when performing a transformation of the
coordinates the way described above, these values of the
parameters $a_0$, $a_1$, $b^\psi$ and $b^\varphi$ give two
coordinate systems in which the metric is smooth at the
degenerate horizon $y=y_0$, and we can therefore locally
analytically extend the PS metric across the horizon in the
degenerate case. One choice corresponds to an extension through
the future event horizon, the other through the past event
horizon.
Note also that the values of the parameters
$a_0$, $a_1$, $b^\psi$ and $b^\varphi$ are all well-defined away from $\nu=1$ and hence the computations of this subsection remain valid for any
member of the PS family with a degenerate horizon and $\nu\neq 1$.

\section{Some local and global properties}

\subsection{Stable causality of the domain of outer communications?}
 \label{ss8IX.1}

There is a well developed theory of black hole
uniqueness~\cite{HY,HY3,ChCo} which requires various global
regularity conditions. In particular the domain of outer
communications should be globally hyperbolic, and the orbits of
the group generated by the periodic Killing vectors should be
spacelike or trivial. We do not know whether or not these
properties hold for the solutions at hand, and we note that the
proof of global hyperbolicity for Emparan-Reall
metrics~\cite{CC} required a considerable amount of work,
including a detailed understanding of causal geodesics. The aim
of this section is to give numerical evidence for stable
causality of the domain of outer communications. It is likely
that a proper understanding of the minimisation algorithms,
used in the calculations reported below, can be invoked to
rigorously establish the result, at least for a ``big" set of
parameters, but this lies beyond the scope of our current
analysis.

We start with the question of the causal character of the
orbits of the Killing vectors $\partial_\varphi$ and
$\partial_\psi$: should a linear combination of those vectors
become null, one would immediately obtain violation of strong
causality, or even causality. To analyze this, one needs to
know whether the determinant of the two by two matrix  obtained
by taking the scalar products of the periodic Killing  vectors
has a sign on the region of interest.

This problem turns out to be closely related to the question of
stable causality of the d.o.c. Indeed, from the form of the
metric together with \eq{7IX.4} one finds
\bean
 g(\nabla t, \nabla t) = g^{tt} &=& \frac{g_{xx}
 g_{yy}}{\det g_{\mu \nu}}\det
 \left( \begin{array}{cc}
          g_{\psi \psi} & g_{\psi \varphi} \\
          g_{\psi \varphi} & g_{\varphi \varphi}
        \end{array}
        \right)
\\
 & = &\frac{   (\nu -1)^2
(x-y)^4}{ 4  k^4 G(x) G(y)   }\det
 \left( \begin{array}{cc}
          g_{\psi \psi} & g_{\psi \varphi} \\
          g_{\psi \varphi} & g_{\varphi \varphi}
        \end{array}
        \right)\nonumber\\
&=&\frac{(1+y)(1-x^2)\Theta(x,y,\lambda,\nu)}{(1-\lambda+\nu)H(x,y)G(x)G(y)}
 \;,
\eeal{7IX.2}
where $\Theta$ is the \emph{polynomial} defined in
(\ref{eq:detgphipsi}). Recall that $G(y)$ is negative for $y\in
(y_h,-1)$ while $G(x)$ is positive for $x\in (-1,1)$, and note
that the zeros of $G$ at $x=\pm 1$ and $y=-1$ are canceled by
factors in the numerator. We conclude that $t$ will be a time
function on the region $y>y_h$, and thus the d.o.c. will be
stably causal, if the polynomial $\Theta$ is \emph{strictly
negative}.  The second line of \eq{7IX.2} shows that the
principal orbits of the group generated by $\partial_\varphi$
and $\partial_\psi$ will then be spacelike, as desired.%
\footnote{The strict negativity of $\Theta$ has been
established in \cite{CSPS} after the research reported on here
has been completed.}

Removing some obvious prefactors, this reduces the study to
that of the sign  of $\Theta(x,y,\lambda,\nu)$  on  the closure
of $ {\Xi}_0\cap \{y>y_h\}$, where
\bel{Omx} \Xi_0\equiv\{(x,y,\lambda,\nu) \in \mathbb{R}^4\ ;\
-1\leq x\leq 1\;,\ y_c\leq y< -1\;,\ 0<\nu<1\;,\
2\sqrt{\nu}\leq \lambda < 1+ \nu \}\;, \ee

Visual inspection of various slices of its graph suggests that
$\Theta$ has no zeros on $\Xi_0$, though there are zeros for
the borderline values of $\nu$ and $\lambda$. Since $\Theta$ is
a continuous function on $\overline{\Xi}_0$, we thus expect
that $\Theta$ has a constant sign in int$(\Xi_0)$, which can be
found by picking an arbitrary point in int$(\Xi_0)$ and
computing the value of $\Theta$ there. In our case we have,
e.g.,
$$
\Theta\left(-\frac{1}{2},\frac{6}{5},\frac{18}{625}+\sqrt{2},\frac{1}{2}\right)\approx
-23.4<0 \;,
$$
which in particular proves stable causality at, and near, those
values of parameters.

\begin{figure}[t]
\begin{center}
 \includegraphics[width=.4\textwidth,bb=0 0 300 174]{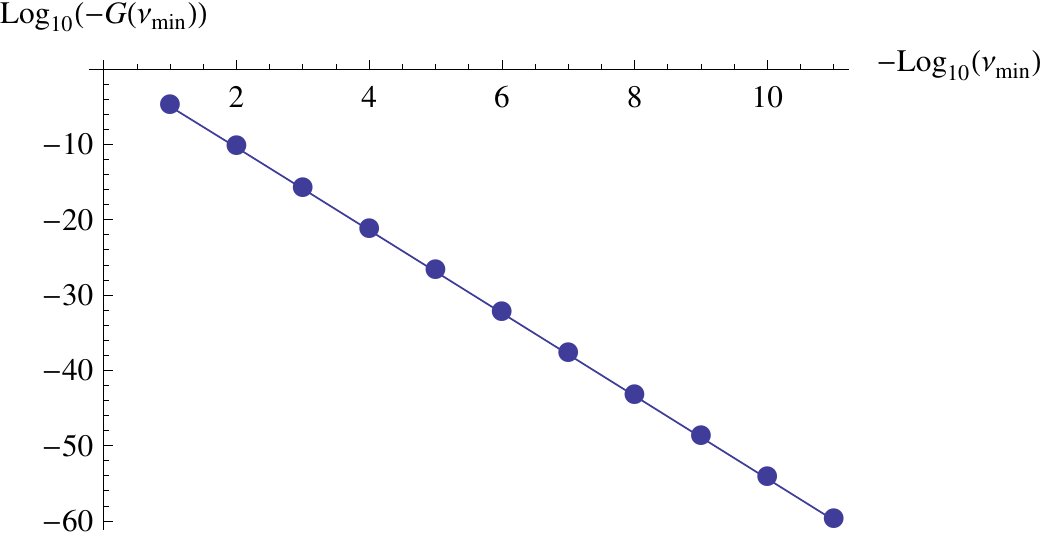}
\end{center}
\caption{
\label{F14IX.1}
Logarithmic graph of $-G(\nu_{\min{}})$ in terms of $\log_{10}\nu_{\min{}}$,
suggesting very strongly that $\Theta$
is strictly negative for admissible $\nu$ and $\lambda$.}
\end{figure}

\begin{figure}[h]
\begin{center}
 \includegraphics[width=.4\textwidth,bb=0 0 300 174]{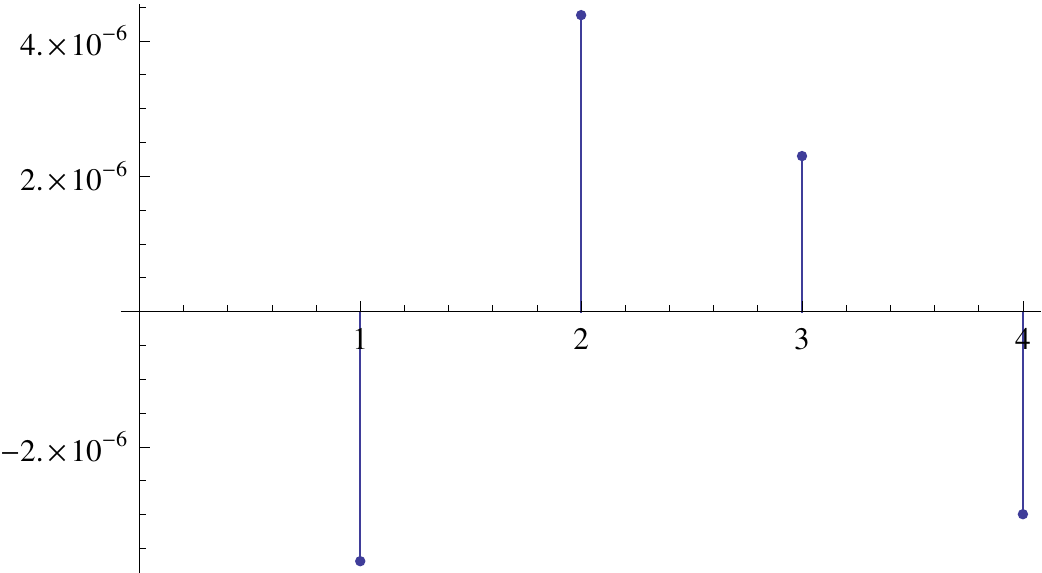}
\end{center}
\caption{
\label{F14IX.0}Graphical representation of
the residuals $r_i$, $i=8,\dots, 11$ for the nonlinear regression of the computed values
of $G(\nu_{\min{}})$ to the model shown in (\ref{eq:regression-model}).}
\end{figure}

To obtain stronger evidence about the sign throughout the
region of interest we proceed as follows: Since $y_c\to
-\infty$ for $\nu\to 0$, $\overline{\Xi}_0$ is unbounded, and
since $\Theta$ has zeros at $\lambda = 1 +\nu$ (compare
\eq{9IX.1} below),  we introduce a cut-off $\nu_{\min}$ and set
$$
 \Xi_{\nu_{\min{}}}:= \overline{\Xi}_0\cap \{\nu \ge \nu_{\min{}}\}\cap
\{\lambda\leq1+\nu-\nu_{\min{}}\}
 \;.
$$
Define
$$
G(\nu_{\min{}}):= \max_{\Xi_{\nu_{\min{}}}} \Theta\;,
$$
We have computed a numeric approximation to the value of the
function $G(\nu_{\min{}})$ for a series of small values of
$\nu_{\min{}}$.  The result is shown in Figure~\ref{F14IX.1}.
For example
$$
G(1/10)=-0.0000106491106406\pm 10^{-17}\;.
$$
The calculation has been done within {\sc Mathematica} using
the function \verb|NMaximize|, and the error above is the value
given to the option \verb|AccuracyGoal|. The small value
obtained is perplexing, but is consistent with our various
tests and the remaining calculations. The requirement of
convergence of {\sc Mathematica}'s minimisation algorithms
forces the values of $G(\nu_{\min{}})$ to be computed with
greater numerical accuracies when $\nu_{\min{}}$ is smaller, to
account for the smaller number, and for $\nu_{\min{}}=1/100$
the accuracy parameter used was approximately  100 digits after
the decimal point. The graph suggests very strongly that
$\Theta$ is strictly negative away from the extreme vales of
$\nu$ and $\lambda$. We have repeated the computations with
accuracy parameter up to 700 digits after the decimal point and
there is a clear convergence in the results, but we did not
produce graphs to illustrate this claim.   One can fit the
points of the graph in Figure~\ref{F14IX.1} to a straight line
defined by the expression
\begin{equation}
\log_{10}(-G(\nu_{\min{}})) = -A\log_{10}\nu_{\min{}}+B  \;.
\label{eq:regression-model}
\end{equation}
We have successively fitted the values of $A$ and $B$ to all
the points on the graph, then to the last ten points on the
graph, etc., with a last fit to the last four points on the
graph. The values of $A$ obtained in this way formed an
increasing sequence, with increasingly smaller variations, with
the fit for the last four points yielding
\bel{15X.1}
 A\approx -5.499993\;,\quad B\approx 0.601987
 \;.
\ee
This suggests very strongly that the correct limiting exponent
is $-5.5$. This half-integer value is somewhat perplexing, but
can probably be explained by the square-root behaviour of the
lower bound for $\lambda$ in terms of $\nu$. The accuracy
required to obtain further points on the graph is beyond the
capabilities of the computing resources available to us. The
values of the {\em residuals}, say $r_i$, $i=8,\ldots, 11$, of
this linear regression are represented in Figure~\ref{F14IX.0}.
From these residuals, we get the following value for the
variance $\sigma = \sqrt{\frac {1}{4} \sum_{i=8}^{11} r_i^2}$:
\nopagebreak
$$
\sigma\approx 3.43\times 10^{-6}.
$$

\begin{figure}[h]
\begin{center}
 \includegraphics[width=.4\textwidth,bb=0 0 300 174]{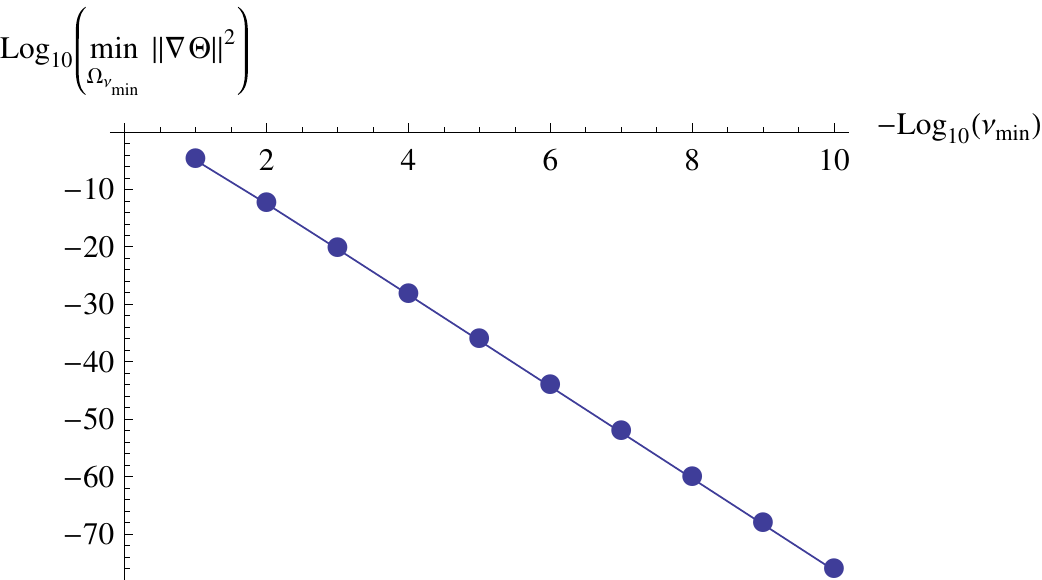}
\end{center}
\caption{\label{fig:gradtheta}A logarithmic graph of
$\min_{\Xi_{\nu_{\min{}}}}\|\nabla\Theta\|^2$ in terms of $-\log_{10}\nu_{\min{}}$, suggesting very strongly that
$\Theta$ has no critical points in the region of interest, for admissible values of $\nu$ and $\lambda$. The
accuracy parameter used for the computed values of $\min_{\Xi_{\nu_{\min{}}}}\|\nabla\Theta\|^2$
was 100 digits after the decimal point for $\nu_{\min{}}=10^{-10}$.}
\end{figure}

The numerical study also indicates that the function $\Theta$
does not have any critical points in ${\Xi}_0$ (we have not
been able to verify this analytically either). This is
illustrated in Figure~\ref{fig:gradtheta}.
If this is indeed the case, then the extrema of $\Theta$ lie on
the boundary of $\overline{\Xi}_0$, which should be simpler to
analyze. Indeed, the polynomial $\Theta(x,y,\lambda,\nu)$
adopts a simpler form in certain regions of the boundary of
$\overline{\Xi}_0$. These are:\nopagebreak%
\begin{eqnarray}
\Theta(x,y,1+\nu,\nu)&=&-8 (\nu +1)^2 (x-y)^2 (x \nu +1) (y \nu +1)\times\nonumber\\
&&\hspace{-2cm}\left(x
   \left(y^2 \nu ^2-\left(y^2+2 y-1\right) \nu
   +1\right)+y^2 \nu ^2+\left(y^2-2 y-1\right) \nu
   +1\right),\label{eq:tildetheta}\\
\Theta(x,y,\lambda,0)&=&-(x\lambda+1)\bigg((2x^2-1)\lambda^3+(6 x^2-4 x+1)
   \lambda ^2+\nonumber\\
&&+(4 x-1) \lambda +y^2 (\lambda +1)^3\lambda +1+\nonumber\\
&&+y \left(2 x^2 (\lambda -1)\lambda^2-4 x \lambda(2 \lambda ^2+\lambda +1)
   -\lambda^4+2\lambda^3-1\right)\bigg),\\
\Theta(1,y,\lambda,\nu)&=&-(y-1) (\lambda +\nu +1)^3
\left(y^2 \nu -y\lambda+1\right)\times\nonumber\\
&&\left(\lambda  \left(y^2 \nu-1\right)+(\nu -1)(y^2 \nu +1)\right)\;,\\
\Theta(-1,y,\lambda,\nu)&=&
\big((\nu -1)(y^2 \nu +1)-\lambda(y^2 \nu -1)\big)\times\nonumber\\
&&\big(A(\lambda,\nu)y^3+B(\lambda,\nu)y^2+C(\lambda,\nu)y+D(\lambda,\nu)\big)\;,
\label{eq:theta_y_minusone}\\
\Theta(x,-1,\lambda,\nu)&=&
2 \bigg(x^2 \nu  \left((\lambda -2) \lambda +\nu ^2-1\right)+2 x
   \lambda  \left(\nu ^2-1\right)-\lambda ^2+2 \lambda  \nu +\nu
   ^2-1\bigg)\times\nonumber\\
&&\bigg(x^2 \left((\nu +2) \left(\lambda ^2+\nu
   ^2\right)+\nu \right)+x \lambda  \left(\lambda ^2+3 (\nu
   +1)^2\right)+\nonumber\\
&&+\lambda ^2 (2 \nu +1)+(\nu +1)^2\bigg)\;,\\
\Theta(x,y_c,\lambda,\nu)&=&\frac{\lambda ^2(\lambda+\nu+1)^2
(x(x\nu +\lambda)+1)}{\nu^3}\times\nonumber\\
&&\bigg(\lambda\nu\left(x^2 \nu\left(\mathcal{D}-1\right)-2 x
\left(\mathcal{D}+3 \nu \right)-2\mathcal{D}+3\right)+\nonumber\\
&&+\lambda ^2 \big(\nu\left(x\left(x \nu +2\mathcal{D}-2\right)-4\right)-
\mathcal{D}\big)+\nonumber\\
&& \nu
   \big(\mathcal{D}-\nu\left(x \left(x
   \mathcal{D}+2 x \nu +2\mathcal{D}
-4\right)-2\right)\big)\nonumber\\
&& +\lambda^3 \left(2 x \nu
   +\mathcal{D}-1\right)+\lambda ^4\bigg)
   \;,
\end{eqnarray}
where
$$
\mathcal{D}:=\sqrt{\lambda^2-4\nu}\;,
$$
and where $A(\lambda,\nu),\dots, D(\lambda,\nu)$ are
polynomials in $\lambda$, $\nu$. We have not been able to find
a simple form of $ \Theta(x,y,2\sqrt\nu,\nu)$.

From the first of the previous set of relations one finds
\bel{9IX.1}
  \Theta(-1,-1,1+\nu,\nu)=0\;,\quad
 \Theta\left(x,-\frac{1}{\nu},1+\nu,\nu\right)=0
 \;.
\ee
The above points, at which $\Theta$ is zero, lie on the
boundary of $\Xi_0$. Our numerical analysis suggests that these
are the only points of $\overline{\Xi}_0$ on which
$\Theta(x,y,\lambda,\nu)$ vanishes. Note that the zeros of the
factor $y^2 \nu -y \lambda+1$ in $ \Theta(1,y,\lambda,\nu)$ are
$-y_h$ and $-y_c$, which are positive and therefore do not lead
to zeros of $\Theta$ on $\overline{\Xi_0}$.

From the expression for $ \Theta(1,y,\lambda,\nu)$ we easily
deduce
$$
 \lim_{\nu\rightarrow 0^+}\Theta(1,y,\lambda,\nu)=
 (y-1)(\lambda +1)^4 (1-y \lambda )
 \;,
$$
and, since $y_c\to-\infty$ when $\nu\to 0$, we find
$$
 \lim_{\nu_{\min{}}\to 0}\inf_{\Xi_{\nu_{\min{}}}}\Theta=-\infty
 \; .
$$

Inspection of graphs indicates that causality violations are a
typical feature of the solutions in the region $y<y_c$, as
illustrated by Figures~\ref{Fsummary} and \ref{Fsummary2}, and
are always present near $y=1$ in any case.

\subsection{No struts}
In this section we verify the regularity of the metric at the
rotation axes.

\subsubsection{$\psi$ axis: $y=-1$}
 \label{sss7IX.1}

First of all, note that
$$
g_{\psi\psi}|_{y=-1}=0
 \;.
$$
The $\psi$-$y$ part of the metric can be cast in the form
$$
ds^2=
-\frac{2 k^2 H(x,y) (y+1)}{(\nu -1)^2 G(y)(x-y)^2}\left(\frac{(\nu -1)^2 G(y)
(x-y)^2\left(\frac{F(x,y)}{H(y,x)}+
\frac{H(y,x)M(x,y)^2}{H(x,y)}\right)}{2 k^2 (y+1)H(x,y)}d\psi^2+\frac{dy^2}{y+1}
\right).
$$
The conformal factor is regular, bounded away from zero,
provided that $x\ne y$ and $y_h<y\ne y_c$. On the other hand
$$
 \lim_{y\to -1}
\frac{(\nu -1)^2 G(y)
(x-y)^2\left(\frac{F(x,y)}{H(y,x)}+
\frac{H(y,x)M(x,y)^2}{H(x,y)}\right)}{2 k^2 (y+1)^2H(x,y)}=4
\;.
$$
This shows that the usual quadratic change of variables,
$y+1=\rho^2$, leads to a smooth metric near a rotation axis
$\rho=0$ provided that $\psi$ is a $2\pi$-periodic angular
coordinate.

\subsubsection{$\varphi$ axis: $x=\pm 1$}
\label{sss7IX.2}

Here we are interested in the behaviour of the metric near
$x=\pm 1$, where again
$$
g_{\varphi\varphi}|_{x=\pm 1}=0 \;.
$$
Similarly to the analysis in Section~\ref{sss7IX.1}, we write
$$
ds^2=
\frac{2 k^2 (x\pm 1)
   H(x,y)}{(\nu-1)^2 G(x) (x-y)^2}\left(\frac{dx^2}{x\pm 1}-
\frac{(\nu-1)^2 G(x) (x-y)^2\left(\frac{H(y,x)
   P(x,y)^2}{H(x,y)}-\frac{F(y,x)}{H(y,x)}\right)}
{2 k^2 (x\pm 1) H(x,y)}d\varphi^2\right) \;.
$$
One finds again a well behaved conformal factor on $\Omega_0$
away from $\{x=y\}$, and
$$
\lim_{x\to \pm 1} \frac{(\nu-1)^2 G(x) (x-y)^2\left(\frac{H(y,x)
   P(x,y)^2}{H(x,y)}-\frac{F(y,x)}{H(y,x)}\right)}{2 k^2 (x\pm 1)^2H(x,y)}
=4 \;.
$$
Imposing $2\pi$-periodicity on $\varphi$, we conclude that, as
long as one stays away from the set $y\in [-1,1]$, the
coordinates $(x,\varphi)$ are  coordinates on two-spheres.

\subsubsection{$\hat \psi$ axis: $y=1$}
 \label{ssshpsi}

The Killing vector
\begin{equation}
 \hat{\xi}:=\frac{\partial}{\partial t}
+\underbrace{\frac{\sqrt{(1+\nu)^2-\lambda^2}}{4k\lambda}}_{=:\alpha}
 \frac{\partial}{\partial\psi}
  \; .
\label{eq:killing-horizon2}
\end{equation}
is spacelike near $\{y=1\}$, and lies in the kernel of $g$ at
$y=1$, in the sense that
\bel{28X.1}
 \lim_{y\to 1}g(\hat \xi,\cdot) =0
 \;.
\ee
If we use  a new coordinate system $(\hat t,\hat x,\hat y,\hat
\psi,\hat  \varphi)$, where
$$
  \hat t =t
  \;,\ \hat x =x
  \;,\ \hat y =y
  \;,\
  \hat \psi = \psi - \alpha t
  \;,\ \hat  \varphi =\varphi \;,
$$
then \eq{28X.1} implies existence of functions $f_{\hat \mu}$,
smooth near $y=1$, such that
$$
 g_{\hat t \hat \mu} =  (y-1)f_{\hat \mu}
 \;,
$$
in particular $g_{\hat t \hat t}$ vanishes at $y=1$. As in the
last two sections, a conical singularity at $y=1$ will be
avoided if and only if
$$
 \frac{ (1+\nu)^2-\lambda^2}{4 k^2 \lambda^2}
 = \lim_{y\to 1} \frac{g_{\hat t \hat t} }{g_{y y} (y-1)^2}
 =4
 \;,
$$
the first equality above resulting from the calculation of the
limit. So, there will be a conical singularity unless $k$ is
chosen to be equal to
\bel{28X.2}
 k = \frac  {  \sqrt{ (1+\nu)^2-\lambda^2}}{4 \lambda}
 \;.
\ee
It will become clear in Section~\ref{S11X09.1} that the axis
$y=1$ lies beyond event horizons, in a region where both
causality violations and naked singularities are present
anyway, and therefore there does not seem to be any significant
reason for imposing \eq{28X.2}.

\subsection{Asymptotics of the Pomeransky \& Senkov solution}
\label{subsec:asymptotic}

In this section we verify  asymptotic flatness. To that end we
need to write down the line element in a suitable coordinate
system. In \cite{EmparanReallReview} a coordinate system
leading to manifest asymptotic flatness was proposed,  related
to the {\em ring coordinates} $x$, $y$ as follows
\begin{equation}
r_1:=L\frac{\sqrt{1-x^2}}{x-y}\;,\quad r_2:=L\frac{\sqrt{y^2-1}}{x-y}\;,
\label{eq:ring-coord}
\end{equation}
where $L$ is a nonzero real constant. If $L$ is positive these
relations establish a diffeomorphism between the open region of
${\mathbb R}^2$ defined by the conditions $-1<x<1$,
$-\infty<y<-1$, and the open positive quadrant  of ${\mathbb
R}^2$ defined as $0<r_1<\infty$, $0<r_2<\infty$. Indeed the
inverse of (\ref{eq:ring-coord}) is
\begin{equation}
 x=\frac{L^2-(r^2_1+r^2_2)}{\Sigma}\;,\quad
y=-\frac{L^2+r^2_1+r^2_2}{\Sigma}\;,
\label{eq:xytoring}
\end{equation}
where
$$
\Sigma:=\sqrt{L^4+2L^2(r^2_1-r^2_2)+(r^2_1+r^2_2)^2}.
$$
Similarly, for $L$ negative one obtains a diffeomorphism with
the region $-1<x<1$ and $y>1$ by changing both signs above
\begin{equation}
 x=-\frac{L^2-(r^2_1+r^2_2)}{\Sigma}\;,\quad
y=\frac{L^2+r^2_1+r^2_2}{\Sigma}\;,
\label{eq:xytoring2}
\end{equation}
with the same function $\Sigma$.

Eq. (\ref{eq:xytoring}) adopts a simpler form if we make the
transformation
$$
r_1=r\sin\theta\;,\quad r_2=r\cos\theta\;,
$$
where $0<r<\infty$ and $0<\theta<\pi/2$. In this case we have
\begin{equation}
 x=\frac{L^2-r^2}{\Sigma}\;,\quad
y=-\frac{L^2+r^2}{\Sigma}\;,
\quad
\Sigma=\sqrt{L^4-2L^2r^2\cos2\theta+r^4}.
\label{eq:xytopolar}
\end{equation}
The Jacobian of this transformation is
$$
 - \frac{8L^4 r^3 \sin(2\theta)}{\Sigma^4}
 \;,
$$
which vanishes at the axes $\theta=0,\pi/2$, and therefore some
care is required there.

We perform the coordinate change (\ref{eq:xytopolar}) in
(\ref{eq:line-element1}) and  study the resulting expression
for large values of $r$. To understand the asymptotic behaviour
of the metric it is convenient to choose $L$ as
%
$$
 L:=  \sqrt{\frac{2k^2(1-\lambda+\nu)}{1-\nu}}
 \;,  \ \mbox{ or } \  L:= -\sqrt{\frac{2k^2(1+\lambda+\nu)}{1-\nu}}
 \;,
$$
and this choice will be made in what follows. Choosing the
positive value (which corresponds to points near $(x=-1,y=-1)$)
one then obtains
%
%
\begin{eqnarray}
 g_{tt}&=&-1+\frac{8k^2\lambda }{ (1-\lambda+\nu)r^2}+
 O\left({r^{-4}}\right)\;,\nonumber
\\
 g_{rr}
 &=&
 1-\frac{4k^2\lambda
 \big((\lambda-4\nu+\lambda\nu)\cos(2\theta)-(-1 + \nu)^{2}\big)}
 {(1-\lambda+\nu)(-1 + \nu)^{2}r^{2}}+O\left({r^{-4}}\right)
 \;,
 \nonumber
\\
 g_{\theta\theta}&=&r^2\left(1
 -\frac{4 k^2 \cos (2\theta ) \left(-3 \lambda  \nu ^2+2 ((\lambda -1)
   \lambda -1) \nu +\lambda +2 \nu ^3\right)}{(\nu -1)^2(-\lambda +\nu +1)r^2}+
 \right.\nonumber
 \\
 &&\left.+\frac{4 k^2 \lambda }{(-\lambda +\nu +1)r^2}\right)+O\left({r^{-2}}\right)\;,
 \nonumber
\\
 g_{\varphi\varphi}&=&
 r^2 \sin ^2\theta\bigg(1+\frac{2 k^2((\nu -1) \cos (2 \theta )
 (\lambda -2 \nu )+3 \lambda  \nu +\lambda -2 (\nu -1)\nu )}
 {r^2(\nu -1)^2}
 \nonumber
\\
 && \phantom{r^2 \sin ^2\theta\bigg(}+ O\left({r^{-4}}\right)\bigg)
 \nonumber\;,\\
g_{t\varphi}&=& \frac{\sin ^2\theta}{r^2 }\bigg(
\frac{16 k^3 \lambda\sqrt{\nu}\sqrt{(\nu +1)^2-\lambda ^2}}
{(\nu -1)^2 (\lambda-\nu-1)}+O\left({r^{-2}}\right)\bigg)\;,\nonumber\\
g_{t\psi}&=&
 \frac{\cos ^2\theta}{r^2}\bigg(-\frac{8 k^3 \lambda
\left(\lambda\nu+\lambda +\nu ^2-6\nu+1\right)
\sqrt{(\nu+1)^2-\lambda ^2}}{(\nu -1)^2 (-\lambda +\nu +1)^2}+O\left({r^{-2}}\right)\bigg)
 \;,
 \nonumber
\\
 g_{\varphi\psi}
  &=&\frac{\sin ^2 \theta\; \cos^2 \theta}{r^2}\bigg(-\frac{32 k^4 \lambda  \sqrt{\nu }
   \left(\lambda ^2 (\nu +1)-4 \lambda  \nu +(\nu -1)^2
   (\nu +1)\right)}{ (\nu -1)^4 (\lambda -\nu -1)}
   +O\left({r^{-2}}\right)\bigg)\;,
 \nonumber\\
 g_{\psi\psi}&=&r^2\cos^2\theta\left(1+
\frac{2 k^2 \left(\lambda ^2 (3 \nu +1)-\lambda(\nu(\nu +10)-3)
+2\nu\left(\nu^2-1\right)\right)}{r^2 (\nu -1)^2 (-\lambda +\nu +1)}+\right.\nonumber\\
&&\left.+\frac{2 k^2 \cos (2 \theta ) (\lambda -2 \nu )}
 {r^2 (\nu-1)}+O\left({r^{-4}}\right)\right)
   \;.
\label{eq:assymptotic-metric}
\end{eqnarray}

The leading powers of $r$ in the diagonal terms in
\eq{eq:assymptotic-metric} correspond to the metric
\begin{equation}
ds^2=-dt^2+dr^2+r^2(d\theta^2+\sin^2\theta d\varphi^2+\cos^2\theta d\psi^2)\;,
\label{eq:flat-metric}
\end{equation}
which is just the 5-dimensional Minkowski spacetime, as one
checks by making the coordinate transformation $x^0=t$ and
\begin{eqnarray}
&&x^1=r\cos\theta\cos\psi\;,\quad
x^2=r\cos\theta\sin\psi\;,\nonumber\\
&&x^3=r\sin\theta\sin\varphi\;, \quad
x^4=r\sin\theta \cos\varphi\;,
\label{eq:cartesian-coordinates}
\end{eqnarray}
which leads to
$$
ds^2=-(dx^0)^2 +(dx^1)^2+(dx^2)^2+(dx^3)^2+(dx^4)^2
 \;.
$$
To avoid ambiguities, we will write $\bar g_{x^\mu x^\nu}$ for
the components of the metric tensor in the manifestly
asymptotically flat coordinates \eq{eq:cartesian-coordinates}.
One can check that the angular prefactors in $g_{\psi\psi}$,
etc., have the right structure so that in the coordinates above
we have
\bel{8X.1}
 \bar g_{x^\mu x^\nu} - \eta_{\mu\nu} = O(r^{-2})
 \;,\quad \partial_{x^\sigma}
 \bar g_{x^\mu x^\nu}   = O(r^{-3})
  \;.
\ee
For instance,  let us define the functions
$h_{\mu\nu}(r,\theta)$ by the equations
$$
 g_{tt} = -1 + \frac{\htt }{r^2}\;, \quad
 g_{rr} =  1 + \frac{\hrr }{r^2}\;,\quad
 g_{\psi \psi} =  \left(1 + \frac{\hpsipsi }{r^2}\right)
r^2 \cos^2 \theta\;,
$$
$$
g_{t \phi}=\frac{\htphi\sin^2 \theta}{r^2} \;,\quad
g_{t \psi}=\frac{\htpsi\cos^2 \theta}{r^2} \;,\quad
g_{\varphi\psi}=\frac{h_{\varphi\psi} \sin^2 \theta\,\cos^2 \theta}{r^2} \;.
$$
$$
 g_{\theta \theta} = r^2 \left(1 + \frac{\hthetatheta }{r^2}\right)\;, \quad
 g_{\varphi\varphi} =
r^2\sin^2 \theta \left(1+\frac{h_{\varphi\varphi}}{r^2}\right) \;,
$$
The prefactors have been chosen so  that the functions
$h_{\mu\nu}$ are rational functions of $x$ and $y$, smooth near
$\{x=y=-1\}$.
 One finds
\begin{eqnarray}
&&\bar g_{x^1x^1}=1+\frac{ (x^2)^{2} r^{2} \,\hpsipsi +(x^1)^{2}
 \big((x^1)^{2}+(x^2)^{2}\big) \,\hrr +
 (x^1)^{2}\big((x^3)^{2}+(x^4)^{2}\big)  \,\hthetatheta  }{r^{4}
 \left((x^1)^{2}+(x^2)^{2}\right)}
 \nonumber
\\
&& \phantom{\bar g_{x^1x^1}}=1+\frac{ \sin^{2}\psi \, \hpsipsi
 + \cos^{2}\theta \cos^{2}\psi \, \hrr +
    \sin^{2}\theta\cos^{2}\psi \, \hthetatheta  }{ r^{2}  }
 \;,
 \nonumber
\\
 &&
  \bar g_{x^1x^2}=\frac{x^1 x^2\bigg(\big((x^1)^{2}+(x^2)^{2}\big) \,\hrr
  +\big((x^3)^{2}+(x^4)^{2}\big)
 \hthetatheta -r^{2}  \,\hpsipsi \bigg)}{r^{4}\left((x^1)^{2}+(x^2)^{2}\right)}
 \nonumber
\\
 &&
  \phantom{\bar g_{x^1x^2}}=\frac{\cos \psi\, \sin \psi \, \big (\cos^{2}\theta \,\hrr
  +\sin^2 \theta\,
 \hthetatheta -   \,\hpsipsi \big )}{r^{2} }
 \;.
 \nonumber
\end{eqnarray}
Continuity  of $\bar g_{x^1x^1}$ and $\bar g_{x^1x^2}$ at the
rotation axis $\theta=\pi/2 $ requires
\bel{7X.2}
 \hthetatheta (r,\pi/2) =   \hpsipsi (r,\pi/2) \quad \Longleftrightarrow
 \quad g_{\theta\theta}(r,\pi/2)= \lim_{\theta\to \pi/2} \frac{g_{\psi\psi}(r,\theta)}{\cos^2\theta}
 \;,
\ee
which can be checked by direct calculations. To obtain
differentiability one writes
\begin{eqnarray}
 \bar g_{x^1x^1}&= & 1+\frac{ (x^2)^{2}  \,\hpsipsi +(x^1)^{2}
 \hrr +\big((x^3)^{2}+(x^4)^{2}\big)\hthetatheta }{r^{4}
 }
 \nonumber
\\
 &&
 +
 \frac{(x^2)^{2}
 \big((x^3)^{2}+(x^4)^{2}\big)  \big(\hpsipsi -\hthetatheta   \big)}{r^{4}
 \left((x^1)^{2}+(x^2)^{2}\right)}
 \;,
  \label{8X.5}
\end{eqnarray}
A {\sc Mathematica} calculation shows that
$$
 \hpsipsi -\hthetatheta = \frac{1+y}{x-y} W(x,y,\nu,\lambda)
 \;,
$$
where $W$ is a rational function of its arguments, smooth near
$\{x=y=-1\}$, which is precisely what is needed to cancel the
factor $(x^1)^{2}+(x^2)^{2}$ in the denominator of the second
line of \eq{8X.5}, and make  this term uniformly well behaved
as claimed in \eq{8X.1}; here it is useful to observe that
$$
 \frac{\partial x}{\partial x^i} = O(x^i r^{-2})
 \;,
 \quad
  \frac{\partial y}{\partial x^i} = O(x^i r^{-2})
  \;.
$$

The equality \eq{7X.2} similarly guarantees uniform derivative
estimates for  $\bar g_{x^1x^2}$  and $\bar g_{x^2x^2}$ at the
rotation axis $\theta=\pi/2 $.

We continue with
\beaa
 {\bar g}_{x^3 x^3} & = & 1+  \frac{\hrr  (x^3)^2
   \left((x^3)^2+(x^4)^2\right)+\hthetatheta
   (x^3)^2 \left((x^1)^2+(x^2)^2\right)+\hphiphi  (x^4)^2
  r^2
   }{r^4 \left((x^3)^2+(x^4)^2\right)}
\\
 & = & 1+
 \frac{ \hrr (x^3)^2+ \hthetatheta (r^2 -(x^3)^2) }{r^4}+ \frac{
 (\hphiphi-
 \hthetatheta)(x^4)^2}{r^4\left((x^3)^2+(x^4)^2\right) }
 \;.
\eeaa
For continuity of $\bar g_{x^3 x^3}$ one thus obtains the
condition
\bel{7X.2a}
 \hphiphi (r,0) =   \hthetatheta (r,0) \quad \Longleftrightarrow
 \quad g_{\theta\theta}(r,0)= \lim_{\theta\to 0}  \frac{g_{\varphi\varphi}(r,\theta)}{\sin^2 \theta}
 \;,
\ee
while uniform differentiability is equivalent to uniform
differentiability of
$$
 \frac{(x^4)^2(\hphiphi-\hthetatheta)}
 {r^4 \big((x^3)^2 + (x^4)^2\big)}
 \;.
$$
This, in turn, requires a factorization of
$\hphiphi-\hthetatheta$ by $(x^3)^2 + (x^4)^2$. The required
regularity ensues from the identity
$$
 \hphiphi-\hthetatheta= \frac{(1-x^2)}{x-y} \hat W(x,y)
 \;,
$$
where $\hat W$ is a rational function regular near
$\{x=y=-1\}$. The same formula takes care of the regularity of
$\bar g_{x^3 x^4}$ and $\bar g_{x^4x^4}$.

The remaining components of the metric are manifestly
asymptotically flat, e.g.
\begin{eqnarray}
 &&\bar g_{x^0 x^0}=-1+\frac{\htt }{r^2}\;,
 \nonumber
\\
 &&\bar g_{x^0 x^1}=-\frac{ x^2\, \htpsi }{r^4} = -\frac{\cos  \theta \, \sin \psi \,\htpsi  }{r^3}\;,
 \nonumber
\\
 &&\bar g_{x^1 x^3}=\frac{(\hrr - \hthetatheta) r^2   x^1 x^3  -
 \hphipsi x^2 x^4}{r^6}
 \;,
\end{eqnarray}
with similar expressions for those non-zero $\bar g_{x^\mu
x^\nu}$'s that have not been listed so far.

Uniform decay estimates on higher order derivatives follow from
\eq{8X.1} using elliptic estimates applied to the stationary
Einstein equations, which establishes asymptotic flatness of
the solutions.

It is well known that the ADM mass $m$  of a stationary
solution equals its Komar mass, independently of dimension. One
can therefore read the ADM mass from the $1/r^2$ term in
$g_{tt}$  and so, perhaps up to normalisation-dependent
factors, the mass is
$$
  \frac{4 k^2\lambda }{ (1-\lambda+\nu)}
 \;.
$$
For positive $\lambda$, positivity of the total mass is then
equivalent to
\bel{13X.1}
 \lambda< 1+ \nu
 \;.
\ee
This proves that, for $\lambda$'s that do not satisfy that
constraint, the domain of outer communications associated to
this asymptotically flat end contains naked singularities, in
the sense that the hypotheses of the positive energy theorem
with horizon boundaries are violated. However, this does not
necessarily prove that the solutions are nakedly singular for
all domains of outer communications for $\lambda$'s that do not
satisfy \eq{13X.1}, as the asymptotically flat end obtained as
above near $\{x=y=-1\}$ could be shielded from other such ends
by a horizon. Equivalently, to show that \eq{13X.1} is
necessary for regularity, one would need to locate all
asymptotically flat regions of all maximal extensions of the
metric (\ref{eq:line-element1}), and analyse the associated
domains of outer communication.

\subsubsection{$(x=1,y=1)$}
 \label{ss29X}

 Near $(x=1,y=1)$, the asymptotics
of those components of the metric which do not carry a $\psi$
index can be obtained by replacing $\lambda$ by $-\lambda$ in
\eq{eq:assymptotic-metric}. This is due to the fact that the
transformation
$$
 (x,y,\lambda,\nu)\mapsto (-x,-y,-\lambda,\nu)
$$
maps all the metric functions into themselves, except for $M$,
and this last function only affects $g(\partial_\psi, \cdot)$.
Those last components of the metric read
\begin{eqnarray}
g_{\psi\psi}&=&r^2 \cos ^2\theta+\frac{k^2}{{2 (\nu -1)^2
   \left((\nu +1)^2-\lambda^2\right)}}\times\nonumber\\
&&\Bigg((\lambda -\nu -1)\bigg((\nu -1)\cos(4\theta)
(\lambda +\nu +1)(\lambda +2 \nu )-\nonumber\\
&&-4\lambda\cos(2\theta)
\left((\lambda +6) \nu +\lambda-\nu^2-1\right)\bigg)+\nonumber\\
&&+\lambda ^3 (-(5 \nu +3))-2 \lambda^2(\nu -1)(13 \nu -12)-\nonumber\\
&&-\lambda(\nu +1)(3 (\nu -8)\nu+5)+2 (\nu -1) \nu  (\nu +1)^2\Bigg)+
O\left({r^{-1}}\right)\;, \\
g_{t\psi}&=&\frac{4 k \lambda }{\sqrt{(\nu +1)^2-\lambda ^2}}+O\left(\cos^2\theta{r^{-1}}\right)
\;,\\
g_{\psi\phi}&=&-\frac{16 k^4 \lambda  \sqrt{\nu } \sin ^2(\theta )}
 {r^2 (\nu -1)^4 (\lambda +\nu +1)}\times
\Bigg(\cos (2 \theta ) \left(\lambda ^2 (\nu +1)+4
 \lambda\nu +(\nu -1)^2 (\nu +1)\right)+
 \nonumber
\\
 &&+\lambda^2(\nu+1)+4 \lambda  \left(\nu ^2-\nu +1\right)+(\nu -1)^2
 (\nu +1)\Bigg)
 \nonumber
\\
 &&
 +O\left({\cos^2\theta\sin^2\theta r^{-3}}\right).
\end{eqnarray}

\subsection{The singular set $\{H(x,y)=0\}$}
 \label{ssNature}

Throughout this section we restrict attention to admissible
pairs $(\nu,\lambda)$.

In order to understand the geometry of the singular set
$$
\Sing:=\{H(x,y)=0\}\;,
$$
 we proceed as follows, keeping in mind the analysis of
Section~\ref{sfunH}, which concerned the region $\{x\in
[-1,1]\;, \ y_c< y <-1\}$: The equation $H(x,y)=0$ can also be
solved as
%
\begin{eqnarray}
 \label{14X.3}
y_\pm(x):=\frac{\lambda  \nu(1-  x^2)
   \pm\sqrt{\tilde W(x) }}{x \nu  \left(2
   \lambda  \nu +x \left(\lambda ^2+\nu
   ^2-1\right)\right)}
   \;,
\end{eqnarray}
where
\bel{14X.2}
 \tilde W:= \nu  \left(\left(x^2-1\right)^2 \nu
   \lambda ^2+x \left(\lambda ^2+2 x \lambda -\nu
   ^2+1\right) \left(2 \lambda  \nu +x \left(\lambda
   ^2+\nu ^2-1\right)\right)\right)
 \;,
\ee
provided the denominator $x \nu  \left(2
   \lambda  \nu +x \left(\lambda ^2+\nu
   ^2-1\right)\right)$ of (\ref{14X.3}) is nonzero.
So, if this condition holds, at each $x\in \R$ there are either
two real values (counting multiplicity) of $y$ for which a
solution exists, or none. Each branch $y_\pm$ is a smooth
function of $x$ near any given point $x_0$ if and only if the
other one is, except at the zeros of denominator.

We are interested in the topology of $\Sing$ in the region $x\in
[-1,1]$, and Section~\ref{sfunH} tells us that the graphs $y_{\pm}$
do \emph{not} meet the region $\{y_c\le y< -1\}$ there. For reasons
that will become clear in Section~\ref{S11X09.1} we need to
understand both the positive and negative branches of $y_\pm$.

\begin{figure}[h]
 \begin{center}
 \includegraphics[width=.2\textwidth]{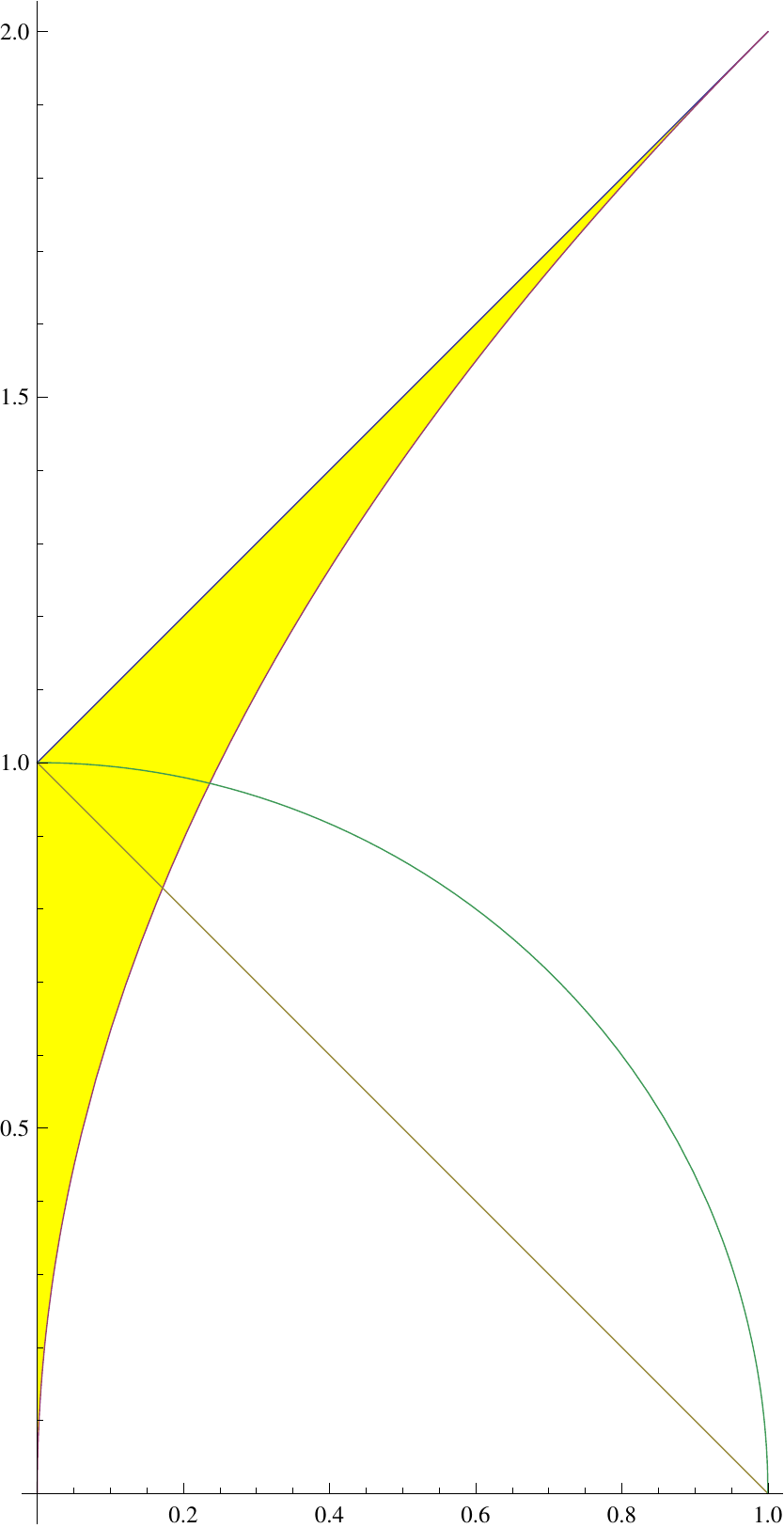}
\caption{\label{F13X.1} The regions of distinct behavior of $x_*$; the set of admissible $(\nu,\lambda)$ has been
shaded in yellow.
}
\end{center}
\end{figure}

The denominator in (\ref{14X.3}) vanishes at $x=0$, and at
$x=x_*$ if $\nu^2+\lambda^2\ne 1$, where
\bel{28X.4}
 x_*:=-\frac{2 \lambda  \nu }{\lambda ^2+\nu ^2-1}
 \;.
\ee
The pole $x_*$ belongs to $(0,1)$ for allowed $\nu,\lambda$
such that $\nu+\lambda<1$; we have $x_*>1$ for $\nu+\lambda>1$
and $\nu^2+\lambda^2<1$, with $x_*\rightarrow \infty$ when
$(\nu,\lambda)$ approaches from inside the unit circle centered
at the origin of the $(\nu,\lambda)$-plane; and finally
$x_*<-1$ for allowed $\nu,\lambda$ such that
$\nu^2+\lambda^2>1$, see Figure~\ref{F13X.1}.
At each zero of the denominator in (\ref{14X.3}) the graphs of
$y_\pm$ split into two components, except if the numerator vanishes
there as well. Given that there are at most two zeros, we conclude
that $\Sing$ can have up to five connected components.

The identity, in obvious notation,
%
$$
 y_\pm = \frac{-b \pm \sqrt{b^2 - 4 ac}}{2a} =  \frac{2c}{-b \mp \sqrt{b^2 - 4 ac}}
$$
shows that  the numerator of one of the $y_\pm$'s necessarily
vanishes at a zero of the denominator.  At $x=0$ the expression
under the square root equals $\lambda^2 \nu^2>0$, so there are
always precisely two real valued solutions for $x\ne 0$ small.
Near $x=0$ we have
$$
y_-= -\frac{ (2 x \lambda +\lambda ^2-\nu ^2+1 )}
  {\lambda  \nu(1-  x^2)
   +\sqrt{\tilde W}}=-\frac{\lambda ^2-\nu ^2+1}{2  \lambda  \nu  }+O\left(x \right)
 \;,
 \qquad
 y_+ = \frac{1}{x \nu }+O\left(1\right)
  \;,
$$
%
%
%
with $y_-$ smaller than $y_c$ for $|x|$ small and for
admissible $\nu$ and $\lambda$;  this follows from the analysis
in Section~\ref{sfunH} in any case, but can also be seen by a
direct calculation. So, near $x=0$, the graph of $y_+$ splits
into two branches, tending to minus infinity at the left, and
plus infinity at the right, while the graph  of $y_-$ continues
smoothly across $x=0$.

In particular the singular set $\Sing$ is never empty.

In the  case $\nu^2+\lambda^2=1$, the denominator
of~(\ref{14X.3}) vanishes only at $x=0$, with the behavior just
described being independent of those particular values of the
parameters.

Assuming  $\nu^2+\lambda^2 \ne 1$, we  have
%
%
$$
\tilde W (x_*) = \left(\frac{\lambda  \nu
 (\lambda -\nu -1) (\lambda -\nu +1) (\lambda +\nu -1) (\lambda +\nu +1)}{\left(\lambda ^2+\nu
   ^2-1\right)^2}\right)^2
   \;,
$$
which is positive for all allowed $\nu$ and $\lambda$, except
when $\nu+\lambda=1$, where $\tilde W (x_*)$ vanishes. It
follows that for $\nu+\lambda\ne 1$, near $x_*$ the branches
behave as:
$$y_+=\frac{(1-\nu-\lambda)(1+\nu+\lambda)(1+\lambda-\nu)(1+\nu-\lambda)}{\nu(x-x_*)(1-\nu^2-\lambda^2)^2}+O(1) \ ,\ y_-=O(1) $$
so that $y_-$ extends smoothly across $x=x_*$, whereas $y_+$
blows up. The case $\nu+\lambda=1$ requires separate attention,
and will be analyzed shortly.
\medskip

Next, in order to understand the behaviour of the components of
the set $\Sing$, it is useful to study the two branches $y_\pm$
for all $x \in \R$. Now, the top order term of the polynomial $
\tilde W(x)$ is $\nu^2 \lambda^2 x^4$, and so $\tilde W$ is
strictly positive for large positive or negative $x$. This
implies that at fixed admissible $\nu$ and $\lambda$ the sets
\bel{12XI.1}
 \tilde \Omega_{\nu,\lambda}:=
 \{x: \tilde W(x)\ge 0\}
\ee
have one, two, or three connected components. Moreover, each
branch $y_\pm$, which exists for $|x|$ large enough, has a
finite limit at infinity provided that $\nu^2+\lambda^2\ne 1$:
$$
y_+ \rightarrow 0\;,\quad  y_-\rightarrow
\frac{2\lambda}{1-\nu^2-\lambda^2}
 \;,
$$
when $x\rightarrow \pm\infty$.
In the case $\nu^2+\lambda^2=1$, we   obtain instead
$$
y_+ \rightarrow 0\ ,\ y_-=-\frac{x}{\nu} + O(1) \rightarrow \mp
\infty
$$
when $x \rightarrow \pm \infty$.
\medskip

We now turn our attention to the roots of $\tilde W$: at those,
both branches $y_\pm$ meet at that side of the root where
$\tilde W$ is positive, and stop existing nearby on the side
where $\tilde W$ is negative, which happens if the order of the
root is odd. Such points will be referred to as \emph{turning
points}. Note that there are at most four such turning points.

Recall that we have
 $\tilde W(0)=\lambda^2 \nu^2
>0$. Further
%
$$
 \tilde W(1)= \nu  (\lambda -\nu +1) (\lambda +\nu -1) (\lambda +\nu +1)^2\;,
$$
so the sign of $\tilde W$ at one is determined by the sign of
$\lambda+\nu-1$. Next,
%
$$
 \tilde W(-1)=
4 \nu  (\lambda -\nu -1)^2 \left(\lambda +(1-\nu)\right)
\left(\lambda -(1-\nu)\right)= 4 \nu  (\lambda -\nu -1)^2
\left(\lambda^2 -(1-\nu)^2\right)
 \;,
$$
so the sign at minus one is the same as that at plus one, both
vanishing for admissible values of parameters if and only if
$\lambda=1-\nu$:
%
$$
\tilde W(-1)\tilde W(1)= \nu ^2 (\lambda -\nu -1)^2 (\lambda -\nu
+1)^2 (\lambda +\nu -1)^2
   (\lambda +\nu +1)^2 \ge 0
    \;.
$$

When $\lambda+\nu \ne 1$, the equations $H(\pm 1 ,y)=0$ have
the following four solutions
$$
x=1: \quad y_{\uparrow,\pm} = \pm \sqrt{
 \frac{1+\lambda - \nu} { \nu (\lambda+\nu-1)}
 }
 \;;
 \qquad
x=-1: \quad y_{\downarrow,\pm} = \pm \sqrt{  \frac{\lambda+\nu-1} {
\nu (1+\lambda - \nu)}
 }
 \;.
$$
The function under the square root in $y_{\uparrow,\pm}$ is
positive for admissible $(\lambda,\nu)$ if and only if
$\lambda+\nu> 1$, and then it is larger than one.
On the other hand, the function under the square root in
$y_{\downarrow,\pm}$ is always smaller than $1$ for the
parameters of interest, non-negative if and only if
$$
 \lambda> 1-\nu
 \;.
$$

We continue with a Lemma about the number of roots of $\tilde
W$:

\begin{Lemma}
 \label{L8.22}
 \begin{enumerate}
     \item There exists a smooth curve $\gamma$, separating
         the set $\mcU$ of admissible  $(\nu,\ \lambda)$
         into two components, which is a graph of a
         function $\chi:[0,\nu_*]\to [0,1]$, satisfying
$$
 1-\nu \le \chi \le 1-\nu^2\;, \qquad \chi(\nu_*)=2 \sqrt{\nu_*}
$$
(see Figure~\ref{F16X.1}),
\begin{figure}[h]
 \begin{center}
 \includegraphics[width=.4\textwidth]{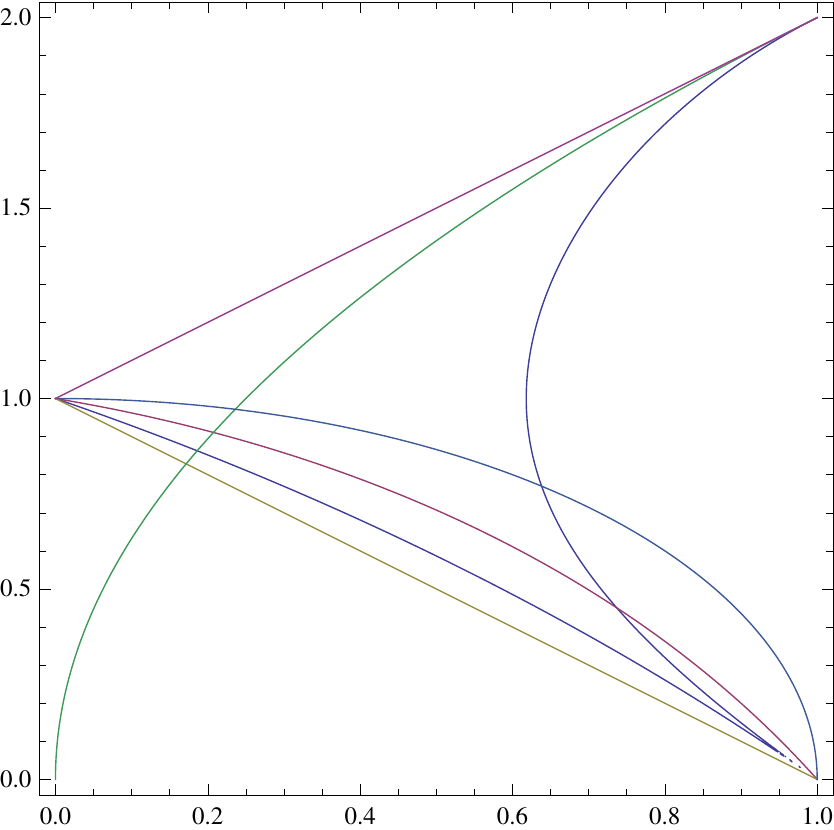}
\caption{\label{F16X.1} The curve of double zeros of $\tilde W$,
separating $\mcU$ in two components, is drawn in dark blue; $\gamma$
is the intersection of that curve with $\mcU$, and is the graph of a
function $\chi$. The function $\lambda_k$ is in magenta, a quarter
of the unit circle is in light blue.}
\end{center}
\end{figure}
such that $\tilde W$ has a multiple root, for admissible
$(\nu,\lambda)$, if and only if
$$
 (\nu,\lambda) \in \gamma \qquad \Longleftrightarrow \qquad \lambda =
 \chi(\nu)
 \;.
$$
Moreover
\begin{enumerate}
     \item In the connected component of $\mcU\setminus
         \gamma$  where $\lambda<\chi(\nu)$ the
         polynomial $\tilde W$ has four distinct real
         roots,  and at least one of them is bigger
         than $x_*$.
  \item  In the remaining connected component the
      polynomial $\tilde W$ has two distinct real roots
      and two distinct roots in $\C\setminus \R$.
\end{enumerate}

 \item $\tilde W$ has no third or fourth order zeros for
     $(\nu,\lambda)\in \mcU$.
 \end{enumerate}
\end{Lemma}

\proof
1.
A necessary condition for existence of a second order zero of
$\tilde W$ is the existence of a joint zero for $\tilde W$  and its
first derivative. This, in turn, is equivalent to the vanishing of
the resultant of the polynomials $x\mapsto
 \tilde W $ and $x\mapsto
\partial_x \tilde W $. This resultant is
$$
{ -2^{18} \lambda ^6 (\nu -1)^2 \nu ^9
 \left(\lambda ^2-(\nu
   +1)^2\right)^4
    f(\nu,\lambda)}
 \;,
$$
where
\beaa
 f(\nu,\lambda)&:=& \lambda ^8+2 \lambda ^6 \left(1-4 \nu +\nu
^2\right)+15 \lambda ^4 (-1+\nu )^2 \nu
 \\
&& -2 \lambda ^2 (-1+\nu )^4 \left(1+4 \nu +\nu ^2\right)-(-1+\nu
)^6 (1+\nu )^2\ . \eeaa
Since $f$ is a polynomial of degree four in $\lambda^2$, we can
define a polynomial $q(\nu,L)=f(\nu,\lambda)$ where $L=\lambda^2$,
and study $q$. We have
$$
\frac{\partial^2 q}{\partial L^2}=6\left(2L^2+2(1-4\nu+\nu^2) L +
5\nu (1-\nu)^2\right)\ ,$$
which is obviously positive for $\nu\in [0,2-\sqrt3\approx
0.26]$ and all $L$, and so $q$ is convex there. For $\lambda\ge
2\sqrt{\nu}$ we find
$$
2L^2+2(1-4\nu+\nu^2) L \ge 32 \nu^2 +8(1-4\nu+\nu^2) \nu =  8(1
+\nu^2) \nu> 0
 \;,
$$
and so $q$ is always convex in $L$ above the graph of $2\sqrt{\nu}$.
Now
$$
q(\nu,0)=-(\nu -1)^6 (\nu +1)^2<0\;, \qquad
\partial_L q(\nu,0)= -2 (\nu -1)^4 \left(\nu ^2+4 \nu +1\right)
 <0
 \;,
$$
$$
q(\nu,(1+\nu)^2) = 27\nu(1-\nu)^2(1+\nu)^4 >0\; .
$$
Convexity of $L\mapsto q$ implies that, for $\nu\in [0,2-\sqrt3]$,
the zero-level set of $q$ is a smooth graph.

For $\nu>1/4$ one can instead argue as follows: We have
$$
\partial_L q(\nu,4\nu)=2 \left(-\nu ^2+10 \nu -1\right) \left(\nu ^4+10 \nu ^3-18 \nu ^2+10 \nu +1\right)
 \;.
$$
The second factor is positive for $\nu\in [5-2\sqrt 6, 5+2\sqrt
6]\supset [ 0.102,1]$. For the last factor, we  write
$$
 10 \nu ^3-18 \nu ^2+10 \nu \ge \nu(10 \nu ^2-20 \nu +10 ) = 10 \nu (1+\nu)^2
  > 0
$$
and the positivity of $\partial_L q(\nu, 4\nu)$ follows. Convexity
gives positivity of $q$ in the admissible region.

\begin{figure}[h]
 \begin{center}
 \includegraphics[width=.5\textwidth]{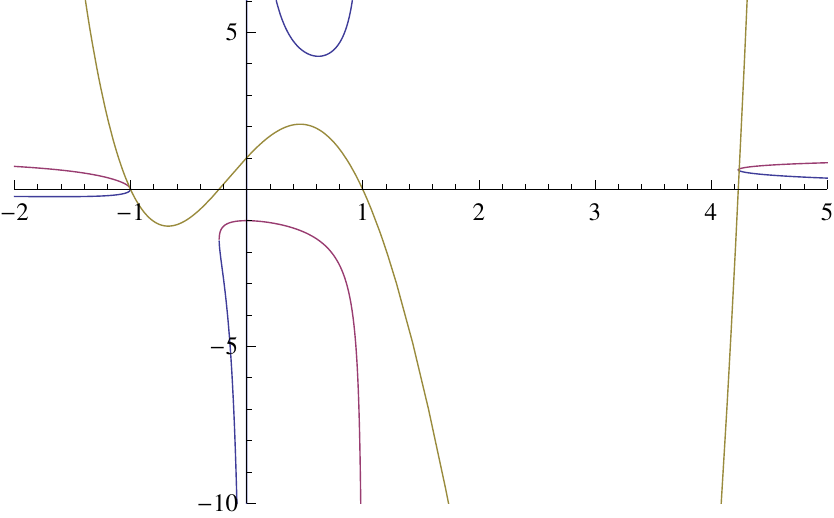}
\caption{\label{F14X.9} The polynomial $\tilde W$ (green) and the
set $\{H(x,y)=0\}$ (blue and magenta) when $\lambda = 1-\nu$. The
vertical axis is $\nu y$.}
\end{center}
\end{figure}

To finish the proof of the   point 1., we note first that the region
$$
 \mcU_1:= \mcU\cap \{\lambda<\chi(\nu)\}
$$
contains the line $\lambda=1-\nu$, where all roots are simple and
real. Indeed, in this special case the function $H$ equals
$$
 H(x,y)|_{\lambda = 1-\nu}= -2 (\nu -1) (x (y \nu -1) ((x-1) y \nu -1)+y \nu +1)
 \;,
$$
and the zeros are
%
$$
\nu y_\pm = \frac{x^2-1\pm\sqrt{\tilde W}}{2 (x-1) x}\;, \quad
 \tilde W = \nu^2 (1-\nu)^2 (x-1) (x+1) \left(x-\sqrt{5}-2\right) \left(x+\sqrt{5}-2\right)
 \;,
$$
see Figure~\ref{F14X.9}.
In this case we have four simple real roots, independently of
$\nu$. Continuity implies that all roots of $\tilde W$ are
simple and real in  $\mcU_1$. Moreover, still in the case
$\lambda=1-\nu$, we have $x_*=1$, and the biggest root of
$\tilde W$ equals $2+\sqrt 5 > 1$. One can conclude the first
point again by a continuity argument: both the function
$(\nu,\lambda)\mapsto x_*$ (compare \eq{28X.4}) and the
function which to $(\nu,\lambda)$ assigns the largest root of
$\tilde W$ are continuous on $\mcU_1$. Since $\tilde W (x_*)
>0$ for $(\nu,\lambda) \in \mcU_1$,  the largest root of $\tilde
W$ cannot become smaller than $x_*$ when moving along paths
contained in $\mcU_1$.

\begin{figure}[h]
 \begin{center}
 \includegraphics[width=.4\textwidth]{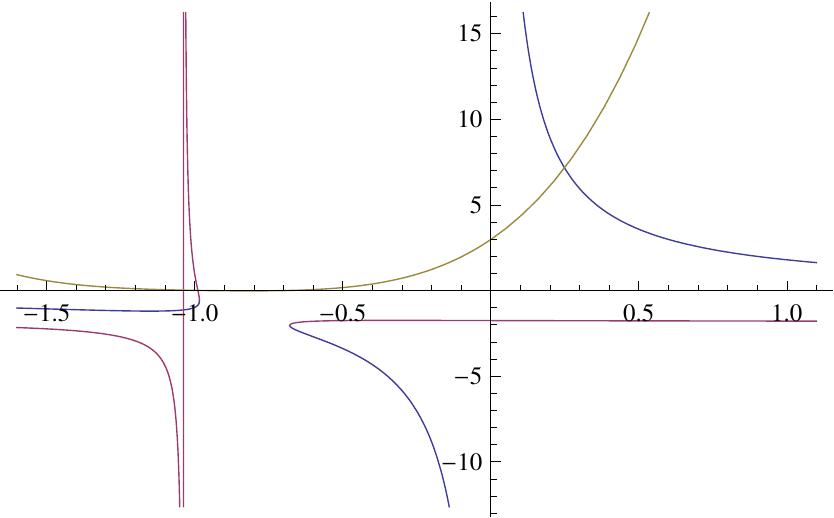}\quad\includegraphics[width=.4\textwidth]{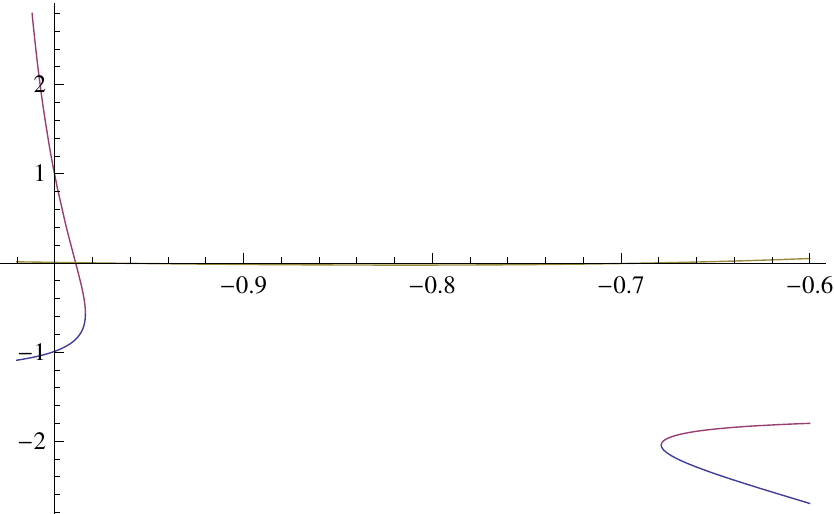}
\caption{\label{F17X.1} The polynomial $\tilde W$ (green) and the
set $\{H(x,y)=0\}$ (blue and magenta) when $\nu=9/16$ and $\lambda =
49/32$, with a zoom to the region where $\tilde W$ is negative.}
\end{center}
\end{figure}

Another continuity argument, using the fact that for $\nu =
9/16$ and $\lambda = 7/4$ the zeros of $\tilde W $ are,
approximately,
$$
 -1.09746 \pm 0.541984 \ii\in \C\setminus\R\;, \quad -0.983642\;, \quad -0.678586
 \;,
$$
finishes the proof (exact formulae for the roots can be given, but
they are not very enlightening). The resulting $\tilde W$ and
singular set $\{H(x,y)=0\}$ are plotted in Figure~\ref{F17X.1}.
%

2. A necessary condition for existence of a fourth-order zero is the
existence of a joint zero for the second and third derivatives.
This, in turn, is equivalent to the vanishing of the resultant of
the polynomials $x\mapsto
\partial_x^2 \tilde W $ and $x\mapsto
\partial_x^3\tilde W $. This resultant is
$$
2^{12} 3^2 \lambda ^4 \nu ^4 \left(\lambda ^4 (2 \nu -3)-2 \lambda
^2
   \left(\nu ^2-3\right)-(2 \nu +3) \left(\nu ^2-1\right)^2\right)
   \;.
$$
The  roots of the last factor are $\lambda=\pm(1+\nu)$ and
$$
 \lambda_\pm=\pm(1-\nu)\sqrt{\frac {2 \nu +3}{ 3-2 \nu}}
 \;,
$$
and so $\lambda_+$ is the only value of interest. When substituted
into $f$, we obtain
$$
f(\nu,\lambda_+)=\frac{27 (\nu -1)^6 \nu  \left(16 \nu ^4-40 \nu
^2+17\right)}{(3-2 \nu )^4}
$$
which is clearly positive for $\nu\in(0,\sqrt{17/40}]\approx
0.65]\supset(0,1/4]$, and so there are no fourth order roots on
$\gamma$.

To exclude the possibility of third order zeros on $\gamma$, we
calculate likewise the resultant of $\partial_x \tilde W$ and
$\partial^2_x \tilde W$, which is
$$
2^{16}\lambda ^4 \nu ^6 \left(\lambda ^2-(\nu +1)^2\right)^2 \hat
f(\nu,\lambda)
 \;,
$$
where
\beaa
   \hat f(\nu,\lambda) &=& \lambda ^8 (8 \nu -9)+4 \lambda ^6 \nu  (4 (\nu -4) \nu +13)
   +6  \lambda ^4 (\nu -1)^2 \left(5 \nu ^2+3\right)
\\
 &&
   -4 \lambda ^2 (\nu
   -1)^4 \nu  (4 \nu  (\nu +4)+13)-(\nu -1)^6 (\nu +1)^2 (8 \nu
   +9)
 \;.
\eeaa
A necessary condition for a third order zero is   the vanishing of
the resultant of the polynomials $\lambda\mapsto f$ and
$\lambda\mapsto \hat f$. That resultant is
$$
 3^{24} (\nu -1)^{40} \nu ^8 (\nu +1)^8 \left(256 \nu ^4-864 \nu
   ^2+513\right)^2
   \;,
$$
with the last factor vanishing at
$$
 \pm \frac{1}{4} \sqrt{27\pm 6 \sqrt{6}}
$$
The only   value in $(0,1)$ is $\frac{1}{4} \sqrt{27- 6
\sqrt{6}}\approx 0.88$.

Another necessary condition for a third order zero is  the vanishing
of the resultant of the polynomials $\nu\mapsto f$ and $\nu\mapsto
\hat f$. That resultant is
\beaa
 \lefteqn{
 -3^{24}   (\lambda -2)^2 (\lambda -1)^2 \lambda ^{44} (\lambda
   +1)^2 (\lambda +2)^2 \left(\lambda ^2+1\right)^2
   }&&
\\
&& \times    \left(65536 \lambda
   ^8-327680 \lambda ^6+526848 \lambda ^4-279296 \lambda ^2+9025\right)
   \;,
\eeaa
with the last factor vanishing at approximately
$$
\pm 1.06033\;,\ \pm 1.51468 \;,\ \pm 0.185775 \;,\ \pm 1.24376
 \;,
$$
the exact values of the positive solutions being
$$
 \frac{1}{4} \sqrt{20+3 \sqrt{6}-\sqrt{3 \left(39- 4 \sqrt{6}\right)}}
 \;, \quad  \frac{1}{4} \sqrt{20-3 \sqrt{6}+\sqrt{3 \left(39+ 4 \sqrt{6}\right)}}
 \;,
$$
$$
 \frac{1}{4} \sqrt{20+3 \sqrt{6}+\sqrt{3 \left(39- 4 \sqrt{6}\right)}}
 \;, \quad  \frac{1}{4} \sqrt{20-3 \sqrt{6}-\sqrt{3 \left(39+ 4 \sqrt{6}\right)}}
 \;.
$$
One checks that $f$ does not vanish at the above values of $\nu$ and
$\lambda$, except at
$$
\left(\frac{1}{4} \sqrt{27-6 \sqrt{6}},\frac{1}{4} \sqrt{20-3
   \sqrt{6}-\sqrt{3 \left(39+4 \sqrt{6}\right)}}\right) \approx (0.88, 0.18)
   \;.
$$
However, $0.19\approx
 \lambda < 2\sqrt \nu\approx1.87$ there, which
is therefore not admissible.
\qed

We wish to show, next, that all zeros of $\tilde W$ in $[-1,1]$
are simple. The proof of this requires understanding of the
behaviour of the branches $y_\pm$ in $[-1,1]$; this is the
purpose of the next lemma:

\begin{Lemma}
 \label{L8.23}
In the region $\{-1\leq x \leq 0\;,\;y<y_c\}$, the two branches
$y_\pm$ which exist for small negative values of $x$ meet smoothly
at some $\bar{x}\in (-1,0)$, where $\bar{x}$ is a simple root of
$\tilde{W}$.
\end{Lemma}
\proof The existence of both $y_\pm$ at $x=-\varepsilon$, for
small enough positive $\varepsilon$, comes from the facts that
$\tilde{W}(0)$ is positive, and that their denominator can
vanish only at $x=0$, or $x=x_*$, and $x_*$ lies always outside
$[-1,0]$, for any $(\nu,\lambda)\in \mcU$. Moreover, the
asymptotics of these branches studied above shows that they are
both below the $\{y=y_c\}$-level set. One should recall from
Section~\ref{sfunH} that these branches can neither enter the
region $\{-1\leq x \leq 1\;,\;y_c\leq y \leq -1\}$, nor cross
the $\{x=-1\}$ -axis below $y_c$. But the functions $x\mapsto
y_\pm(x)$ are continuous on each connected component of the set
$\tilde \Omega_{\nu,\lambda}\cap (-1,0)$, since this set does
not intersect the lines $x=0$ and $x=x_*$. As a consequence,
there exists $\bar{x}\in (-1,0)$ at which $\tilde{W}$ has a
change of sign, that is to say an odd-order zero of
$\tilde{W}$. Since $\tilde{W}$ has degree four, the only
possibilities for the order are $1$ and $3$. But the
existence of triple-zeros has already been excluded in
Lemma~\ref{L8.22}.
\qed
\begin{remark}
\label{xbardef}
From the proof of the lemma above, $\bar x$ can be defined as the largest negative root of $\tilde W$,
which is also the second lowest root of $\tilde W$. Indeed, if $\bar x$ were the biggest (of the four) root of $\tilde W$ for some $(\nu_1,\lambda_1) \in U_1$, then by connectedness of the region $\mcU_1$ defined earlier, it would the case for $\lambda=1-\nu$
. But the the biggest root of $\tilde W$ in this case is positive (see the proof of Lemma~\ref{L8.22}).
Moreover, since $\bar x(\nu,\lambda)$ is always a simple root for $(\nu,\lambda)$ in $\mcU$, $\bar
x$ is a smooth function of the coefficients of
$\tilde W(x)$, and therefore the map $(\nu,\lambda) \mapsto \bar x (\nu,\lambda)$ defined in $\mcU$ is continuous.
\end{remark}
We now have:

\begin{Proposition}
For all admissible $(\nu,\lambda)$, those roots of $\tilde{W}$
which belong to $[-1,1]$ are simple.
\end{Proposition}

\begin{figure}[h]
 \begin{center}
 \includegraphics[width=.4\textwidth]{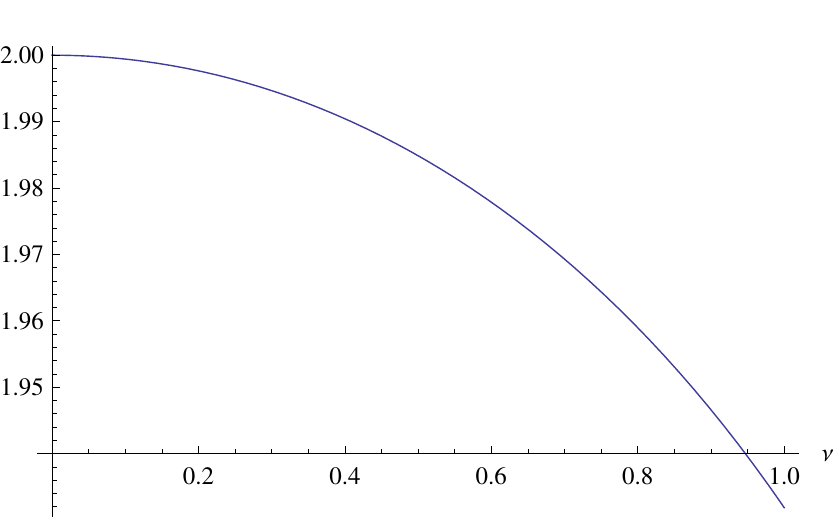}
\caption{\label{F2graphs} One of the  real zeros of the last factor in
\eq{24X.1} as a function  of $\nu$; the other one is the negative of the first.}
\end{center}
\end{figure}

\proof We start by a proof based on {\sc Mathematica} plots, an
alternative analytic argument will also be given. Another
necessary condition for a double zero of $\tilde W$ is the
vanishing of the resultant  of the polynomials $\lambda\mapsto
\tilde W$ and $\lambda\mapsto \partial_x \tilde W$. This
resultant  is
\bel{24X.1}
 2^{20} x^4 \left(x^2-1\right)^4 \nu ^{10} \left(\nu
^2-1\right)^5 \left(x^8-4 x^6+4 x^2 \nu ^2-\nu ^2\right)
 \;.
\ee
So, zeros of this resultant, with a $\lambda$ which is   a zero
of $f$, provide the only candidates for solutions of the two
equations $\tilde W=\partial_x \tilde W=0$.  The relevant zeros
are of course those of the last factor; it is a quadratic
polynomial in $x^2$, so explicit formulae can be given. {\sc
Mathematica} plots show that, in the relevant range of $\nu$'s,
only two out of the eight possible roots are real, and lie
outside of the range of interest, as can be seen on the graph
in Figure~\ref{F2graphs}.

The proof that does not appeal to graphs proceeds as follows:

The study of the resultants in the proof of the
Lemma~\ref{L8.22}  shows that a non-simple root $x_d$ of
$\tilde{W}$ occurs if and only if $f(\nu,\lambda)=0$, and that
in this case, it is exactly a double root. Then, we saw that
the admissible values of $\nu$ and $\lambda$ which satisfy
$f(\nu,\lambda)=0$ are such that $\nu+\lambda > 1$.
 Hence we are in the case $\tilde{W}(-1)>0$, with
$y_\pm(-1) \in (-1,1)$; moreover, the pole $x=x_*$ lies in
$(1,+\infty)$. From
 Lemma~\ref{L8.23}, $\tilde{W}$ has to change sign at some
$\hat{x} \in (-1,\bar{x})$, since the branches $y_\pm$ cannot
enter the region $\{-1\leq x \leq 1\;,\;y_c\leq y\leq -1\}$;
more precisely $\hat{x}$ is another simple root of $\tilde{W}$
in $(-1,\bar{x})$. So we already have two simple roots of
$\tilde W$ in $(-1,0)$. But we know from point 1. of
Lemma~\ref{L8.22} that one of the roots of $\tilde W$ must be
greater than $x_*$ whenever $f(\nu,\lambda)<0$: this still
holds by continuity for $f(\nu,\lambda)\leq 0$. Thus so is the
double root $x_d$, which finishes the proof.
\qed

Now, as announced previously, we show that $H(x,1)$ does not
vanish for any $x \in [-1,1]$:
\begin{Lemma}
No component of the singular set $\Sing$ can cross the segment
$[-1,1]\times \{1\}$ in the $(x,y)-$plane.
\end{Lemma}

\proof Let us go back to the expressions of $y_\pm(x)$
in~(\ref{14X.3}). If we examine the numerator of $y_\pm(x)$, we
always have:
$$
|y_+(x)|=\frac{\lambda \nu (1-x^2)+\sqrt{\tilde W (x)}}{\nu |x|
\left|2\nu\lambda  x(\lambda^2+\nu^2-1)\right|}\geq
\frac{\left|\lambda \nu (1-x^2)-\sqrt{\tilde W (x)}\right|}{\nu |x|
\left|2\nu\lambda x(\lambda^2+\nu^2-1)\right|}=|y_-(x)|\;.
$$
Then, from Section~\ref{sfunH}, we know that if $x \in [-1,1]$
is in the connected component of $\tilde \Omega_{\nu,\lambda}$
which contains $0$, then we have $y_-(x) < y_c < -1$. Hence, we
have $|y_\pm(x)|>1$ for $x\in [-1,1]$ in that connected
component of $\tilde \Omega_{\nu,\lambda}$. But from the
analysis of the singular set above, we know that for admissible
values of the parameters $\nu,\lambda$ such that $\nu+\lambda
\geq 1$, we have another connected component of $\tilde
\Omega_{\nu,\lambda}$ in $[-1,1]$, which contains $-1$, and is
contained in $[-1,0)$. Then, the denominator $\nu x
\left(2\nu\lambda + x(\nu^2+\lambda^2-1)\right)$ of $y_\pm$ is
always negative for $x \in [-1,0)$. Indeed:
\begin{itemize}
\item if $\nu^2+\lambda^2 \leq 1$, then $2\nu\lambda +
    x(\nu^2+\lambda^2-1)$ is obviously positive since
    $x<0$, hence the denominator is negative.
\item if $\nu^2+\lambda^2 > 1$, then one has the set of
    inequalities:
$$
2\nu\lambda -(\nu^2+\lambda^2-1)=(1+\nu-\lambda)(1+\lambda-\nu)\leq
2\nu\lambda + x(\nu^2+\lambda^2-1)\leq 2\nu \lambda\;,
$$
for any $x$ in $[-1,0)$. Since the term at the far left is
positive for allowed $\nu$ and $\lambda$, the negativity of
the denominator follows.
\end{itemize}
Therefore, we obtain the inequality $y_+(x) \leq y_-(x)$ for
any $x \in [-1,0)$ as long as they exist. Moreover, at $x=-1$,
we already noticed that $y_\pm(-1)$ exist and are in $(-1,1)$.
Hence, again from Section~\ref{sfunH} and by continuity,
$y_+(x)$ has to be above $-1$ for all $x$ in $[-1,0)$
and in the connected component of $\tilde \Omega_{\nu,\lambda}$
which contains $-1$. In conclusion, and from the fact that
$|y_+(x)|\geq |y_-(x)|$, we have $-1 < y_+(x)\leq y_-(x) \leq -
y_+(x)<1$ for such $x$, and the Lemma follows.
\qed

\bigskip

From what has been said so far we conclude:

\newcommand{\mygamm}{\gamma}
\begin{Theorem}
 \label{Tsing10X}
 For all admissible $(\nu,\lambda)$ let $\Sing$ denote the set
 $$\{H(x,y)=0\;,\  x\in [-1,1]\}
 \;.
 $$
Then
$$
 \Sing \cap\{y\not \in (-1,1)\} = \mygamm _+\cup \mygamm _-
 \;.
$$
 where  $\mygamm _\pm$ are two connected differentiable curves, with $\mygamm _-$ included within
 the region $\{y\in(-\infty,y_c)\}$, and $\mygamm _+$ included in the region $\{y\in (1,\infty)\}$, separating
 each of those regions  in two
 connected components, such that:
\begin{enumerate}
  \item for $\lambda+ \nu < 1$ the curves $\mygamm _\pm$
      stay away from the axes $x=\pm1$ and asymptote, both
      at plus and minus infinity, to the vertical lines
      $x=0$ and $x=x_*$;
  \item for $\lambda+ \nu >1$ each of the curves $\mygamm
      _\pm$ intersects the vertical line $\{x=1\}$
      precisely once, stays away from the vertical line
      $x=-1$, and asymptotes to the axis $ x=0 $ as $|y|$
      tends to infinity.

  \item for $\lambda+ \nu =1$ each of the curves $\mygamm
      _\pm$ stays away from the vertical line $x=-1$,
      asymptotes to the vertical lines $ x=0 $ and $x=1$ as
      $|y|$ tends to infinity, without intersecting
      $\{x=1\}$.
\end{enumerate}
\end{Theorem}

\begin{figure}
 \begin{center}
 \includegraphics[width=.6\textwidth]{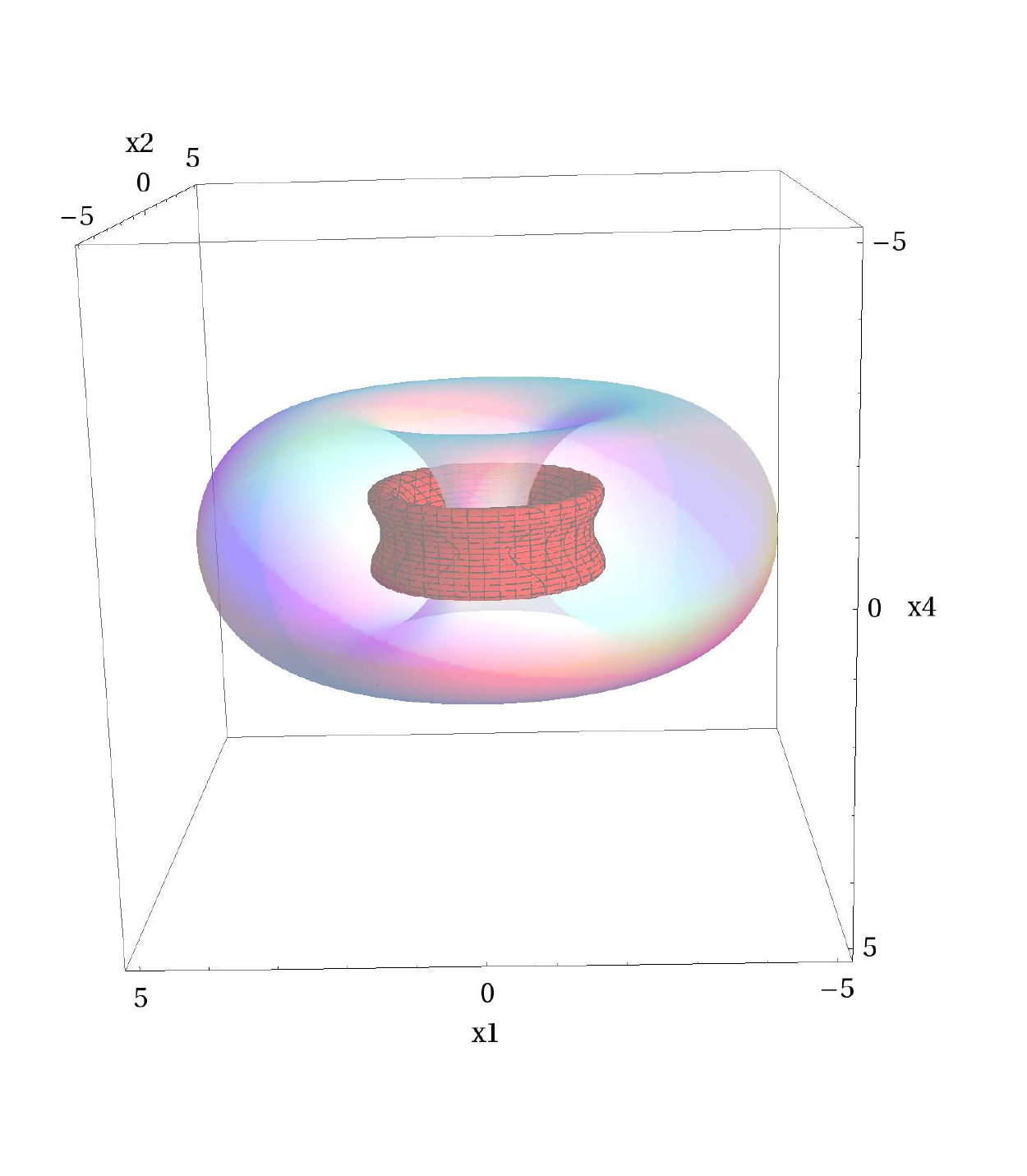}
\caption{\label{fig:singularity}
3-dimensional cut of the 4-dimensional set $\{t=const.\;, \ H(x,y)=0\}$ (red surface) for
$\lambda= 1.27$, $\nu= 0.38$, $k= 2.75$
in the asymptotically Euclidean coordinates $(x^1,x^2,x^3,x^4)$.
The corresponding event horizon (in translucid blue) has also been added to the picture.  }
\end{center}
\end{figure}

We finish this section by a short discussion of  the special
case $\lambda^2+\nu^2=1$, where we set
$$
 \nu= \cos \alpha \;, \quad \alpha \in (0,\pi/2)
 \;.
$$
We then have
$$
 H(x,z/\nu) =
-\left(x^2-1\right) y \sin (2 \alpha )-2 \cos ^2(\alpha ) \left(x y^2
   \sin (\alpha )+1\right)+2 x \sin (\alpha )+2
 \;,
$$
$$
 \tilde W(x) = \sin ^2(2 \alpha ) \left(\left(x^2+1\right)^2+4 x \sin (\alpha )\right)
 \;,
$$
and
$$
 \nu y_\pm = \frac{\pm\sqrt{ \left(x^2+1\right)^2+4 x \sin (\alpha
   ) }-x^2+1}{2 x}
   \;.
$$

\begin{figure}[h]
 \begin{center}
 \includegraphics[width=.4\textwidth]{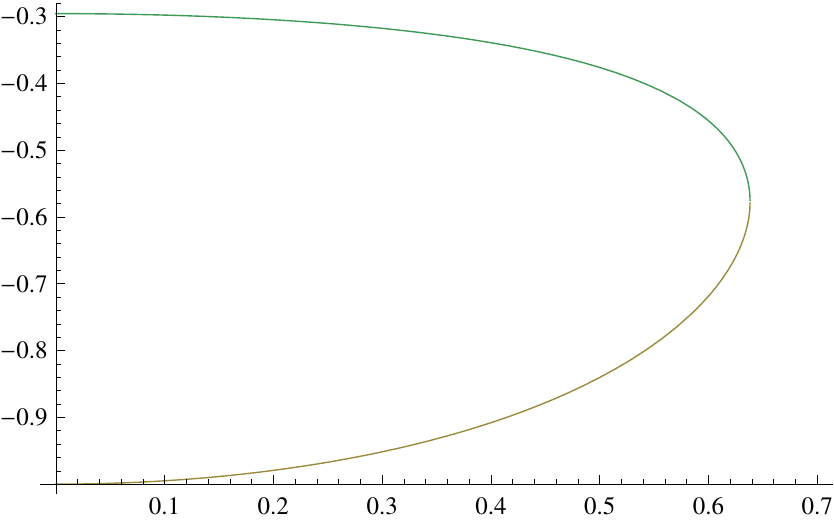}
\caption{\label{Circleroots}
The two real roots of  $\tilde W$  as functions of $\nu$ when $\lambda^2+\nu^2 = 1$,
whenever they exist. Admissible $\nu$'s belong to the interval
$[0,\sqrt 5 -2\approx 0.236]$,
the upper bound
being determined by the intersection of the circle with the lower limit $\lambda= 2 \sqrt \nu$, therefore
the double root at $\nu=\sqrt{11/27}\approx 0.638$
corresponds to a non-admissible value of $(\nu,\lambda)$.}
\end{center}
\end{figure}

We have already see that two of the roots are  imaginary when
$\lambda^2 + \nu^2 =1$. The two remaining ones are graphed as
functions of $\nu$ in Figure~\ref{Circleroots}.

\begin{figure}[h]
 \begin{center}
 \includegraphics[width=.6\textwidth]{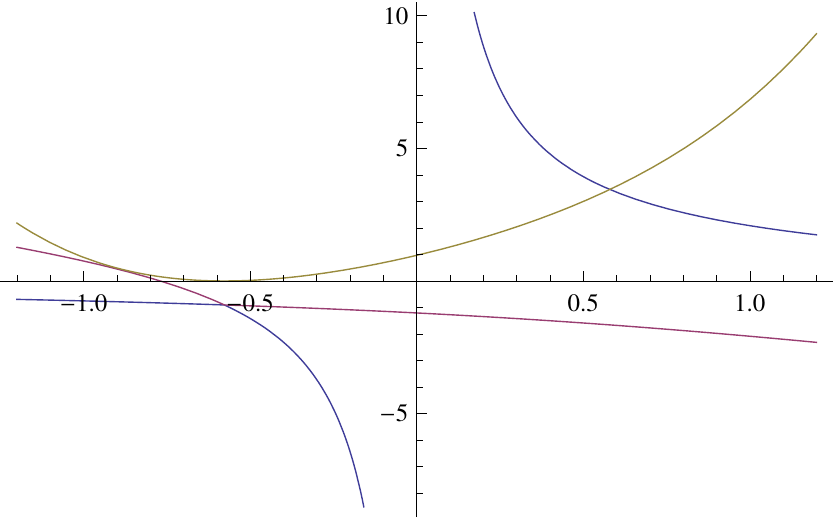}
\caption{\label{FNotAdmissible}
The polynomial $\tilde W$ (green) and the set $\{H(x,y)=0\}$ (the blue and the magenta curves), illustrating
the behaviour of the singular set at a (non-admissible) double root solution
with $\nu=\sqrt{11/27}$ and $\lambda=\sqrt{16/27}$.}
\end{center}
\end{figure}

 \subsection{The Kretschmann scalar}
 \label{SKretschmann}
Now, one expects existence of a curvature singularity on
$\{H(x,y)=0\}$. In order to test this, we consider the
Kretschmann scalar
$$
K=R_{abcd}R^{abcd}.
$$
An {\sc xAct} \cite{xAct} calculation gives%
\footnote{We are
grateful to Jos\'e M. Mart\'{\i}n-Garc\'{\i}a for his
assistance in this computation.}
\begin{eqnarray}
&
 \displaystyle
 K=\frac{3\lambda^2(\nu-1)^4(x-y)^4\Pi(x,y,\lambda,\nu)}{2k^4
 H(x,y)^6} \;,
 \label{eq:Kretschman}
\end{eqnarray}
where $\Pi(x,y,\lambda,\nu)$ is a huge
polynomial, still tractable by computer algebra manipulations.
Indeed we can write this polynomial in a shorter form if
we introduce the quantities
\begin{eqnarray}
&&\tilde{G}(x):=\frac{G(x)}{1-x^2}\;,\quad
\tilde{J}(x,y):=\frac{(x-y)(1-\nu )^2 J(x,y)}{2 k^2 \left(1-x^2\right) \left(1-y^2\right)
\lambda  \sqrt{\nu }}\;,\nonumber\\
&&\tilde{F}(x,y):=\frac{(-1 + \nu  x y)(-1 + \nu )^{2}(x -y)^{2}F(x,y)}{2 k^2 (-1 + y^{2})}
 \;.
\end{eqnarray}
These quantities are
polynomials in $x$, $y$, $\lambda$, $\nu$ as is easily checked and their explicit expressions are
\begin{eqnarray}
&&\tilde{G}(x)\equiv \nu  x^2+\lambda  x+1\;,\nonumber\\
&&\tilde{J}(x,y)\equiv\lambda ^2+2 (x+y) \nu  \lambda
   -\nu ^2-x y \nu(-\lambda ^2-\nu^2+1)+1\;,\nonumber\\
&&\tilde{F}(x,y)\equiv
\lambda  \nu x^{2}
(-1 + y^{2})(-x + y) (-\lambda ^2 +(1 + \nu)^{2}) (-1 + \nu )+H(x, y)\times\nonumber\\
&&\bigg(1+ \lambda  y +
\nu\Big(-1 + \lambda  x (-1 + \nu x^{2}-x y) + x (-1 + \nu)\big(y + x (-1 + \nu  x y)\big)\Big)\bigg)
 \;,
 \nonumber\\
\label{eq:poly-basis}
\end{eqnarray}

To shorten the final form of $\Pi(x,y,\lambda,\nu)$ one computes
a Gr\"obner basis from $\tilde{G}$, $H$, $\tilde{J}$ and
$\tilde{F}$ and then one uses this basis to show that
$\Pi(x,y,\lambda,\nu)$ must take the form
\begin{eqnarray}
&&\Pi(x,y,\lambda,\nu)=P_1(x,y,\lambda,\nu)H(x,y)+P_2(x,y,\lambda,\nu)
\tilde{G}(x)+\nonumber\\
&&+P_3(x,y,\lambda,\nu) \tilde{J}(x,y)+
 P_4(x,y,\lambda,\nu)\tilde{F}(x,y)
\end{eqnarray}
for some polynomials
$P_1$, $P_2$, $P_3$, $P_4$.
in $x$, $y$, $\lambda$, and $\nu$.
The Kretschmann scalar was computed by simplifying an
expression of 120Mb size for about 12 hours on a desktop
computer.

\begin{figure}[h]
\begin{center}
\includegraphics[width=.6\textwidth]{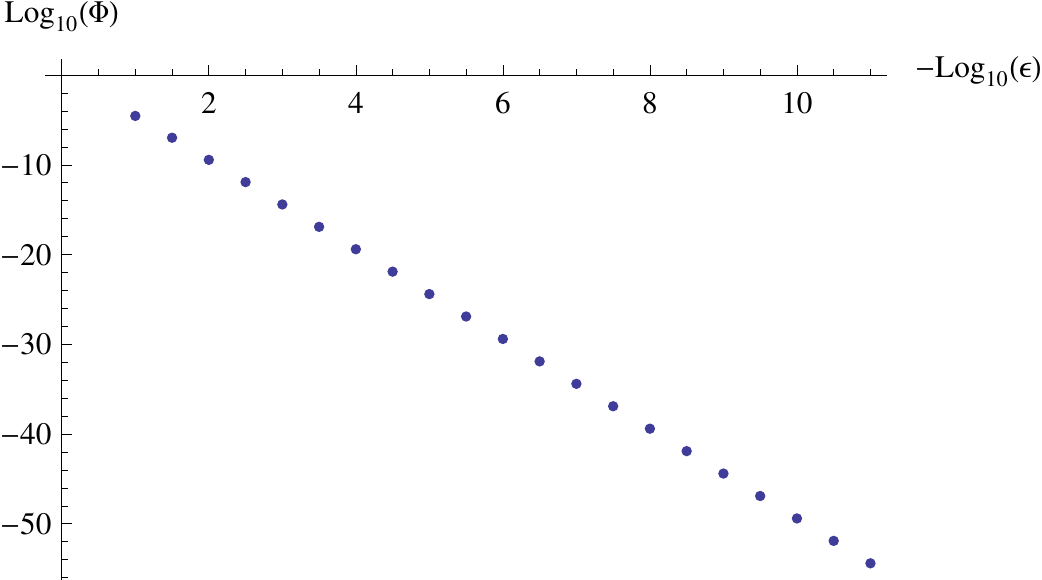}
\caption{\label{fig:PhiMin}
Logarithmic representation of the numerically calculated minimum of $\Phi(x,y,\lambda,\nu)$
on ${\mathcal N}_{\epsilon}$ (denoted by $\Phi_{\min}$) against the parameter $\epsilon$.
The graph strongly suggests that the Kretschmann scalar is singular everywhere on the set $\{H(x,y)=0\}$.
}
\end{center}
\end{figure}

The formula for the Kretschmann scalar shows that a curvature
singularity is present in those points where the polynomial
$H(x,y)$ vanishes (the set of these points is studied in detail
in subsection \ref{ssNature}) and $\Pi(x,y,\lambda,\nu)$ is
different from zero. We give below necessary and sufficient
conditions for this to happen. First of all, we define the
polynomial
$$
\Phi(x,y,\lambda,\nu):=H(x,y)^2+\Pi(x,y,\lambda,\nu)^2.
$$
Clearly, $\Phi(x,y,\lambda,\nu)$ is non-negative at any point.
Next we compute numerically the minimum of
$\Phi(x,y,\lambda,\nu)$ in the set
$$
{\mathcal N}_{\epsilon}:=\{-1\leq x\leq 1\;,2\sqrt{\nu}\leq\lambda< 1+\nu-\epsilon\;,
0\leq \nu\leq 1\;,|y|\geq 1\}
$$
for different values of $\epsilon$. The results of these
computations are represented in Figure~\ref{fig:PhiMin}.
The graph shown in this picture suggests that the polynomials
$H(x,y)$ and $\Pi(x,y,\lambda,\nu)$ do not have common zeros in
the set
$$
{\mathcal N}_{0}:=\{-1\leq x\leq 1\;,2\sqrt{\nu}\leq \lambda<1+\nu\;,
0\leq\nu\leq 1\;,|y|\geq 1\},
$$
and when we consider the closure of this set, the polynomials
$H(x,y)$ and $\Pi(x,y,\lambda,\nu)$ will have a common set of
zeros if and only if $\lambda=1+\nu$ holds. We have not been
able to obtain an analytical proof of the necessity of this
condition and we only rely on the numerical evidence shown in
Figure~\ref{fig:PhiMin} for the claim. However, the sufficiency
of the condition $\lambda=1+\nu$ is easily proven analytically
and to that end we just set $\lambda=1+\nu$ in the definition
of $H(x,y)$, obtaining
\begin{equation}
H(x,y,1+\nu,\nu)=-2 (x+1) (\nu +1) (y \nu +1) (x y \nu -1).
\end{equation}
Therefore, if $H(x,y,1+\nu,\nu)=0$ at some point in
$\overline{\mathcal{N}}_0$, then, either $x=-1$, $y=-1/\nu$ or
$y=1/(x\nu)$. We report the formulae
%
\begin{eqnarray}
&&\Pi(x,-1/\nu,1+\nu,\nu)=0\;,\quad \Pi(-1,y,1+\nu,\nu)=0\;,\\
&&\Pi(x,1/(x\nu),1+\nu,\nu)=\frac{64 (x+1)^{10} (\nu +1)^4 \left(x^3 \nu ^2+x^2 \nu
   -x \nu -1\right)^2}{x^8 \nu ^2}.\nonumber\\
&&
\end{eqnarray}
The last expression vanishes at some point in
$\overline{\mathcal{N}}_0$ when either $x=-1$ or
$$
x^3 \nu^2+x^2 \nu-x \nu -1=0\Rightarrow
x=-\frac{1}{\nu}\;,
x=\frac{1}{\sqrt{\nu}}\;,
x=-\frac{1}{\sqrt{\nu}}.
$$
These values of $x$ are in $\overline{\mathcal{N}}_0$ only if
$\nu=1$ (and thus $x=\pm 1$).

As a summary, we conclude that the polynomials
$\Pi(x,y,\lambda,\nu)$ and $H(x,y)$ both vanish at the
following points of $\overline{\mathcal{N}}_0$
%
$$
(x,-1/\nu,1+\nu,\nu)=0\;,\quad
(-1,y,1+\nu,\nu)=0\;,\quad
(1,1,2,1)=0 \;,
$$
and these are probably the only such points.

We conclude this section by a discussion of the behavior of the
Kretschmann scalar for large $|y|$. Both the numerator and the
denominator of the Kretschmann scalar are polynomials of order
twelve in $y$. The limit $|y|\to \infty$ of the Kretschmann
scalar is a rational function with denominator
\begin{eqnarray}
   {2 k^4 x^6 \nu ^3 \left(x
   \left(\lambda ^2+\nu ^2-1\right)+2 \lambda  \nu
   \right)^6}
   \;.
\label{eq:kretschmann-denominator}
\end{eqnarray}
The value of the numerator  at $x=0$ is
$$
192 \lambda ^4 (\nu -1)^4 \nu ^3 (\nu +1)^2
$$
The other zero of the denominator is located at
$$
x=-\frac{2 \lambda  \nu }{\lambda ^2+\nu ^2-1}
\;,
$$
and the value of the numerator there reads
$$
\frac{192 \lambda ^4 (\nu -1)^6 \nu ^3 (\lambda
   -\nu +1)^2 (-\lambda +\nu +1)^4 (\lambda +\nu
   -1)^2 (\lambda +\nu +1)^4 \left(\lambda ^2 (2
   \nu +1)+\nu ^2-1\right)^2}{\left(\lambda ^2+\nu
   ^2-1\right)^8}
   \;.
$$
Continuity implies that the Kretschmann scalar is singular on
$\{H(x,y)=0\}$ for all $y$ sufficiently large positive or
negative, except possibly at the zeros of the last factor
above, which for admissible parameters occur when
\begin{equation}
 \lambda=\frac{\sqrt{1-\nu ^2}}{\sqrt{2 \nu +1}}
 \;.
\label{eq:special-lambda}
\end{equation}
The limit when $|y|$ goes to infinity of  the Kretschmann
scalar for this value of $\lambda$ is a rational expression
whose denominator zero-set coincides with the zero-set of the
polynomial shown in (\ref{eq:kretschmann-denominator}) also
when $\lambda$ is set to the value of
(\ref{eq:special-lambda}). Therefore the Kretschmann scalar
will be singular too for all $y$ sufficiently large positive or
negative when $\lambda$ takes the special value
(\ref{eq:special-lambda}).

\subsection{The event horizon has $S^2\times S^1\times \R$
 topology}

In this Section we wish to prove that the set $\{y=y_h\}$ forms
the boundary of the d.o.c., both in the original domain of
definition of the metric of~\cite{PS}, and in our extension
here. The arguments are a (succinct) adaptation to the metric
at hand of those in~\cite[Section~4.1]{CC}, the reader is
referred to that last reference for more detailed arguments.

\begin{figure}
\begin{center}
%
%
%
%
 \psfrag{i1}{}
 \psfrag{e1}{}
 \psfrag{i3}{}
 \psfrag{d12}{}
 \psfrag{d11}{}
 \psfrag{scri2+}{}
 \psfrag{srci2-}{}
 \psfrag{i4}{}
 \psfrag{skraj2+}{}
 \psfrag{skraj2-}{}
 \psfrag{d1}{\Huge $I $}
 \psfrag{d2}{\Huge $I\!I $}
 \psfrag{d3}{\Huge $I\!I\!I $}
 \psfrag{d4}{\Huge $I\!V $}
 \psfrag{d5}{\Huge $V $}
 \psfrag{d6}{\Huge $V\!I $}
 \psfrag{d7}{\Huge $V\!I\!I $}
 \psfrag{d8}{\Huge $V\!I\!I \!I$}
 \psfrag{d9}{\Huge $I\!X $}
 \psfrag{d10}{\Huge $X $}
 \psfrag{r1}{\Large $y_c $}
 \psfrag{rzero}{$\!\!\!\!\!\!\!\!\!\!\!\!\!\!\!\!\!\!\!\!\!\! \{H(x,y) = 0\}\cup \{y=1\}$}
 \psfrag{r2}{\Large $y_h$}
 \psfrag{i2}{\Huge $i_0 $}
 \psfrag{skraj1+}{\Huge $\!\!\!\!\scrip\phantom{xx }$}
 \psfrag{skraj1-}{\Huge $\!\!\!\!\!\!\!\!\scrim\phantom{xxxx}$}
 \psfrag{skraj+}{\Huge $\!\!\!\!\scrip\phantom{xxx}$}
 \psfrag{skraj-}{\Huge $\!\!\!\!\!\!\!\!\scrim\phantom{xxxxx}$}
 \psfrag{scri+}{\Huge $\scrip$}
 \psfrag{srci-}{\Huge $\scrim$}
\resizebox{1.7in}{!}{\includegraphics[scale=.5]
{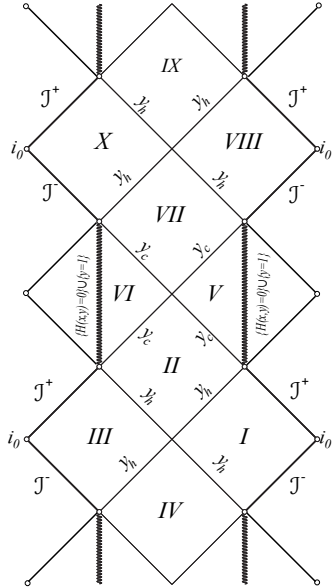}}
\caption{A visualisation of the global structure of an
extension obtained by iterating our procedure when $\lambda \ne 2 \sqrt \nu$, very similar to that of the non-extreme
Kerr space-time. The singular set in the (isometric) regions $V$ and $V\!I$   does \emph{not} separate this region in two. This is
\emph{neither} a conformal diagram, \emph{nor} is the manifold
a topological product of the diagram with some three
dimensional manifold. However, the picture depicts correctly
the causal relations between various regions, when the
light-cones are thought to have forty-five degrees slopes. We are grateful to M.~Eckstein for providing the figure.
 \label{F7IX.1}}
\end{center}
\end{figure}

We start by noting that
$$
 g(\nabla y, \nabla y) =
 g^{yy}= -\frac{(\nu-1)^2 G(y) (x-y)^2}{2 k^2 H(x,y)}
$$
is negative for $y_h<y < y_c$, hence $y$ is a time function
there. This implies that $y$ is monotonous along causal curves
through this region, and hence points for which $y_h<y < y_c$
lie within a black hole or a while hole region, unless some
topological identifications are introduced (for example,
consider the manifold consisting of the union of the closures
of the blocs $I$ to $V\!I\!I$ in Figure~\ref{F7IX.1}, in which
blocs $IV$ and $V\!I\!I$ are identified; in this space-time
there is no black hole region).

Next, consider the determinant of the three-by-three matrix of
scalar products of Killing vectors (compare
\eq{8IX.1}-\eq{8IX.3})
\bel{8IX.4} \det({g_{ij}})=\frac{4 k^4 G(x)G(y)}{(\nu
-1)^2(x-y)^4}
 \;.
\ee
This is negative in the region $\{y>y_h\}$, which implies that
neither the black hole event horizon $\partial J^-(\Mext)$, nor
the white hole event horizon $\partial J^-(\Mext)$ can
intersect this region. We conclude that $\{y=y_h\}$ forms the
boundary of the d.o.c., as claimed.

Keeping in mind that $x$ and $\varphi$ are coordinates on
$S^2$, and $\psi$ is a coordinate on $S^1$ as long as one stays
away from the rotation axes $y=\pm 1$, we  conclude that the
topology of cross-sections of the event horizon $\{y=y_h\}$, as
well as that of the Killing horizon $\{y=y_c\}$, is $S^2\times
S^1$.

\section{An extension across $y=-\infty$}
 \label{S11X09.1}

We have seen in Section~\ref{SKretschmann} that there always
exist  one or two intervals of $x$'s, say $I_a$, included in
the region for which $H(x,y)$ is positive, such that the metric
is defined for all $y\in(-\infty, -1]$. The metric is
Lorentzian throughout this region, which follows from
\eq{7IX.4}. It turns out that the metric can be analytically
extended on those intervals across ``the set $\{y=-\infty\}$"
to a Lorentzian metric by introducing a new variable
$$
 Y = -1/y
 \;.
$$
To see that this is the case, we start by noting that
$$
 g_{yy} dy^2 = \frac 1 {Y^4}  g_{yy} dY^2= y^4  g_{yy} dY^2
 \;.
$$
Since
$$
\lim_{y\rightarrow -\infty} y^4 g_{yy}= -\frac{2 k^2 x \left(x
 \left(\lambda ^2+\nu ^2-1\right)+2\lambda  \nu \right)} {(\nu
 -1)^2}
 \;,
$$
we see that the function
$$
 g_{YY}(x,Y):=  \big(y^4 g_{yy}\big)(x, y= 1/Y)
 \,
$$
defined for $x\in I_a$ and for $Y<1$, is a rational function of
$Y$ which analytically extends to negative values across the
set $Y=0$.

Likewise, the remaining metric functions analytically extend
across $Y=0$, except possibly at $x=0$ and $x=x_*$   defined by
\eq{28X.4}, which follows immediately from
\begin{eqnarray*}
\lim_{y\to-\infty} g_{tt}&=&\frac{2 \lambda  (\nu -1)}
{x\left(\lambda ^2+\nu^2-1\right)+2 \lambda\nu}-1
 \;,
\\
 \lim_{y\to-\infty}
g_{t\psi}&=&-\frac{2k\lambda\sqrt{(\nu
+1)^2-\lambda ^2} (x (\lambda +\nu-1)+2)}{(\lambda -\nu -1)
\left(x \left(\lambda ^2+\nu
   ^2-1\right)+2 \lambda  \nu
   \right)}
 \;,
\\
 \lim_{y\to-\infty}
g_{t\phi}&=&0
 \;,
\\
 \lim_{y\to-\infty}
g_{xx}&=&\frac{2 k^2 x \nu\left(x \left(\lambda
^2+\nu^2-1\right)+2 \lambda\nu\right)} {\left(x^2-1\right)(\nu
-1)^2 (x (x \nu +\lambda
   )+1)}
 \;,
\\
 \lim_{y\to-\infty}
g_{yy}&=&0
 \;,
\\
 \lim_{y\to-\infty}
g_{\psi\psi}&=&-\frac{ 2 k^2}{x (\nu -1)^2
   (-\lambda +\nu +1) \left(x \left(\lambda ^2+\nu
   ^2-1\right)+2 \lambda  \nu
   \right)} \times
\\
 &&
    \Bigg(x^4(\nu -1)\nu(-\lambda +\nu
+1) \left(\lambda ^2+\nu^2-1\right)
\\
 &&
+x^3 \lambda(\lambda -\nu
-1) \left(\lambda^2-(\nu -1)^2 (4\nu +1)\right)
\\
 &&
 +x^2
\bigg(\lambda ^3
   (\nu  (2 \nu -1)+1)+\lambda ^2 (3 \nu -1) (\nu  (2 \nu
   -5)+1)
\\
 &&
 -\lambda  (\nu -1)^2 (\nu +1)+\nu ^4-2 \nu
   ^2+1\bigg)
\\
 &&+x \lambda  \left(-\lambda ^3+\lambda ^2 (\nu
   +1)+5 \lambda  (\nu -1)^2+3 (\nu -1)^2 (\nu +1)\right)+2
   \lambda ^2 \nu  (-\lambda +\nu +1)\Bigg)
 \;,
\\
 \lim_{y\to-\infty}
g_{\psi\phi}&=&-\frac{2 k^2 \left(x^2-1\right) \lambda
\sqrt{\nu}}
{x (\nu-1)^2}
 \;,
\\
 \lim_{y\to-\infty}
g_{\phi\phi}&=&\frac{2 k^2 \left(x^2-1\right)\nu(x (\nu
-1)+\lambda)} {x (\nu -1)^2}
 \;,
\end{eqnarray*}
the remaining components of the metric being identically zero.

The signature remains Lorentzian, which can be seen by
calculating of the limit of the determinant of the metric in
the new coordinates, equal to  $y^4\det g_{\mu \nu}$, where
$g_{\mu \nu}$ refers to the original coordinates
$(t,x,y,\psi,\varphi)$:
$$
\lim_{y\rightarrow -\infty} y^4\det g_{\mu \nu}=-\frac{16 k^8 x^2 \nu ^2 \left(x \left(\lambda
^2+\nu^2-1\right)+2\lambda\nu \right)^2}{(\nu -1)^6}
 \;.
$$
We also note the limit
$$
\lim_{y\rightarrow -\infty} y g_{t\phi}=\frac{2 k
\left(x^2-1\right) \lambda  \sqrt{(\nu
   +1)^2-\lambda ^2}}{x \sqrt{\nu } \left(x \left(\lambda
   ^2+\nu ^2-1\right)+2 \lambda  \nu \right)}
   \;,
$$
which shows that $g_{t\phi}$ has a first order zero at $Y=0$.

In the region $Y<0$ one can introduce a new $y$ variable by the
formula
$$
 y = -1/Y
 \;,
$$
which brings the metric back to the original form
(\ref{eq:line-element1}), except that $y$ is positive now.

We have $H(x,-1/Y)= Y^{-2}\hat H(x,Y)$, where
$$
 \hat H(x,Y) =  2 \left(x^2-1\right) Y \lambda  \nu +Y^2 \left(2 x \lambda +\lambda
   ^2-\nu ^2+1\right)-x \nu  \left(x \left(\lambda ^2+\nu ^2-1\right)+2
   \lambda  \nu \right)
   \;.
$$
It follows that the singular set $\{H(x,y)=0\}$ meets the
hypersurface $Y=0$ at
$$
 x =0 \ \mbox{ and } \  x_*=-\frac{2 \lambda  \nu }{\lambda ^2+\nu ^2-1}
 \;.
$$
We have
$$
 \partial_Y \hat H(0,0)= -2\lambda \nu\;, \qquad  \partial_x \hat H(x_*,0) = 2 \lambda \nu^2
 \;,
$$
which shows that both branches of the singular set
$\{H(x,y)=0\}$ form a manifold when crossing $\{Y=0\}$.

The set $\{H(x,y)=0\}$ can be thought as being timelike, in the
following sense:
For any level set $\{H(x,y)=\epsilon\}$ the norm of the
gradient of $H$ is
$$
 g^{\mu\nu}\partial_\mu H \partial_\nu H = \frac{(\nu-1)^2 (x-y)^2}{2k^2 H(x,y)}
  \big( G(x) (\partial_x H)^2 - G(y) (\partial_y H)^2
 \big)
 \;.
$$
Both for $y<y_c$ and for $y>1$ the function $G(y)$ is negative,
which shows that the normal to the level sets of $H$ is
spacelike in that region. Note, however, that this discussion
leaves open the possibility of a null limiting hypersurface; we
have not attempt to quantify this any further.

A topological space will be called a \emph{pinched $S^1\times
S^2$}
if it is homeomorphic to the set obtained by rotating, around
the $z$-axis in $\R^3$, a disc lying in the $(x,z)$ plane and
tangent to the $z$ axis.
What has been said so far can now be summarized as follows:

\begin{Theorem}
 \label{T29X.1}
Suppose that
$$
 0<\nu<1\;,\quad  2\sqrt{\nu}\le \lambda < 1+\nu
 \;.
$$
Consider the set {\rm Coords} obtained by replacing the
coordinate $y\le y_c$
by $Y\in (-1,-1/y_c]$  and smoothly adjoining the hypersurface
$Y=0$, with the remaining coordinates $x\in [-1,1]$, $t\in \R$,
$\varphi, \psi \in [0,2\pi]$, $(x,Y)\ne (-1,1)$. Set
$$
 \hat H(x,Y) = Y^2 H(x,-1/Y)
 \;.
$$
Then the Pomeransky-Senkov metric extends analytically from the
region $0<Y<-1/y_c$ to an analytic Lorentzian vacuum metric on
$$
 \mbox{{\rm Coords}$\setminus \{\hat H(x,Y)\le 0\}$}
 \;,
$$
with an asymptotically flat region near $(x=-1,Y=1)$, and with
strong causality violation near $Y=-1$.  The boundaries $Y=-1$
and $x=\pm 1$ are rotation axes for suitable Killing vectors
with, however, a conical singularity at $Y=-1$ unless
\bel{28X.5}
 k = \frac  {  \sqrt{ (1+\nu)^2-\lambda^2}}{4 \lambda}
 \;.
\ee
The metric has a $C^2$--singularity at $\{\hat H(x, Y)=0\}$,
and, when viewed as a subset in space-time,  the singular set
$\{\hat H(x, Y)=0\}$  has precisely one component homeomorphic
to:
\begin{enumerate}
 \item     $\R\times S^1\times S^1 \times S^1$ when
     $\lambda+\nu <1$;
 \item   a pinched $\R \times S^1\times S^2$ when
     $\lambda+\nu = 1$;
 \item   $\R\times S^1\times S^2$ when $\lambda+\nu > 1$.
\end{enumerate}

\end{Theorem}

\section{The global structure}
 \label{S29X.1}

To pass from Theorem~\ref{T29X.1} to an analytic extension of
an asymptotically flat region of the PS solution, one needs to
keep in mind that every extension across a bifurcate Killing
horizon $y=y_h$ or $y=y_c$ as in Section~\ref{PSExt} leads to
three distinct new regions near that horizon. For $\lambda \ne
2 \sqrt{\nu}$ one is then led to a space-time with global
structure resembling somewhat that of the usual maximal
extension of a non-extreme Kerr black hole (cf.,
e.g.,~\cite{HE,CarterKerr}), with the following notable
differences: whereas the extended Kerr space-time contains
asymptotically Minkowskian regions with  naked singularities,
in our case the corresponding asymptotic regions are \emph{not}
asymptotically Minkowskian, and their exact nature  has  yet to
be analysed. Further, except when $k$ is appropriately chosen,
in our case the singular set has two components, corresponding
to $\hat H(x,Y)=0$, and to the conical singularity at $Y=-1$,
with the topology of the former depending upon the parameters.
\begin{figure}
\begin{center}
 \psfrag{d1}{\Huge $I $}
 \psfrag{d2}{\Huge $I\!I $}
 \psfrag{d5}{\Huge $I\!I\!I $}
 \psfrag{d7}{\Huge $I\!V $}
 \psfrag{d8}{\Huge $V $}
 \psfrag{r1}{\Large $y_0 $}
 \psfrag{rzero}{$\!\!\!\!\!\!\!\!\!\!\!\!\!\!\!\!\!\!\!\!\!\! \{H(x,y) = 0\}\cup \{y=1\}$}
 \psfrag{r2}{\Large $y_0$}
 \psfrag{i2}{\Huge $i_0 $}
 \psfrag{skraj1+}{\Huge $\!\!\!\!\scrip\phantom{xx }$}
 \psfrag{skraj1-}{\Huge $\!\!\!\!\!\!\!\!\scrim\phantom{xxxx}$}
 \psfrag{skraj+}{\Huge $\!\!\!\!\scrip\phantom{xxx}$}
 \psfrag{skraj-}{\Huge $\!\!\!\!\!\!\!\!\scrim\phantom{xxxxx}$}
 \psfrag{scri+}{\Huge $\scrip$}
 \psfrag{srci-}{\Huge $\scrim$}
\resizebox{1.7in}{!}{\includegraphics[scale=.5]
{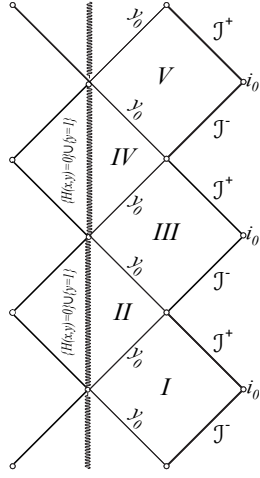}}
\caption{Graphical representation of the global structure of the extension
for the case $\lambda=2\sqrt{\nu}$ (extremal case) obtained by iteration of
the procedure explained in subsection \ref{xPSExt} (recall that $y_0=-1/\sqrt{\nu}$).
Similar considerations as in
Figure~\ref{F7IX.1} apply.}
\end{center}
\end{figure}

\appendix

\section{The Pomeransky-Senkov metric}
\label{sec:properties-metric}

We give here some formulae needed for explicit calculations
with the Pomeransky-Senkov (PS) metric \eq{eq:line-element1}.
The expression for $F(x,y)$ can be shortened when written in
terms of $H(x,y)$:
\begin{eqnarray}
&&F(x, y)=\frac{2 k^2 (-1 + y^{2})}{(-1 + \nu  x y)(-1 + \nu )^{2}(x -y)^{2}}\times\nonumber\\
&&\Bigg(\lambda  \nu x^{2}
(-1 + y^{2})(-x + y) (-\lambda ^2 +(1 + \nu)^{2}) (-1 + \nu )+H(x, y)\times\nonumber\\
&&\bigg(1+ \lambda  y +
\nu\left(-1 + \lambda  x (-1 + \nu x^{2}-x y) + x (-1 + \nu)\big(y + x (-1 + \nu  x y)\big)\right)\bigg)\Bigg)\nonumber\\
&&\label{eq:new-F}
\end{eqnarray}
The non-zero components of the metric tensor are
\bel{4IX.1}
\begin{array}{l}
 g_{tt}=-\frac{H(y,x)}{H(x,y)} \\
 g_{t\psi}=-\frac{   M(x,y) H(y,x)}{H(x,y)} \\
 g_{t\varphi}=-\frac{   P(x,y) H(y,x)}{H(x,y)} \\
 g_{xx}=\frac{2 k^2 H(x,y)}{(\nu -1)^2 G(x)(x-y)^2} \\
 g_{yy}=
   -\frac{2 k^2 H(x,y)}{(\nu -1)^2 G(y)
   (x-y)^2} \\
 g_{\psi\psi}=-\frac{H(y,x)^2    M(x,y)^2 - H(x,y) F(x,y)}{H(x,y)H(y,x)}
   \\
 g_{\varphi\psi}=
   -\frac{H(x,y)J(x,y)-
  H(y,x)^2 M(x,y)  P(x,y)}{H(x,y)H(y,x)} \\
 g_{\varphi\varphi}=
   -\frac{
    H(y,x)^2 P(x,y)^2+H(x,y)F(y,x)}{H(x,y)H(y,x)}
\end{array}
\ee

We have the identity
\begin{equation}
-\frac{F(x,y) F(y,x)+J(x,y)^2}{H(x,y)
   H(y,x)}=-\frac{4k^4 G(x) G(y)}{(\nu
   -1)^2 (x-y)^4}
    \;,
\label{jidentity}
\end{equation}
which allows us to rewrite the metric (\ref{eq:line-element1})
as
\begin{eqnarray}
 ds^2&=&\frac{(F(y,x)d\varphi-J(x,y)d\psi)^2}{H(y,x)F(y,x)}-
 \frac{4k^4G(x)G(y)H(x,y)d\psi^2}{(\nu-1)^2F(y,x)(x-y)^4}+\nonumber
\\
 &&+\frac{2k^2H(x,y)}{(1-\nu)^2(x-y)^2}
 \left(\frac{dx^2}{G(x)}-\frac{dy^2}{G(y)}\right)-
 \frac{H(y,x)(dt-\Omega)^2}{H(x,y)}
 \;.
  \phantom{xxxx}
 \label{eq:orthogonal-frame}
\end{eqnarray}
This provides an orthonormal frame, and is in particular
convenient for studying the signature of the metric.

There are other interesting identities fulfilled by the
rational functions involved in the definition of the PS metric.
To write them we need to add to the rational functions the
explicit dependencies on the variables $\lambda$, $\nu$ in
addition to $x$ and $y$. If $\mathcal{Q}(x,y,\lambda,\nu)$
denotes any of the rational functions $H(x,y,\lambda,\nu)$,
$F(x,y,\lambda,\nu)$ and $J(x,y,\lambda,\nu)$, then an explicit
computation shows the property
\begin{equation}
\mathcal{Q}(x,y,\lambda,\nu)=-x^2 y^2\nu^3\mathcal{Q}\left(\frac{1}{x},\frac{1}{y},\frac{\lambda}{\nu},\frac{1}{\nu}\right).
\label{eq:q-scaling}
\end{equation}
In addition we also have the relations
\begin{eqnarray}
&& M(x,y,\lambda,\nu)=M\left(\frac{1}{x},\frac{1}{y},\frac{\lambda}{\nu},\frac{1}{\nu}\right)\;,\quad
P(x,y,\lambda,\nu)=P\left(\frac{1}{x},\frac{1}{y},\frac{\lambda}{\nu},\frac{1}{\nu}\right)\;,\nonumber\\
&& G(x,\lambda,\nu)=-x^4G\left(\frac{1}{x},\frac{\lambda}{\nu},\frac{1}{\nu}\right)\;,
\label{eq:mpg-scaling}
\end{eqnarray}
which again are shown by expanding explicitly all the functions
involved. A straightforward consequence of these properties is
contained in the following result:

\begin{prop}
The PS family of metrics is invariant under the transformation
\begin{equation}
x\mapsto\frac{1}{x}\;,\quad y\mapsto\frac{1}{y}\;,\quad
\lambda\mapsto\frac{\lambda}{\nu}\;,\quad\nu\mapsto\frac{1}{\nu}\;,\quad
k\mapsto k\nu.
\label{eq:ps-symmetry}
\end{equation}
\label{prop:ps-invariance}
\end{prop}

A direct consequence of Proposition~\ref{prop:ps-invariance} is
that all the properties of the PS metric which have been proven
in this paper under the assumption that
$(x,y,\lambda,\nu)\in\Omega_0$ hold {\em mutatis mutandi} when
$(x,y,\lambda,\nu)$ belong to the set $\tilde{\Omega}_0$
defined by
$$
\tilde{\Omega}_0:=\{(x,y,\nu,\lambda) \in \mathbb{R}^4\ ;\
|x|\geq 1\;,\ -1\leq y\leq y_h\;,\ 1<\nu\;,\
2\sqrt{\nu}\leq \lambda < 1+ \nu \},
$$
which is the image of $\Omega_0$ under the map
(\ref{eq:ps-symmetry}). Since $\nu>1$ then we have $|y_h|<1$
(see Proposition \ref{prop:p-roots}) and therefore for points
in $\tilde{\Omega}_0$ we always have $|y|<1$. We can now
localise the asymptotically flat ends, horizons and so on by
just applying the map (\ref{eq:ps-symmetry}) to the
corresponding regions in $\Omega_0$ which were discussed in the
paper. In this way we find that the point
$(-1,-1,\lambda,\nu)\in\tilde{\Omega}_0$ corresponds to an
asymptotically flat end and
$(x,y_c,\lambda,\nu)\in\tilde{\Omega}_0$ is the event horizon
associated to the d.o.c. of this asymptotically flat end (note
that $y_c$ is mapped to $y_h$, $y_h$ is mapped to $y_c$ under
(\ref{eq:ps-symmetry}) and their absolute value is less than
unity when $\nu>1$).

The determinant of the metric reads
$$
\det(g_{\mu\nu})=
-\frac{4 k^4 H(x,y) \left(J(x,y)^2+F(x,y)
   F(y,x)\right)}{(\nu -1)^4 G(x) G(y) H(y,x)
   (x-y)^4}
    \;.
$$
Using the identity \eq{jidentity} one obtains
\bel{7IX.4} \det(g_{\mu\nu})=-\frac{16 k^8 H(x,y)^2}{(\nu -1)^6
(x-y)^8}
 \;.
\ee

The restriction of the metric to the hyperplanes
$\Span\{\partial_t,\partial_\varphi,\partial_\psi\}$ is
\begin{eqnarray}
 \nonumber
&&ds^2=-\frac{2 d\psi dt M(x,y)
H(y,x)}{H(x,y)}-\frac{2 d\varphi d t P(x,y)
H(y,x)}{H(x,y)}-
\frac{dt^2H(y,x)}{H(x,y)}-\\
 \nonumber
&&-d\psi^2
\left(\frac{H(y,x)M(x,y)^2}{H(x,y)}+
\frac{F(x,y)}{H(y,x)}
\right)-\\
 \nonumber
&&-2 d\psi d\varphi
\left(\frac{J(x,y)}{H(y,x)}+\frac{H(y,x)M(x,y)
P(x,y)}{H(x,y)}\right)-\\
&&-d\varphi^2
\left(\frac{H(y,x)P(x,y)^2}{H(x,y)}-
\frac{F(y,x)}{H(y,x)}\right)
 \;.
  \label{8IX.1}
\end{eqnarray}

The determinant of the restricted metric is given by
\bel{8IX.2}
 \det({g_{ij}})=\frac{J(x,y)^2+F(x,y) F(y,x)}{H(x,y)
   H(y,x)}\;,
\ee
which, after replacing all the functions simplifies to
\bel{8IX.3}
\det({g_{ij}})=\frac{4 k^4 G(x)G(y)}{(\nu
-1)^2(x-y)^4}
 \;.
\ee
Clearly $\det({g_{ij}})<0$ if $y_h<y<-1$ or $1<y$, where
$\{y=y_h\}$ is a Killing horizon with respect to the following
Killing vector
\begin{eqnarray*}
 {\xi}=\frac{\partial}{\partial t}+\frac{\sqrt{(\nu
   +1)^2-\lambda^2}}{2k(\lambda+\nu+1)}\frac{\partial}{\partial\psi}
+\frac{\left(\lambda ^2+(\nu
   -1)^2\right) \sqrt{\nu } \sqrt{(\nu +1)^2-\lambda ^2}}
{k \lambda (\lambda+\nu +1) \left(\lambda(\nu +1)-
\sqrt{\lambda ^2-4 \nu }(\nu -1)\right)}\frac{\partial}{\partial\varphi}
 \;,
\end{eqnarray*}
and $\{y=y_c\}$ is a Killing horizon with respect to the
Killing vector
\begin{eqnarray*}
 \tilde{\xi}=\frac{\partial}{\partial t}+\frac{\sqrt{(\nu
   +1)^2-\lambda^2}}{2k(\lambda+\nu+1)}\frac{\partial}{\partial\psi}
+\frac{\left(\lambda ^2+(\nu
   -1)^2\right) \sqrt{\nu } \sqrt{(\nu +1)^2-\lambda ^2}}
{k \lambda (\lambda+\nu +1) \left(\lambda(\nu +1)+
\sqrt{\lambda ^2-4 \nu }(\nu -1)\right)}\frac{\partial}{\partial\varphi}
 \;.
\end{eqnarray*}

From the expression for the restriction of the metric to
$\{t,\varphi ,\psi\}$ one easily gets the expression from the
restriction of the metric to $\{\varphi ,\psi\}$. We denote
this restricted metric by $g_{AB}$ and its determinant is
\begin{eqnarray}
&&\det g_{AB}=\nonumber\\
&&\frac{1}{H(x, y)(-1+\lambda -\nu)H(y, x)^{2}}\times\nonumber\\
&&\Bigg[F(y, x)
\Bigg\{F(x, y) H(x, y) (1-\lambda + \nu ) + \nonumber\\
&&\left.4 k^2 \lambda ^2
(1 + y)^{2}(1 + \lambda  + \nu )\Big(1 +\lambda -\nu  + 2 \nu  x-\nu  x y
\big(2 + x (-1 + \lambda  +\nu )\big)\Big)^{2}\Bigg\}
+\right.\nonumber\\
&&\left.+(1 -\lambda  + \nu)
\Bigg\{H(x, y)J(x,y)^{2} +\right.\nonumber\\
&&\left.+4 k^2 \lambda ^2 y\sqrt{\nu}(-1+x^{2}) (1 + \lambda  + \nu )\Bigg( y \sqrt{\nu}
F(x, y)(-1 + x^{2}) (-1 + \lambda-\nu )+\right.\nonumber\\
&&+ 2 J(x, y) (1 + y)\bigg(-1 -\lambda  + \nu  -2 \nu  x + \nu  x y
 \big(2 + x (-1 + \lambda  + \nu )\big)\bigg)\Bigg)\Bigg\}
 \Bigg]
  \;,
  \phantom{xxxxxx}
\label{eq:detg2}
\end{eqnarray}
which can be rewritten in the form
\begin{equation}
\det(g_{AB})=\frac{4 k^4(-1 + x^{2}) (1 + y)}
{(-1 + \lambda-\nu )(-1 + \nu)^{2}(x -y)^{4}H(x,y)}\Theta(x,y,\lambda,\nu).
\label{eq:detgphipsi}
\end{equation}
Clearly $\Theta(x,y,\lambda,\nu)$ is a rational function of
$x$, $y$, $\lambda$ and $\sqrt{\nu}$. However, a {\sc
Mathematica} calculation shows that $\Theta(x,y,\lambda,\nu)$
is a polynomial in $x$, $y$, $\lambda$ and ${\nu}$.

One can give an alternative form for $\det(g_{AB})$ if we use
the parameterization of the angular components of the metric
tensor given in (\ref{eq:detgphipsi}):
\begin{eqnarray}
&&\det(g_{AB})=\frac{-4 k^{4}\lambda ^{2}\nu(-1 + x^{2})^{2}(1 + y)^{2}
\Theta_{\phi\psi }(x, y)^{2}}{(-1 + \nu)^{4}H(x,y)^{2}(x - y)^{2}}-\nonumber\\
&&-\frac{4 k^{4}(1 + y)(-1 + x^{2})\Theta _{\phi\phi}(x, y)\Theta _{\psi \psi }(x, y)}{(-1 + \nu)^{4}(1-\lambda +\nu)H(x, y)^{2}(x-y)^{4}}
.\label{eq:detgAB2}
\end{eqnarray}
Comparing (\ref{eq:detgphipsi}) and (\ref{eq:detgAB2}) we
deduce the relation
\begin{eqnarray}
&&(\nu-1)^2 H(x, y)\Theta(x,y,\lambda,\nu)=\nonumber\\
&&=\lambda^{2}\nu(x-y)^{2}(1 + y)(1-x^{2})(1 +\nu -\lambda)
\Theta_{\phi \psi}(x, y)^{2}-\Theta_{\phi\phi}(x,y)\Theta_{\psi\psi}(x, y).\nonumber\\
&&\label{eq:thetatimesh}
\end{eqnarray}

The non-zero components of $g^{\mu\nu}$ read
$$
\begin{array}{c}
 g^{tt}=
   -\frac{M(x,y) (F(y,x) M(x,y)+2 J(x,y)
   P(x,y)) H(y,x)^2+H(x,y) J(x,y)^2+F(x,y)
   \left(F(y,x) H(x,y)-H(y,x)^2
   P(x,y)^2\right)}{H(y,x) \left(J(x,y)^2+F(x,y)
   F(y,x)\right)} \\
 g^{t\psi}=
   \frac{H(y,x) (F(y,x) M(x,y)+J(x,y)
   P(x,y))}{J(x,y)^2+F(x,y) F(y,x)} \\
 g^{t\varphi}=-
   \frac{H(y,x) (F(x,y) P(x,y)-J(x,y)
   M(x,y))}{J(x,y)^2+F(x,y) F(y,x)} \\
 g^{xx}=\frac{(\nu
   -1)^2 G(x) (x-y)^2}{2 k^2 H(x,y)} \\
 g^{yy}= -\frac{(\nu-1)^2 G(y) (x-y)^2}{2 k^2 H(x,y)} \\
 g^{\psi\psi}=
   -\frac{F(y,x) H(y,x)}{J(x,y)^2+F(x,y)
   F(y,x)} \\
 g^{\psi\varphi}=
   -\frac{H(y,x) J(x,y)}{J(x,y)^2+F(x,y)
   F(y,x)} \\
 g^{\varphi\varphi}=
   \frac{F(x,y) H(y,x)}{J(x,y)^2+F(x,y)
   F(y,x)}
\end{array}
$$
Alternative forms for some of the above can be obtained using
the identity (\ref{jidentity})
$$
\begin{array}{l}
 g^{tt}=
   \frac{(\nu-1)^2 \left(-\frac{4 k^4 G(x) G(y) H(x,y)^2}
   {16 H(y,x) (x-y)^4}-F(y,x)
   M(x,y)^2+F(x,y) P(x,y)^2-2 J(x,y)
   M(x,y) P(x,y)\right) (x-y)^4}{4 k^4
   G(x) G(y) H(x,y)} \\
 g^{t\psi}=
   \frac{(\nu-1)^2 (F(y,x) M(x,y)+J(x,y) P(x,y))
   (x-y)^4}{4 k^4 G(x) G(y) H(x,y)} \\
 g^{t\varphi}=
   -\frac{(\nu-1)^2 (F(x,y) P(x,y)-J(x,y) M(x,y))
   (x-y)^4}{4 k^4 G(x) G(y) H(x,y)} \\
 g^{\psi\psi}=
   -\frac{(\nu-1)^2 F(y,x) (x-y)^4}{4 k^4 G(x)
   G(y) H(x,y)} \\
 g^{\psi\varphi}=
   -\frac{(\nu-1)^2 J(x,y) (x-y)^4}{4 k^4 G(x)
   G(y) H(x,y)} \\
 g^{\varphi\varphi}=
   \frac{(\nu-1)^2 F(x,y) (x-y)^4}{4 k^4 G(x)
   G(y) H(x,y)}
\end{array}
$$

The $x-y$ part of the Pomeransky \& Senkov metric,
$$
ds^2_{x,y}=\frac{2k^2H(x,y)}{(\nu-1)^2(x-y)^2}\left(\frac{dx^2}{G(x)}-\frac{dy^2}{G(y)}\right)
 \;,
$$
can be written in a form which is conformally flat~\cite{PS} by
introducing new coordinates $\rho$, $z$ defined as
$$
\rho^{2}=-\frac{4 k^4 G(x)G(y)}{(-1 + \nu)^{2}\left(x -y\right)^{4}}\;,\quad
z=\frac{k^2 (1-x y) (\lambda  (x+y)+2 x y \nu+2)}{(1-\nu ) (x-y)^2}
 \;.
$$
The line element becomes
$$
ds^2_{\rho,z}=\Lambda(x,y)(d\rho^2+dz^2),
$$
where
\begin{eqnarray}
&&\Lambda(x,y)=\frac{2 (y-x) H(x,y)}{k^2 (x y \lambda +(\nu +1)
   (x+y)+\lambda )}\times\nonumber\\
&&\frac{1}{\left(\lambda ^2 \left(x^2
   \left(-\left(y^2-1\right)\right)+4 x y+y^2-1\right)+4 (\nu
   +1) \left(x^2 y^2 \nu +1\right)+4 \lambda  (x+y)
(x y \nu+1)\right)}.\nonumber\\
\label{eq:conf-factor}
\end{eqnarray}

Finally, we mention a property of one of the polynomial factors
in $G(x)$:

\begin{prop}
Let $p(\xi)\equiv \nu \xi^2+\lambda\xi+1$ and assume that
$2\sqrt{\nu}\le |\lambda|<1+\nu$. Then the real roots
$\xi_{-}\leq\xi_{+}$ of the polynomial $p(\xi)$ fulfill the
inequalities $|\xi_{\pm}|>1$ if $0<\nu<1$ and $|\xi_{\pm}|<1$
if $1<\nu$. \label{prop:p-roots}
\end{prop}

\Proof The roots of $p(\xi)$ are given by
$$
\xi_{\pm}=\frac{-\lambda\pm\sqrt{\lambda^2-4\nu}}{2\nu}\;,
$$
and therefore, these are real if and only if
$2\sqrt{\nu}\leq|\lambda|$, as assumed. The roots $\xi_{\pm}$
can be regarded as functions of $\lambda$ and $\nu$ and these
functions are continuous on the set ${\mathcal
Z}:=\{(\lambda,\nu):\nu>0,\;2\sqrt{\nu}\leq|\lambda|<1+\nu\}$.
Now the equations $\xi_{\pm}=1$, $\xi_{\pm}= -1$ admit as
respective solutions the values $\nu=-1-\lambda$ and
$\nu=\lambda-1$, and therefore no point lying in ${\mathcal Z}$
has the property that  $|\xi_{\pm}|=1$.  The conclusion is then
that for any $(\lambda,\nu)\in{\mathcal Z}$ we have that either
$|\xi_{\pm}|>1$ or $|\xi_{\pm}|<1$ and given the continuity of
$\xi_{\pm}$ on ${\mathcal Z}$ as a function of $\lambda$, $\nu$
only one of these alternatives will hold on each connected
component of ${\mathcal Z}$. These connected components are
\begin{eqnarray*}
&&{\mathcal Z}_1:=\{(\lambda,\nu):0<\nu<1,\;0<\lambda,\;2\sqrt{\nu}\leq|\lambda|<1+\nu\},\\
&&{\mathcal Z}_2:=\{(\lambda,\nu):0<\nu<1,\;0>\lambda,\;2\sqrt{\nu}\leq|\lambda|<1+\nu\},
\\
&&{\mathcal Z}_3:=\{(\lambda,\nu):1<\nu,\;0<\lambda,\;2\sqrt{\nu}\leq|\lambda|<1+\nu\},\\
&&{\mathcal Z}_4:=\{(\lambda,\nu):1<\nu,\;\lambda<0,\;2\sqrt{\nu}\leq|\lambda|<1+\nu\}.
\end{eqnarray*}
Again, by continuity, it is enough to check the values of
$\xi_{\pm}$ at particular points of ${\mathcal Z}_1$,
${\mathcal Z}_2$, ${\mathcal Z}_3$, ${\mathcal Z}_4$ to draw the desired
conclusions. Our explicit choices are as follows
\begin{eqnarray*}
&& \left(\lambda=1,\nu=\frac{1}{9}\right)\in\mathcal{Z}_1
\Rightarrow \left|\xi_{\pm}\left(1,\frac{1}{9}\right)\right|>1,\\
&& \left(\lambda=\frac{16}{5},\nu=\frac{9}{4}\right)\in\mathcal{Z}_3
\Rightarrow \left|\xi_{\pm}\left(\frac{16}{5},\frac{9}{4}\right)\right|<1,\\
&&\left(\lambda=-\frac{16}{5},\nu=\frac{9}{4}\right)\in\mathcal{Z}_4
\Rightarrow \left|\xi_{\pm}\left(-\frac{16}{5},\frac{9}{4}\right)\right|<1,\\
&& \left(\lambda=-1,\nu=\frac{1}{9}\right)\in\mathcal{Z}_2
\Rightarrow \left|\xi_{\pm}\left(-1,\frac{1}{9}\right)\right|>1,
\end{eqnarray*}
Therefore $|\xi_{\pm}|>1$ on $\mathcal{Z}_1$, $\mathcal{Z}_2$ and
$|\xi_{\pm}|<1$ on $\mathcal{Z}_3$, $\mathcal{Z}_4$, which finishes the proof.\qed

\section{Emparan-Reall limit of the Pomeransky \& Senkov
metric}
 \label{AERPS}

In this section we verify that the Emparan-Reall solutions are
a special case of the PS metrics. We take the Emparan-Reall
metric in the form given in \cite{EmparanReallReview},
\begin{equation}
ds^2=\frac{R^2 F(x)}{(x-y)^2}\left(\frac{dx^2}{G(x)}-\frac{dy^2}{G(y)}+
\frac{G(x)}{F(x)}d\varphi^2-
\frac{G(y)}{F(y)}d\psi^2\right)-\frac{F(y)}{F(x)}
\left(dt-\frac{CR(1+y)}{F(y)}d\psi\right)^2
 \;,
\label{emp-reall}
\end{equation}
where 
$$
F(z)=1+\lambda z,\quad G(z)=(1-z^2)(1+\nu z),\quad
C=\sqrt{\frac{\lambda(1+\lambda)(\lambda-\nu)}{1-\lambda}}
 \;.
$$
The parameters are assumed in that paper to range over 
$$
0<\nu\leq\lambda<1
 \;.
$$
However, the requirement that there are no struts imposes the
supplementary relation 
$$
\lambda=\frac{2\nu}{1+\nu^2}.
$$
The coordinates of (\ref{emp-reall}) are not the ones of the
original paper of Emparan \& Reall~\cite{EmparanReall}. If we
denote by
$\{\hat{t},\hat{x},\hat{y},\hat{\psi},\hat{\varphi}\}$ the
original coordinates of Emparan \& Reall, then we have the
relation
\begin{equation}
t=\hat{t},\quad x=\frac{\hat{\lambda}-\hat{x}}{-1+\hat{\lambda}\hat{x}},\quad
y=\frac{\hat{\lambda}-\hat{y}}{-1+\hat{\lambda}\hat{y}},\quad
\varphi=\frac{1-\hat{\lambda}\hat{\nu}}{\sqrt{1-\hat{\lambda}^2}}
\hat{\varphi},\quad
\psi=\frac{1-\hat{\lambda}\hat{\nu}}{\sqrt{1-\hat{\lambda}^2}}
\hat{\psi},
\label{hatchange}
\end{equation}
where
$$
\hat{\nu}=\frac{\nu-\lambda}{\lambda\nu-1},\quad \hat{\lambda}=\lambda,\quad
\nu=\frac{\hat{\nu}-\hat{\lambda}}{\hat{\lambda}\hat{\nu}-1}.
$$
The transformation (\ref{hatchange}) brings the metric
(\ref{emp-reall}) into the form (as in \cite{EmparanReall} with
hats on all the coordinates and functions and $R$ replaced by
$\hat{A}$)
\begin{eqnarray}
ds^2=-\frac{\hat{F}(\hat{x})}{\hat{F}(\hat{y})}(d\hat{t}+\hat{A}
\sqrt{\hat{\lambda}\hat{\nu}}(1+\hat{y})d\hat{\psi})^2+\nonumber\\
+\frac{\hat{A}^2}{(\hat{x}-\hat{y})^2}
\left(\hat{F}(\hat{y})^2\left(\frac{d\hat{x}^2}{\hat{G}(\hat{x})}+
\frac{\hat{G}(\hat{x})}{\hat{F}(\hat{x})}d\hat{\varphi}^2\right)
-\hat{F}(\hat{x})\left(\frac{\hat{F}(\hat{y})}{\hat{G}(\hat{y})}d\hat{y}^2+
\hat{G}(\hat{y})d\hat{\psi}^2\right)\right)\;,
\end{eqnarray}
where
$$
\hat{F}(z)=1-\hat{\lambda}z,\quad \hat{G}(z)=(1-z^2)(1-\hat{\nu}z),\quad
\hat{A}=-R\sqrt{\frac{(1-\hat{\lambda}\hat{\nu})}{1-\hat{\lambda}^2}}\;.
$$

To check the limit as $\nu\to 0$ of PS metrics, we start by
rewriting the Pomeransky \& Senkov solution in the form
\begin{eqnarray}
 ds^2 &= &\frac{Q\left(\frac{dx^2}{G(x)}-
\frac{dy^2}{G(y)}\right) H(x,y)}
{(x-y)^2}-2\frac{d\varphi d\psi J(x,y)}{H(y,x)}-
\frac{H(y,x)(dt+\Omega)^2}{H(x,y)}
 \nonumber
\\
 &&
 -\frac{d\psi^2F(x,y)}{H(y,x)}+
\frac{d\varphi^2 F(y,x)}{H(y,x)}
 \;.
\label{eq:metric-g}
\end{eqnarray}
Here $Q>0$ is a constant which can be eliminated by a rescaling
of the metric together with an appropriate rescaling of the
coordinates $t$, $\psi$, $\varphi$.  If we set
$Q=2k^2/(1-\nu)^2$ in (\ref{eq:metric-g}), we recover
(\ref{eq:line-element1}). One checks that the metric
(\ref{eq:metric-g}) reduces to the Emparan-Reall solution
(\ref{emp-reall}) if we set $\nu=0$ and $Q=2k^2/(1+\lambda^2)$,
and if
$$
\nu_e=\lambda\;,\quad
R_e=-\sqrt{2}k
$$
where  the Emparan-Reall independent parameters are denoted by
$\nu_e$ and $R_e$.

\def\polhk#1{\setbox0=\hbox{#1}{\ooalign{\hidewidth
  \lower1.5ex\hbox{`}\hidewidth\crcr\unhbox0}}}
  \def\polhk#1{\setbox0=\hbox{#1}{\ooalign{\hidewidth
  \lower1.5ex\hbox{`}\hidewidth\crcr\unhbox0}}} \def\cprime{$'$}
  \def\cprime{$'$} \def\cprime{$'$} \def\cprime{$'$}
\providecommand{\bysame}{\leavevmode\hbox to3em{\hrulefill}\thinspace}
\providecommand{\MR}{\relax\ifhmode\unskip\space\fi MR }
\providecommand{\MRhref}[2]{%
  \href{http://www.ams.org/mathscinet-getitem?mr=#1}{#2}
}
\providecommand{\href}[2]{#2}

\end{document}